\documentclass[twocolumn, reprint, amsmath,amssymb, superscriptaddress,floatfix]{revtex4-2}

\usepackage{graphicx}
\usepackage{dcolumn}
\usepackage{bm}
\usepackage{hyperref}
\usepackage[modulo]{lineno}

\def\*#1{\mathbf{#1}}
\def\^#1{\boldsymbol{#1}}
\def\br#1{\left[ #1 \right]}
\def\pr#1{\left( #1\right)}

\def\ang#1{\left\langle #1\right\rangle}
\newcommand{\iu}{\mathrm{i}\mkern1mu}

\newcommand{\pd}[2]{\dfrac{\partial {#1}}{\partial {#2}}}

\def\SIGen{Supplementary Note 1}
\def\SIPclassical{Supplementary Note 2}
\def\SIPt{Supplementary Note 3}
\def\SIPgaussian{Supplementary Note 4}
\def\SIPpairwise{Supplementary Note 5}
\def\SISKmodel{Supplementary Note 6}
\def\SIcomparison{Supplementary Note 7}

\makeatletter
\g@addto@macro{\UrlBreaks}{\UrlOrds}
\makeatother
\begin{document}

\preprint{APS/123-QED}

\title{A unifying framework for mean-field theories of asymmetric kinetic Ising systems}

\author{Miguel Aguilera}
 \email{sci@maguilera.net}
 \affiliation{IAS-Research Center for Life, Mind, and Society, Department of Logic and Philosophy of Science, University of the Basque Country, Donostia, Spain.
}
\affiliation{Department of Informatics \& Sussex Neuroscience, University of Sussex, Brighton, UK.
}
\affiliation{ISAAC Lab, Arag\'on Institute of Engineering Research, University of Zaragoza, Zaragoza, Spain.}

\author{S. Amin Moosavi}
\affiliation{Graduate School of Informatics, Kyoto University, Kyoto, Japan.}
\affiliation{Present address: Department of Neuroscience, Brown University, USA}
\author{Hideaki Shimazaki}%
\affiliation{Graduate School of Informatics, Kyoto University, Kyoto, Japan.}
\affiliation{Present address: Center for Human Nature, Artificial Intelligence, and Neuroscience (CHAIN), Hokkaido University, Japan}


\date{\today}

\begin{abstract} 
Kinetic Ising models are powerful tools for studying the non-equilibrium dynamics of complex systems. As their behavior is not tractable for large networks, many mean-field methods have been proposed for their analysis, each based on unique assumptions about the system's temporal evolution. This disparity of approaches makes it challenging to systematically advance mean-field methods beyond previous contributions. Here, we propose a unifying framework for mean-field theories of asymmetric kinetic Ising systems from an information geometry perspective. The framework is built on Plefka expansions of a system around a simplified model obtained by an orthogonal projection to a sub-manifold of tractable probability distributions. This view not only unifies previous methods but also allows us to develop novel methods that, in contrast with traditional approaches, preserve the system's correlations. We show that these new methods can outperform previous ones in predicting and assessing network properties near maximally fluctuating regimes. 
\end{abstract}

\maketitle

\section*{Introduction}

Advances in high-throughput data acquisition technologies for very large biological and social systems are providing unprecedented possibilities to investigate their complex, non-equilibrium dynamics. For example, optical recordings from genetically modified neural populations make it possible to simultaneously monitor activities of the whole neural network of behaving C. elegans \cite{nguyen_whole-brain_2016} and zebrafish \cite{ahrens_whole-brain_2013}, as well as thousands of neurons in the mouse visual cortex \cite{stringer_high-dimensional_2019}. Such networks generally exhibit out-of-equilibrium dynamics \cite{nicolis_self-organization_1977}, and are often found to self-organize near critical regimes at which their fluctuations are maximized \cite{tkacik_thermodynamics_2015, mora_dynamical_2015}. Evolution of such systems cannot be faithfully captured by methods assuming an asymptotic equilibrium state. Therefore, in general, there is a pressing demand for mathematical tools to study the dynamics of large-scale, non-equilibrium complex systems and to analyze high-dimensional data sets recorded from them.



The kinetic Ising model with asymmetric couplings is a prototypical model for studying such non-equilibrium dynamics in biological \cite{hertz_ising_2013, roudi_multi-neuronal_2015} and social systems \cite{bouchaud_crises_2013}. 
It is described as a discrete-time Markov chain of interacting binary units,
resembling the nonlinear dynamics of recurrently connected neurons. The model exhibits non-equilibrium behavior when couplings are asymmetric or when model parameters are subject to rapid changes, ruling out quasi-static processes. These conditions induce a time reversal asymmetry in dynamical trajectories, leading to positive entropy production (the second law of thermodynamics) as revealed by the fluctuation theorem  \cite{evans_probability_1993, jarzynski_nonequilibrium_1997, crooks_nonequilibrium_1998, crooks_entropy_1999, lebowitz_gallavotticohen-type_1999, ito_unified_2020} (see \cite{evans_fluctuation_2002, seifert_stochastic_2012} for reviews).
This time-asymmetry is characteristic of non-equilibrium systems as it can only be displayed by systems in which energy dissipation takes place \cite{gaspard_time_2007}.
In the case of symmetric connections and static parameters, the model converges to an equilibrium stationary state. Therefore, it is a generalization of its equilibrium counterpart known as the (equilibrium) Ising model \cite{salinas_ising_2001}. 

The forward Ising problem refers to calculating statistical properties of the model, such as mean activation rates (mean magnetizations of spins) and correlations, given the parameters of the model. In contrast, inference of the model parameters from data is called the inverse Ising problem \cite{ackley_learning_1985}.
In this regard, kinetic Ising models \cite{witoelar_neural_2011,donner_inverse_2017} and their equilibrium counterparts \cite{schneidman_weak_2006,cocco_neuronal_2009,shimazaki_state-space_2012} have become popular tools for modeling and analyzing biological and social systems. In addition, they capture memory retrieval dynamics in classical associative networks. Namely, they are equivalent to the Boltzmann machine, extensively used in machine learning applications \cite{ackley_learning_1985}.
Unfortunately, exact solutions of the forward and inverse problems often become computationally too expensive due to the combinatorial explosion of possible patterns in large networks or the  high volume of data, and applications of exact or sampling-based methods are limited in practice to around a hundred of neurons \cite{shimazaki_state-space_2012,tyrcha_effect_2013,tkacik_thermodynamics_2015}.
Therefore, analytical approximation methods are necessary for analysing large systems. In this endeavour, mean-field methods have emerged as powerful tools to track down otherwise intractable statistical quantities.

The standard mean-field approximations to study equilibrium Ising models are the classical naive mean-field (nMF) and  the more accurate Thouless-Anderson-Palmer (TAP) approximations \cite{thouless_solution_1977}. These methods have also been employed to solve the inverse Ising problem \cite{kappen_efficient_1998,roudi_statistical_2009, roudi_ising_2009, donner_approximate_2017}.
Plefka demonstrated that the nMF and TAP approximations for the equilibrium model can be derived using the power series expansion of the free energy around a model of independent spins, a method which is now referred to as the Plefka expansion \cite{plefka_convergence_1982}.
This expansion up to the first and second orders leads to the nMF and TAP mean-field approximations respectively. The Plefka expansion was later formalized by Tanaka and others in the framework of information geometry
\cite{tanaka_mean-field_1998, tanaka_theory_1999, bhattacharyya_information_2000, tanaka_information_2001, amari_information_2001}.

In non-equilibrium networks, however, the free energy is not directly defined, and therefore it is not obvious how to apply the Plefka expansions. Kappen and Spanjers \cite{kappen_mean_2000} proposed an information geometric approach to mean-field solutions of the asymmetric Ising model with asynchronous dynamics. They showed that their second-order approximation for an asymmetric model in the stationary state is equivalent to the TAP approximation for equilibrium models. Later, Roudi and Hertz derived TAP equations for non-stationary states using a Legendre transformation of the generating functional of the set of trajectories of the model \cite{roudi_dynamical_2011}.
Another study by Roudi and Hertz extended mean-field equations to provide expressions for the non-stationary delayed correlations assuming the presence of equal-time correlations at the previous step \cite{roudi_mean_2011}.
Yet another interesting method proposed by M\'ezard and Sakellariou approximates the local fields by a Gaussian distribution according to the central limit theorem, yielding more accurate results for fully asymmetric networks \cite{mezard_exact_2011}. This method was later extended to include correlations at the previous time-step, improving the results for symmetric couplings \cite{mahmoudi_generalized_2014}. 
More recently, Bachschmid-Romano et al. extended the path-integral methods in \cite{roudi_dynamical_2011} with Gaussian effective fields \cite{bachschmid-romano_variational_2016}, not only recovering \cite{mezard_exact_2011} for fully asymmetric networks but also proposing a method that better approximates mean rate dynamics by conserving autocorrelations of units.
Although many choices of mean-field methods are available, the diversity of methods and assumptions makes it challenging to advance systematically over previous contributions.


Here, we propose a unified approach for mean-field approximations of the Ising model. While our method is applicable to symmetric and equilibrium models, we focus for generality on asymmetric kinetic Ising models. Our approach is defined as a family of Plefka expansions in an information geometric space.
This approach allows us to unify and relate existing mean-field methods and provide expressions for other statistics of the systems such as pairwise correlations.
Furthermore, our approach can be extended beyond classical mean-field assumptions to propose novel approximations. Here, we introduce an approximation based on a pairwise model that better captures network correlations, and we show that it outperforms existing approximations of kinetic Ising models near a point of maximum fluctuations. We also provide a data-driven method to reconstruct and test if a system is near a phase transition by combining the forward and inverse Ising problems, and demonstrate that the proposed pairwise model more accurately estimates the system's fluctuations and its sensitivity to parameter changes. These results confirm that our unified framework is a useful tool to develop methods to analyze large-scale, non-equilibrium biological and social dynamics operating near critical regimes. In addition, since the methods are directly applicable to Boltzmann machine learning, the geometrical framework introduced here is relevant in machine learning applications.


The paper is organized as follows. First, we introduce the kinetic Ising model and its statistical properties of interest. Second, we introduce our framework for the Plefka approximation methods from a geometric perspective. To explain how it works, we derive the classical naive and TAP mean-field approximations under the proposed framework. Third, we show that our approach can unify other known mean-field approximation methods. We then propose a novel pairwise approximation under this framework. Finally, we compare different mean-field approximations in solving the forward and inverse Ising problems, as well as in performing the data-driven assessment of the system's sensitivity. The last section is devoted to discussion.  
\section*{Results}
\subsection*{The kinetic Ising model}

The kinetic Ising model is the least structured statistical model containing delayed pairwise interactions between its binary components (i.e., a maximum caliber model \cite{presse_principles_2013}). The system consists of $N$ interacting binary variables (down or up of Ising spins or inactive or active of neural units) $s_{i,t}\in\{-1,+1\},i=1,2,..,N$, evolving in discrete time steps $t$ with parallel dynamics. 
Given the configuration of spins at $t-1$, $\*{s}_{t-1}=\{s_{1,t-1},s_{2,t-1},\dots,s_{N,t-1}\}$, 
spins $\*{s}_{t}$ at time $t$ are conditionally independent random variables, updated as a  discrete-time Markov chain, following
\begin{align}
    P\left(\*s_{t}|\*{s}_{t-1}\right)=& \prod_i \frac{\mathrm{e}^{ s_{i,t} h_{i,t}}}{2 \cosh h_{i,t}},
 \label{eq:Ising}
    \\ h_{i,t}=& H_{i}+\sum_j J_{ij} s_{j,t-1}.
 \label{eq:Ising-field}
\end{align}
The parameters $\*{H}=\{H_{i}\}$ and $\*{J}=\{J_{ij}\}$ represent local external fields at each spin and couplings between pairs of spins respectively. When the couplings are asymmetric (i.e, $J_{ij} \neq J_{ji}$), the system is away from equilibrium because the process is irreversible with respect to time.
Given the probability mass function of the previous state $P(\*{s}_{t-1})$, the distribution of the current state is:
\begin{equation}
P\left(\*{s}_t\right) = \sum_{\*{s}_{t-1}} P\left(\*{s}_{t}|\*{s}_{t-1}\right) P\left(\*{s}_{t-1}\right).
\label{eq:marginalP}
\end{equation}
This marginal distribution $P(\*{s}_{t})$ is not factorized (except at $\*{J}=\*{0}$), but it rather exhibits a complex statistical structure, generally containing higher-order spin interactions. 
We can apply this equation recursively, e.g. decomposing $P\left(\*{s}_{t-1}\right)$ in terms of the distribution $P(\*s_{t-2})$, and trace the evolution of the system from the initial distribution $P(\*s_0)$.

In this article, we use variants of the Plefka expansion to calculate some statistical properties of the system. Namely, we investigate the average activation rates $\*{m}_{t}$, correlations between pairs of units (covariance function) $\*{C}_{t}$, and delayed correlations $\*{D}_{t}$ given by
\begin{align}
    m_{i,t} = &\sum_{\*{s}_{t}} s_{i,t}P(\*{s}_{t}),
    \label{eq:mean}
    \\C_{ik,t} = &\sum_{\*{s}_{t}} s_{i,t}s_{k,t}P(\*{s}_{t}) - m_{i,t} m_{k,t},
    \label{eq:correlation}
    \\D_{il,t} = &\sum_{\*{s}_{t},\*{s}_{t-1}} s_{i,t} s_{l, t-1} P(\*{s}_{t},\*{s}_{t-1}) - m_{i,t}m_{l, t-1}.
    \label{eq:D_correlation}
\end{align}
Note that $\*m_t$ and $\*D_t$ are sufficient statistics of the kinetic Ising model. Therefore, we will use them in solving the inverse Ising problem (see Methods). We additionally consider the equal-time correlations $\*C_t$ as they are commonly used to describe the systems, and are investigated by some of the mean-field approximations in the literature \cite{roudi_mean_2011}. Calculation of these expectation values is analytically intractable  and computationally very expensive for large networks, due to the combinatorial explosion of the number of possible states.
To reduce this computational cost, we approximate the marginal probability distributions (Eq.~\ref{eq:marginalP}) by the Plefka expansion method that utilizes an alternative, tractable distribution.

\subsection*{Geometrical approach to mean-field approximation}
Information geometry \cite{amari_information_2001, amari_methods_2007, amari_information_2016} provides clear geometrical understanding of information-theoretic measures and probabilistic models \cite{amari_information_1992,oizumi_unified_2016,ito_unified_2020}. Using the language of information geometry, we introduce our method for mean-field approximations of kinetic Ising systems.

\begin{figure}
 \begin{center}
 \includegraphics[width=8.66cm]{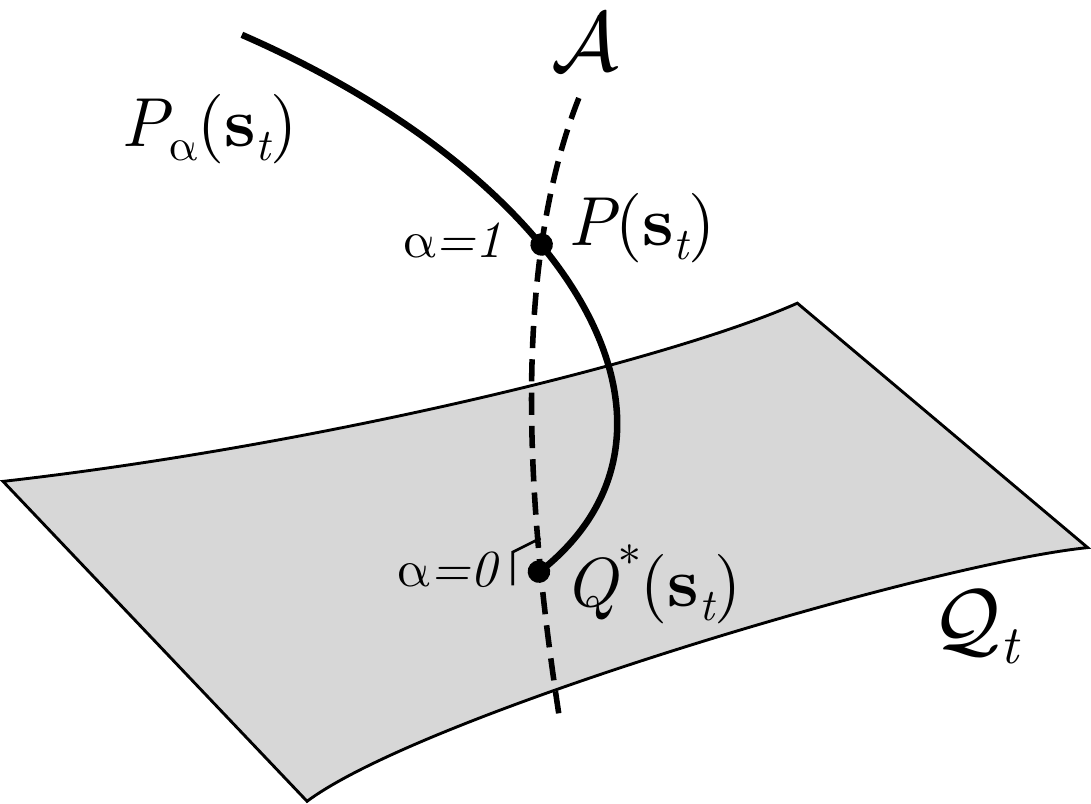}
 \end{center}
 \caption{\textbf{A geometric view of the approximations based on Plefka expansions}. 
 The point $P(\*{s}_{t})$ is the marginal distribution of a kinetic Ising model at time $t$.
 The submanifold $\mathcal{Q}_t$ is a set of tractable distributions, for example a manifold of independent models. The points in $\mathcal{A}$ correspond to a m-geodesic, that is a linear mixture of $P(\*{s}_{t})$ and  $Q^*(\*{s}_{t})$ on $\mathcal{Q}_t$, where for independent $\mathcal{Q}_{t}$ all points on $\mathcal{A}$ share the same mean values $\*{m}_t$. Geometrically, $\mathcal{A}$  constitutes the m-projection from $P(\*{s}_{t})$ to $\mathcal{Q}_t$, defining  $Q^*(\*{s}_{t})$  as the closest point in the submanifold $\mathcal{Q}_{t}$ to the point $P(\*s_{t})$ \cite{amari_information_1992}.
 The Plefka expansion is defined by expanding an $\alpha$-dependent distribution $P_{\alpha}(\*{s}_{t})$ that satisfies $P_{\alpha=0}(\*{s}_{t}) = Q^*(\*{s}_{t})$ and $P_{\alpha=1}(\*{s}_{t}) = P(\*{s}_{t})$.
 } 
 \label{fig:Plefka}
 \end{figure}

Let $\mathcal{P}_{t}$ be the manifold of probability distributions at time $t$ obtained from Eq.~\ref{eq:marginalP}. Each point on the manifold corresponds with a set of parameter values. The manifold $\mathcal{P}_{t}$ contains sub-manifolds $\mathcal{Q}_{t}$ of probability distributions with analytically tractable statistical properties (See Fig.~\ref{fig:Plefka}). 
We use this tractable manifold, i.e., a reference model, to approximate a target point $P(\*{s}_{t}|\*{H},\*{J})$ in the manifold $\mathcal{P}_{t}$ and its statistical properties $\*m_t,\*C_t,\*D_t$. 

The simplest sub-manifold $\mathcal{Q}_{t}$ is the manifold of independent models, used in classical mean-field approximations to compute average activation rates. Each point on this sub-manifold corresponds to a distribution
\begin{equation}
	  Q\left(\*{s}_{t} | \*{\Theta}_{t}\right) = \prod_{i} \frac{\mathrm{e}^{ s_{i,t} \Theta_{i,t}}}{2 \cosh \Theta_{i,t}},
\label{eq:Independent}	  
\end{equation}
where $\^\Theta_t = \{ \Theta_{i,t} \}$ is the vector of parameters that represents a point in $\mathcal{Q}_{t}$. This distribution does not include couplings between units, and its average activation rate is immediately given as $m_{i,t} = \tanh \Theta_{i,t}$. 

Our first goal is to find the average activation rates of the target distribution $P(\*{s}_{t} | \*{H},\*{J})$. 
It turns out that they can be obtained from the independent model $Q(\*{s}_{t}|\^\Theta_t)$ that minimizes the following Kullback-Leibler (KL) divergence from $P(\*{s}_{t})$:
\begin{align}
	D(P||Q) = &  \sum\limits_{\*{s}_{t}} P(\*{s}_{t})
	 \log \frac{P(\*{s}_{t})}{Q(\*{s}_{t})}.
	 \label{eq:distance}
\end{align}
The independent model $Q(\*{s}_{t}|\^\Theta_ t^*) (\equiv Q^*(\*{s}_{t}))$ that minimizes the KL divergence has activation rates $\*m_t$ identical to those of the target distribution $P(\*{s}_{t})$ \cite{kappen_mean_2000} because the minimizing points $\Theta^{*}_{i,t}$ satisfy (for $i=1,\ldots, N$)
\begin{align}
	\frac{\partial D(P||Q)}{\partial \Theta_{i,t}}\bigg\rvert_{\^\Theta_t=\^\Theta_t^*} = &  - \sum\limits_{\*{s}_{t}} s_{i,t} P(\*{s}_{t}) + \tanh \Theta^{*}_{i,t}  
	 \nonumber\\ = &  m_{i,t}^{Q^*} - m_{i,t}^{P} = 0,
	\label{eq:min-distance}
\end{align}
where $m_{i,t}^{P}$ and $m_{i,t}^{Q^*}$ are respectively expectation values of $s_{i,t}$ by $P(\*{s}_{t})$ and $Q(\*{s}_{t}|\*\Theta_t^*)$. 
As these values are equal, for the rest of the paper we will drop their superscripts and just write $m_{i,t}$ for simplicity. Note that the result of this approximation is indifferent of the system's correlations. Later in the paper we will consider approximations that take into account pairwise correlations.

From an information geometric point of view, given $\*{m}_{t}$ (or $\*\Theta_t^*$), we may consider a family of points defined as a linear mixture of $P(\*{s}_{t})$ and $Q(\*{s}_{t}|\*\Theta_t^*)$ for which $\*{m}_{t}$ is kept constant (the dashed line $\mathcal{A}$ in Fig.~\ref{fig:Plefka}).
This is known as an m-geodesic, and it is orthogonal to the e-flat manifold $\mathcal{Q}_{t}$, constituting an m-projection to this manifold 
\cite{amari_information_1992, amari_information_2001}. 
Thus, the previous search of $Q(\*{s}_{t}|\*\Theta_t^*)$ given $P(\*{s}_{t}|\*{H},\*{J})$ is equivalent to finding the orthogonal projection point from $P(\*{s}_{t}|\*{H},\*{J})$ to the manifold $\mathcal{Q}_{t}$ of independent models \cite{tanaka_information_2001,amari_information_2001}.



\subsection*{The Plefka expansion}
\label{sec:Approximation-by-Plefka-expansion}

Although the m-projection provides the exact and unique average activation rates, its calculation in practice requires the complete distribution $P(\*s_{t})$. 
In the Plefka expansion, we relax the constraints of the m-projection, and introduce another set of more tractable distributions that passes only through $P(\*{s}_{t}|\*{H},\*{J})$ and $Q(\*{s}_{t}|\*\Theta_t^*)$ (the solid line in Fig.~\ref{fig:Plefka}). This distribution is defined using a new conditional distribution introducing a parameter $\alpha$ that connects a distribution on the manifold $\mathcal{Q}_t$ with the original distribution $P(\*{s}_{t})$:
\begin{align}
	  P_{\alpha}(s_{i,t}|\*{s}_{t-1}) =&  \frac{\mathrm{e}^{ s_{i,t} {h}_{i,t}(\alpha)}}{2 \cosh {h}_{i,t}(\alpha)},
	\label{eq:ising-alpha}
	  \\ h_{i,t}(\alpha) 
	  =& (1-\alpha) \Theta_{i,t} + \alpha (H_{i} + \sum_j J_{ij} s_{j,t-1}).
\end{align}
At $\alpha=0$, $P_{\alpha=0}(\*{s}_{t}|\*{s}_{t-1}) = Q(\*{s}_{t}|\*\Theta_t)$, and $\alpha=1$ leads to $P_{\alpha=1}(\*{s}_{t}|\*{s}_{t-1}) =P(\*{s}_{t}|\*{s}_{t-1})$. Using this alternative conditional distribution $P_{\alpha}(s_{i,t}|\*{s}_{t-1})$, we construct an approximate marginal distribution $P_{\alpha}(\*{s}_{t})$. Consequently, expectation values with respect to $P_{\alpha}(\*{s}_{t})$ are functions of $\alpha$. We thus write the statistics of the approximate system as $\*{m}_{t}(\alpha)$, $\*{C}_{t}(\alpha)$, and $\*{D}_{t}(\alpha)$. 

The Plefka expansions of these statistics are defined as the Taylor series expansion of these functions around $\alpha=0$. In the case of the mean activation rate, the expansion up to the $n^{\rm }$th-order leads to:
\begin{equation}
	 \*{m}_t(\alpha) = \*{m}_t(\alpha=0) + \sum_{k=1}^n  \frac{\alpha^k}{k!}  \frac{\partial^k  \*{m}_t(\alpha=0)}{\partial \alpha^k} + \mathcal{O}(\alpha^{(n+1)}),
	\label{eq:taylor-mean}
\end{equation}
where $\mathcal{O}(\alpha^{(n+1)})$ stands for the residual error of the approximation of order $n+1$ and higher. For the $n^{\rm }$th-order approximation, we neglect the residual terms as $\mathcal{O}(\alpha^{(n+1)})\big\rvert_{\alpha=1} \approx 0$. Note that all coefficients of expansion are functions of $\*{\Theta}_{t}$. The mean-field approximation is computed by setting $\alpha=1$ and finding the value of $\*{\Theta}_{t}^*$ that satisfies Eq.~\ref{eq:taylor-mean}. Note that since the original marginal distribution is recovered at $\alpha=1$, the equality of Eq.~\ref{eq:min-distance} holds: $\*{m}_t(\alpha=1) = \*{m}_t(\alpha=0)$. Then, we have
\begin{equation}
     \sum_{k=1}^n  \frac{1}{k!}  \frac{\partial^k  \*{m}_t(\alpha=0)}{\partial \alpha^k}  = 0,
    \label{eq:taylor-mean-alpha-1}
\end{equation}
which should be solved with respect to the parameters $\*{\Theta}_{t}$. Since we neglected the terms higher than the $n$-th order, the solution may not lead to the exact projection, $Q(\*{s}_{t}|\*\Theta_t^*)$. In this study, we investigate the first ($n=1$) and second ($n=2$) order approximations. Moreover we can apply the same expansion to approximate the correlations $\*C_t$ and $\*D_t$, using Eq.~\ref{eq:ising-alpha}. 

What is the difference between this approach and other mean-field methods? 
Conventionally, naive mean-field approximations are obtained by minimizing $D(Q||P)$ as opposed to $D(P||Q)$ (Eq.~\ref{eq:distance}) \cite{saul_exploiting_1996, tanaka_information_2001}. 
This approach is typically used in variational inference to construct a tractable approximate posterior in machine learning problems. 
Following the Bogolyubov inequality, minimizing this divergence is equivalent to minimizing the variational free energy. Geometrically, it comprises an e-projection of $P(\*{s}_{t}|\*{H},\*{J})$ to the sub-manifold $\mathcal{Q}_t$, which does not result in $Q(\*{s}_{t}|\*\Theta_t^*)$. Namely, minimizing $D(Q||P)$, as well as minimization of other $\alpha$-divergences except for $D(P||Q)$, introduces a bias in the estimation of the mean-field approximation \cite{tanaka_information_2001, amari_information_2001}. In contrast, if we consider the m-projection point that minimizes $D(P||Q)$, we can approximate the exact value of $\*m_t$ using Eq.~\ref{eq:taylor-mean} up to an arbitrary order. 


In the subsequent sections we show that different approximations of the marginal distribution $P(\*s_{t})$ in Eq.~\ref{eq:marginalP} can be constructed by replacing $P(s_{i,\tau}|\*{s}_{\tau-1})$ with $P_{\alpha}(s_{i,\tau}|\*{s}_{\tau-1})$ for different pairs $i,\tau$ (here we will explore the cases of $\tau=t$ and $\tau=t-1$).
More generally, we show in \SIGen{} that this framework can be extended to a marginal path of arbitrary length $k$, $P(\*s_{t-k+1},\dots,\*s_{t})$. 
In addition, we are not restricted to manifolds of independent models. The independent model is adopted as a reference model to approximate the average activation rate, but one can also more accurately approximate correlations using this method. In this vein, we can extend our framework to use reference manifolds $\mathcal{Q}_{t-k+1:t}$ (of models $Q(\*s_{t-k+1},\dots,\*s_{t})$) that include interactions, e.g., pairwise couplings between elements at two different time points, to more accurately approximate the delayed correlations (see \SIGen{}).
By systematically defining these reference distributions, we will provide a unified approach to derive Plefka approximations of $\*{m}_{t}$, $\*{C}_{t}$, and $\*{D}_{t}$, including the one that utilizes a pairwise structure. 

\subsection*{Plefka[$t-1,t$]: Expansion around independent models at times $t-1$ and $t$}
Before elaborating different mean-field approximations, we demonstrate our method by deriving the known results of the classical nMF and TAP approximations for the kinetic Ising model \cite{kappen_mean_2000, roudi_dynamical_2011}. 
In order to derive these classical mean-field equations, we make a Plefka expansion around the points $\*{\Theta}^{*}_{t}$ and $\*{\Theta}^{*}_{t-1}$ that are respectively obtained by orthogonal projection to the independent manifolds $\mathcal{Q}_t$ and $\mathcal{Q}_{t-1}$, computed as in Eq.~\ref{eq:min-distance}. Note that assuming an approximation where previous distributions (e.g., $t-2, t-3,\dots$) are also independent yields exactly the same result. 
In this way, we derive the nMF and TAP equations of a model defined by a marginal probability distribution $P_\alpha^{[t-1:t]}$.
Using Eqs.~(\ref{eq:marginalP}) and (\ref{eq:ising-alpha}), we write
\begin{equation}
    P_\alpha^{[t-1:t]}(\*{s}_t)=\sum_{\substack{\*{s}_{t-1} \\ \*{s}_{t-2}}}  P_{\alpha}(\*{s}_{t}|\*{s}_{t-1})P_{\alpha}(\*{s}_{t-1}|\*{s}_{t-2})P(\*{s}_{t-2}),
\end{equation}
where $P_{\alpha=0}^{[t-1:t]}(\*{s}_{t})=Q(\*{s}_{t})$ and the original distribution is recovered for $P_{\alpha=1}^{[t-1:t]}(\*{s}_{t})=P(\*{s}_{t})$. 

Following Eq.~\ref{eq:taylor-mean-alpha-1}, for the first order approximation we have $\frac{\partial m_{i,t}(\alpha=0)}{\partial \alpha}=0$. Since the derivative of the first order moment is
\begin{align}
   \left. \frac{\partial  m_{i,t}(\alpha=0)}{\partial \alpha} \right. = & (1-m_{i,t}^2) (-\Theta_{i,t} + H_{i} + \sum_j J_{ij} m_{j,t-1}),
   \label{eq:dmda_first}
\end{align}
by solving the equation, we find $\Theta^{*}_{i,t}\approx H_{i}+\sum_j J_{ij} m_{j,t-1}$ that leads to the naive mean-field approximation:
\begin{equation}
	m_{i,t} \approx \tanh [ H_{i} + \sum_j J_{ij} m_{j,t-1} ].
    \label{eq:Plefka_m_nMF_t-1_t}
\end{equation}

We apply the same expansion to approximate the correlations, expanding $C_{ik,t}(\alpha)$ and $D_{il,t}(\alpha)$ around $\alpha=0$ up to the first order using $\Theta_{i,t}=\Theta^{*}_{i,t}$. Then we obtain 
\begin{align}
	C_{ik,t} \approx & 0, \quad i\neq k,
    \label{eq:Plefka_C_nMF_t-1_t}
	\\  D_{il,t} \approx & J_{il}  (1 - m_{i,t}^2)  (1-m_{l,t-1}^2).
    \label{eq:Plefka_D_nMF_t-1_t}
\end{align}
Detailed calculations are presented in \SIPclassical{}.

To obtain the second order approximation, we need to solve $\frac{\partial m_{i}(\alpha=0)}{\partial \alpha}+\frac{1}{2}\frac{\partial^{2} m_{i}(\alpha=0)}{\partial \alpha^{2}}=0$ from Eq.~\ref{eq:taylor-mean-alpha-1}. Here the second order derivative is given as 
\begin{align}
    \frac{\partial^{2} m_{i,t}(\alpha=0)}{\partial \alpha^{2}}
    &\approx -2 m_{i,t} (1-m_{i,t}^2) \sum_{j} J_{ij}^2 (1-m_{j,t-1}^2),
    \label{eq:stationary-second-o-derivative}
\end{align}
where terms of the order higher than quadratic were neglected (see \SIPclassical{} for further details). From these equations, we find $\Theta^{*}_{i,t} \approx H_{i} + \sum_j J_{ij} m_{j,t-1} - m_{i,t}\sum_{j} J_{ij}^2 (1-m_{j,t-1}^2)$ leading to the TAP equation:
\begin{equation}
    	m_{i,t} \approx \tanh [ H_{i} + \sum_j J_{ij} m_{j,t-1} - m_{i,t}\sum_{j} J_{ij}^2 (1-m_{j,t-1}^2) ].
    	\label{eq:Plefka_m_TAP_t-1_t}
\end{equation}
Having $\Theta^{*}_{i,t}$, we can incorporate TAP approximations of the correlations by expanding $C_{ik,t}(\alpha)$ and $D_{il,t}(\alpha)$ (see \SIPclassical{} for details) as:
\begin{align}
	 C_{ik,t} \approx & (1 - m_{i,t}^2)  (1 - m_{k,t}^2) \sum\limits_j J_{ij} J_{kj}(1-m_{j,t-1}^2), 
	 \nonumber\\ \quad i\neq k,
	 \label{eq:Plefka_C_TAP_t-1_t}
	\\  D_{il,t} \approx & J_{il} (1 - m_{i,t}^2)   (1-m_{l,t-1}^2)(1 + 2 J_{il} m_{i,t} m_{l,t-1}). 
	\label{eq:Plefka_D_TAP_t-1_t}
\end{align}
In these approximations, Eqs.~\ref{eq:Plefka_m_nMF_t-1_t} and \ref{eq:Plefka_m_TAP_t-1_t} of activation rates $\*m_t$ correspond to the classical nMF and TAP equations of the kinetic Ising model \cite{kappen_mean_2000, roudi_dynamical_2011}. The mean-field equations for the equal-time and delayed correlations (Eqs.~\ref{eq:Plefka_C_nMF_t-1_t},~\ref{eq:Plefka_D_nMF_t-1_t} and \ref{eq:Plefka_C_TAP_t-1_t},~\ref{eq:Plefka_D_TAP_t-1_t}) are novel contributions from applying the Plefka expansion to correlations. 

We note that, using the equations above, we can compute the approximate statistical properties of the system at $t$ ($\*{m}_{t},\*{C}_{t},\*{D}_{t}$) from $\*{m}_{t-1}$. Therefore, the system evolution is described by recursively computing $\*{m}_{t}$ from an initial state $\*{m}_{0}$ (for both transient and stationary dynamics), although approximation errors accumulate over the iterations. 
After we introduce a unified view of mean-field approximations in the subsequent sections, we will numerically examine approximation errors of these various methods in predicting statistical structure of the system.

\subsection*{Generalization of mean-field approximations}

In the previous section, we described a Plefka approximation that uses a model containing independent units at time $t-1$ and $t$ to construct a marginal probability distribution $P_{\alpha}^{[t-1:t]}(\*{s}_{t})$. 
This is, however, not the only possible choice of approximation. As we mentioned above, other approximations have been introduced in the literature. In \cite{roudi_mean_2011}, expressions are provided for the non-stationary delayed correlations $\*{D}_t$ as a function of $\*{C}_{t-1}$. In \cite{mezard_exact_2011}, an approximation is derived by assuming that units at state $\*{s}_{t-1}$ are independent while correlations of $\*{s}_{t}$ are preserved.

 \begin{figure*}
 \begin{center}
 \includegraphics[width=13cm]{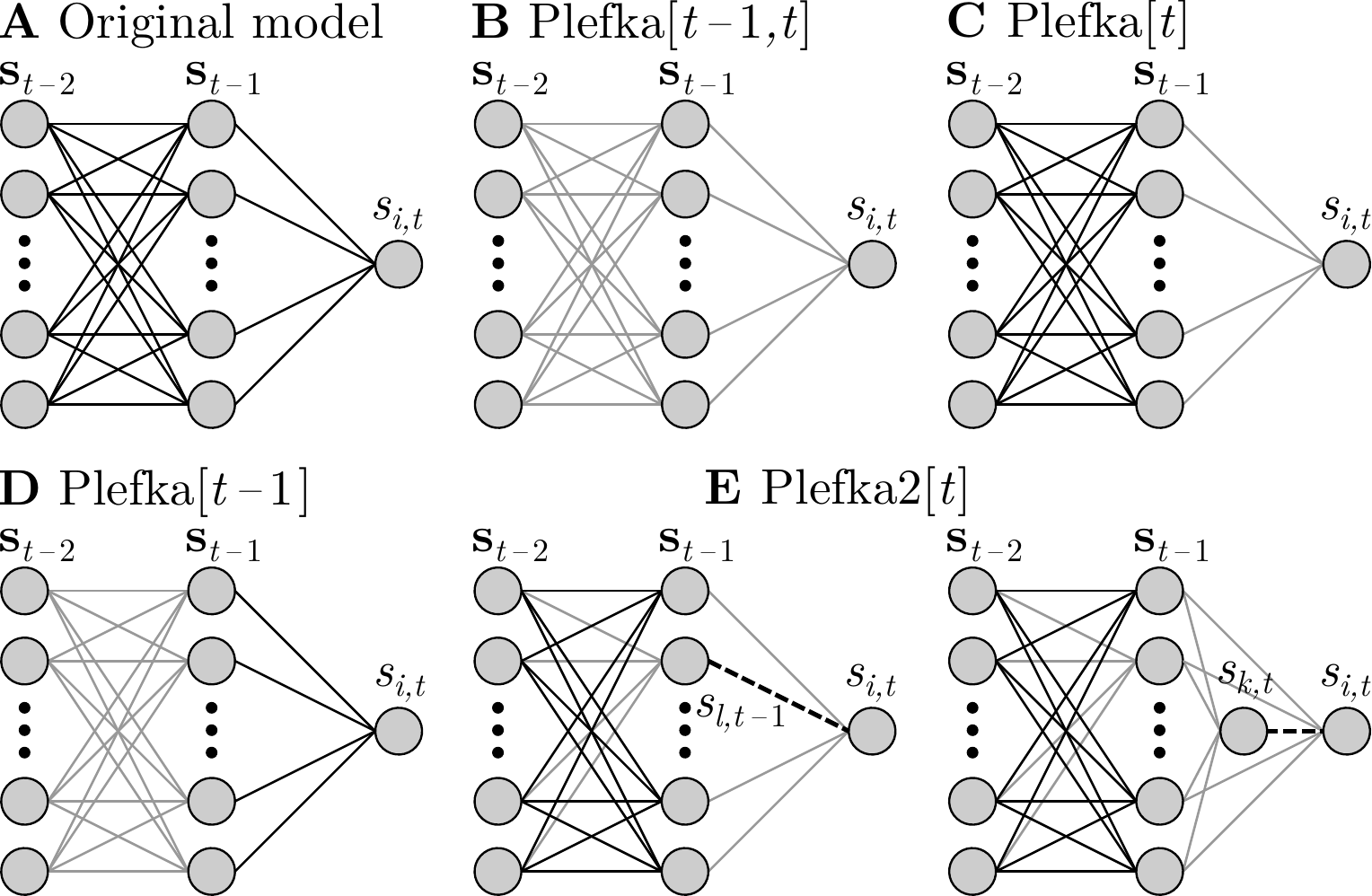}
 \end{center}
 \caption{\textbf{Unified mean-field framework}. Original model (A) and family of generalized Plefka expansions (B-E). Gray lines represent connections that are proportional to $\alpha$ and thus removed in the approximated model to perform the Plefka expansions, while solid black lines are conserved and dashed lines are free parameters. Plefka[$t-1,t$] (B) retrieves the classical mean-field equations \cite{kappen_mean_2000, roudi_dynamical_2011}. Plefka[$t$] (C) results in a novel method which preserves correlations of the system at $t-1$, incorporating equations similar to \cite{roudi_mean_2011}. Plefka[$t-1$] (D) assumes independent activity at t-1, and in its first order approximation reproduces the results in \cite{mezard_exact_2011}. Plefka2[$t$] (E) represents a novel pairwise approximation which performs better in approximating correlations.}
 \label{fig:Plefka-family}
 \end{figure*}
 
In the following sections, we show that various approximation methods, including those mentioned above, can be unified as Plefka expansions. Each method of the approximation corresponds to a specific choice of the sub-manifold $\mathcal{Q}_{t}$ at each time step. Fig.~\ref{fig:Plefka-family} show the corresponding sub-manifolds $\mathcal{Q}_{t-1:t}$ of possible approximations, where gray lines represent interactions that are affected by $\alpha$ in the Plefka expansion. 
The mean-field approximations in the previous section were obtained by using the model represented in Fig.~\ref{fig:Plefka-family}B, where the couplings at time $t-1$ and $t$ are affected by $\alpha$. Below, we present systematic applications of the Plefka expansions around other reference models in order to approximate the original distribution (Fig.~\ref{fig:Plefka-family}C-E). By doing so, we not only unify the previously reported mean-field approximations but also provide novel solutions that can provide more precise approximations than known methods.

\subsection*{Plefka[$t$]: Expansion around an independent model at time $t$}

For the Plefka[$t-1,t$] approximation, explained above, the system becomes independent for $\alpha=0$ at $t$ as well as $t-1$. This leads to approximations of $\*{m}_{t},\*{C}_{t},\*{D}_{t}$ being specified by $\*{m}_{t-1}$, while being independent of $\*{C}_{t-1}$ and $\*{D}_{t-1}$. In \cite{roudi_mean_2011}, the authors describe a mean-field approximation by performing new expansion over the classical nMF and TAP equations that takes into account previous correlations $\*{C}_{t-1}$. 
Here, our framework allows us to obtain similar results by considering only a Plefka expansion over manifold $\mathcal{Q}_{t}$ while assuming that we know the properties of $P(\*{s}_{t-1})$ (Fig.~\ref{fig:Plefka-family}C). Therefore, we denote this approximation as $P^{[t]}_{\alpha}$ and consider  
\begin{equation}
    P_{\alpha}^{[t]}(\*{s}_{t})=\sum_{\*{s}_{t-1}} P_{\alpha}(\*{s}_{t}|\*{s}_{t-1})P(\*{s}_{t-1}).
\end{equation}
In \SIPt{} we derive the equations for this approximation.
For the first order, we obtain
\begin{align}
	 m_{i,t} \approx & \tanh[ H_{i} + \sum_j J_{ij} m_{j,t-1} ],
	 \label{eq:Plefka_m_nMF_t}
	 \\ C_{ik,t} \approx & 0, \quad i\neq k,
	 \label{eq:Plefka_C_nMF_t}
	 \\  D_{il,t} \approx & (1 - m_{i,t}^2) \sum\limits_{j} J_{ij} C_{jl,t-1}.
	 \label{eq:Plefka_D_nMF_t}
\end{align}
Note that Eqs.~\ref{eq:Plefka_m_nMF_t} and \ref{eq:Plefka_C_nMF_t} are the same as the nMF Plefka[$t-1,t$] equations. Eq.~\ref{eq:Plefka_D_nMF_t} includes $\*{C}_{t-1}$, being exactly the same result obtained in \cite[Eq.~4]{roudi_mean_2011}.
The second order approximations leads to:
\begin{align}
	m_{i,t} \approx & \tanh [ H_{i} + \sum_j J_{ij} m_{j,t-1} - m_{i,t}\sum_{jl} J_{ij} J_{il} C_{jl,t-1} ],
	\\ C_{ik,t} \approx & (1 - m_{i,t}^2)  (1 - m_{k,t}^2) \sum_{jl} J_{ij} J_{kl}  C_{jl,t-1}, \quad i\neq k,
	\\  D_{il,t}
	\approx & (1 - m_{i,t}^2) \sum\limits_{j} J_{ij} C_{jl,t-1} (1
    + 2 J_{il} m_{i,t}  m_{l,t-1} ).
\end{align}
All update rules include the effect of $\*{C}_{t-1}$. We can see that if we use the covariance matrix of the independent model at $t-1$, we recover the results of $P_{\alpha}^{[t-1:t]}$ in the previous section. 
In contrast with \cite{roudi_mean_2011}, we provide a novel approximation method that depends on previous correlations using a single expansion (instead of two subsequent expansions), and additionally present approximated equal-time correlations.

\subsection*{Plefka[$t-1$]: Expansion around an independent model at time $t-1$}

In \cite{mezard_exact_2011}, a mean-field method is proposed by approximating the effective field $\*{h}_t$ as the sum of a large number of independent spins, approximated by a Gaussian distribution, yielding exact results for fully asymmetric networks in the thermodynamic limit. In our framework, we describe this approximation as an expansion around the projection point from $P(\*s_{t-1})$ to the submanifold $\mathcal{Q}_{t-1}$, using a model where only $\*{s}_{t-1}$ are independent (Fig.~\ref{fig:Plefka-family}D). In this case (see \SIPgaussian{}), the effective field $\*{h}_t$ at the submanifold is a sum of independent terms, which for large $N$ yields a Gaussian distribution. 

By defining
\begin{equation}
    P_{\alpha}^{[t-1]}(\*{s}_{t},\*{s}_{t-1})=\sum_{\*{s}_{t-2}} P(\*{s}_{t}|\*{s}_{t-1})P_{\alpha}(\*{s}_{t-1}|\*{s}_{t-2})P(\*{s}_{t-2}),
\end{equation}
we see that now the expansion is defined for the marginal distribution of the path $\{\*s_{t-1},\*s_{t}\}$  (see \SIGen{}).
The first order equations for this method are
\begin{align}
	 m_{i,t} \approx & \int \mathrm{D}_x \tanh [H_{i} + \sum_j J_{ij} m_{j,t-1} + x \sqrt{\Delta_{i,t}} ] ,
	 \\  C_{ik,t} \approx &  \int \mathrm{D}_{xy}^{\rho_{ik}} \tanh [H_{i} + \sum_j J_{ij} m_{j,t-1} 
	   + x \sqrt{\Delta_{i,t}} ]
       \nonumber\\ &\cdot\tanh [H_k + \sum_l J_{kl} m_{l,t-1} 
       + y \sqrt{\Delta_{j,t}} ]
       \nonumber\\&- m_{i,t}m_{k,t},
    \quad i\neq k,
	 \\ D_{il,t} \approx & \sum\limits_{j} J_{ij} C_{jl,t-1}  \int \mathrm{D}_x  \big( 1 
	 \nonumber\\ & - \tanh^2 [H_{i} + \sum_j J_{ij} m_{j,t-1} + x \sqrt{\Delta_{i,t}} ] \big).
\end{align}
Here we use $\mathrm{D}_x = \frac{\mathrm{d}x }{\sqrt{2 \pi}}\exp\pr{-\frac{1}{2}x^2}$, $\mathrm{D}_{xy}^{\rho_{ik}} = \frac{\mathrm{d}x \mathrm{d}y}{2 \pi \sqrt{1-\rho_{ik}^2}} \exp\pr{-\frac{1}{2}\frac{(x^2+y^2) - 2\rho_{ik}  x y}{1-\rho_{ik}^2} }$, $\Delta_{i,t} = \sum_j J_{ij}^2 (1- m_{j,t-1}^2)$ and $\rho_{xy} = \sum_j J_{ij}J_{kj} (1- m_{j,t-1}^2)/\sqrt{\Delta_{i,t}\Delta_{j,t}}$. Derivations are described in \SIPgaussian{}.
These results are exactly the same as those presented for $\*{m}_t,\*{D}_t$ in \cite{mezard_exact_2011}, adding an additional expression for $\*{C}_t$.
For this approximation, we do not consider the second order equations since they are computationally much more expensive than the other approximations.

\subsection*{Plefka2[$t$]: Expansion around a pairwise model}
The proposed framework is also a powerful tool to develop novel Plefka expansions. 
To make the expansions more accurately approximate target statistics, we can consider a reference manifold composed of multiple time steps while maintaining some of the parameters in the system (see \SIGen{}). Motivated by this idea, here we propose new methods that directly approximate pairwise activities of the units by choosing a reference manifold that preserves a coupling term.

Let us first consider the joint probability of any arbitrary pair of units at time $t-1$ and $t$ to compute the delayed correlations (Fig.~\ref{fig:Plefka-family}E, left). Namely, we consider the joint probability of spins $s_{i,t}$ and $s_{l,t-1}$:
\begin{align}
    P(s_{i,t},s_{l,t-1}) 
   = & \sum_{\substack{\*s_{\setminus l,t-1}\\ \*{s}_{t-2}}}  P(s_{i,t}|\*{s}_{t-1})  
 P(\*{s}_{t-1}|\*{s}_{t-2}) P(\*{s}_{t-2}),
\end{align}
with $\*s_{\setminus l,t-1}$ containing all elements of $\*s_{t-1}$ except $s_{l,t-1}$.
As a reference manifold $\mathcal{Q}_{t-1:t}$, we consider the dependency among only the units $i$ and $l$:
\begin{align}
	  Q(s_{i,t},s_{l,t-1}) = &  Q(s_{i,t}|s_{l,t-1})Q(s_{l,t-1})
	  \nonumber\\ =&
	  \frac{\mathrm{e}^{s_{i,t} \theta_{i,t}(s_{l,t-1})} }{2 \cosh \theta_{i,t}(s_{l,t-1})}
	  \frac{\mathrm{e}^{s_{l,t-1} \Theta_{l,t-1}}}{2 \cosh \Theta_{l,t-1}},
	  \label{eq:pairwise-manifold}
\end{align}
where $\theta_{i,t}(s_{l,t-1})=\Theta_{i,t} + \Delta_{il,t} s_{l,t-1}$. Orthogonal projection to $\mathcal{Q}_{t}$ is equivalent to minimizing the KL divergence $D(P||Q)$ with respect to the parameters:
\begin{align}
	 \frac{\partial D(P||Q)}{\partial \Theta_{i,t}}\bigg\rvert_{\begin{subarray}{l}\^\theta_t=\^\theta_t^*\\\^\Theta_{t-1}=\^\Theta_{t-1}^*\end{subarray}} =&   m_{i,t}^{Q^*} - m_{i,t}^P = 0,
	\\ \frac{\partial D(P||Q)}{\partial \Theta_{l,t-1}}\bigg\rvert_{\begin{subarray}{l}\^\theta_t=\^\theta_t^*\\\^\Theta_{t-1}=\^\Theta_{t-1}^*\end{subarray}} =&   m_{l,t-1}^{Q^*} - m_{l,t-1}^P = 0,
	\\ \frac{\partial D(P||Q)}{\partial \Delta_{il,t}}\bigg\rvert_{\begin{subarray}{l}\^\theta_t=\^\theta_t^*\\\^\Theta_{t-1}=\^\Theta_{t-1}^*\end{subarray}} =&  \langle s_{i,t} s_{l,t-1} \rangle_{(Q^*\cdot P)}- \langle s_{i,t} s_{l,t-1} \rangle_P 
	\nonumber\\&= 0.
	\label{eq:min-distance3}
\end{align}
with 
\begin{equation}
    \langle x \rangle_{(Q^*\cdot P)} = \sum_{\substack{s_{i,t}\\s_{l,t-1}}}  x \, Q(s_{i,t}|s_{l,t-1},\^\theta_t^*,\^\Theta_{t-1}^*) P(s_{l,t-1}).
\end{equation}
As in the previous approximations, $P(s_{i,t},s_{l,t-1})$ is connected to $Q(s_{i,t},s_{l,t-1}|\^\theta_t^*,\^\Theta_{t-1}^*)$ through an $\alpha$-dependent probability
\begin{align}
    P_{\alpha}^{2[t]}(s_{i,t}, s_{l,t-1}) =& \sum_{\substack{\*s_{\setminus l,t-1}\\\*{s}_{t-2}}} P_{\alpha}(s_{i,t}|\*{s}_{t-1})  P_{\alpha}(s_{l,t-1}|\*{s}_{t-2}) 
    \nonumber\\ &\cdot P(\*s_{\setminus l,t-1}|\*{s}_{t-2}) P(\*{s}_{t-2}),
\end{align}
with conditional probabilities given by
\begin{align}
	  P_\alpha(s_{i,t}| \*{s}_{t-1}) =&  \frac{\mathrm{e}^{ s_{i,t} h_{i,t}(\alpha) }}{2\cosh h_{i,t}(\alpha)}, \label{eq:ising-alpha_D1}
	 \\ h_{i,t}(\alpha) =&    (1-\alpha) \theta_{i,t}(s_{l,t-1}) 
	 \nonumber\\ & + \alpha  (H_{i} + \sum_j J_{ij} s_{j,t-1}), \nonumber
	\\ P_{\alpha}(s_{l,t-1}|\*{s}_{t-2}) =& \frac{\mathrm{e}^{ s_{l,t-1} {h}_{l,t-1}(\alpha)}}{2 \cosh {h}_{l,t-1}(\alpha)}, \label{eq:ising-alpha_D2}
	  \\ h_{l,t-1}(\alpha) = & (1-\alpha) \Theta_{l,t-1}
	  + \alpha (H_l + \sum_n J_{ln} s_{n,t-2}). \nonumber
\end{align}
As in the cases above, we can calculate the equations for the first and second order approximations (see \SIPpairwise{}).
Here, for the second order approximation (which is more accurate than the first order) we have that:
\begin{align}
    \theta_{i,t}^{*}(s_{l,t-1}) \approx&  H_{i} + \sum_{j} J_{ij} m_{j,t-1} 
    +J_{il} (s_{l,t-1} - m_{l,t-1})
    \nonumber\\ & +\big( \sum_{j \neq l, n} J_{ij}J_{ln} D_{jn,t-1} \big)(s_{l,t-1} - m_{l,t-1})
    \nonumber\\ &-\tanh \theta_{i,t}^{*}(s_{l,t-1})  \sum_{jn\neq l} J_{ij}J_{in} C_{jn,t-1}, 
    \label{eq:PlefkaD_1}
    \\ \Theta_{l,t-1}^* \approx& H_l + \sum_n J_{ln} m_{n,t-2} 
    \nonumber\\ & - m_{l,t-1} \sum_{mn} J_{lm}J_{ln} C_{mn,t-2},
    \label{eq:PlefkaD_2}
\end{align}
which directly leads to calculation of means and delayed correlations as: 
\begin{align}
	m_{i,t} \approx &\sum_{s_{l,t-1}} \tanh \theta_{i,t}^*(s_{l,t-1})  Q^*(s_{l,t-1}),
	\\ m_{l,t-1} \approx &\tanh \Theta_{l,t-1}^*,
	\\ D_{il,t} \approx &  \sum_{s_{l,t-1}} \tanh \theta_{i,t}^*(s_{l,t-1})  s_{l,t-1} Q^*(s_{l,t-1})
	\nonumber\\ & - m_{i,t} m_{l,t-1}.
\end{align}
These results are related to previous work \cite{bachschmid-romano_variational_2016} that included autocorrelations as one of the constraints to derive the Plefka approximation. Instead, here we provide a Plefka approximation that includes delayed correlations between any pair of units. 

To compute the above approximations, we need to know $\*C_{t-1}$ and $\*C_{t-2}$. Here, we provide similar pairwise Plefka approximations for the pairwise distribution at time $t$, $P(s_{i,t}, s_{k,t})$.
Since $s_{i,t}, s_{k,t}$ are conditionally independent, we can construct a model in which first $s_{k,t}$ is computed from $\*s_{t-1}$, and then $s_{i,t}$ is computed conditioned on $s_{k,t}, \*s_{t-1}$ (Fig.~\ref{fig:Plefka-family}E, right):
\begin{align}
    Q(s_{i,t},s_{k,t}) =& Q(s_{i,t}|s_{k,t}) Q(s_{k,t}),
    \\     P_{\alpha}^{2[t]}(s_{i,t}, s_{k,t}) =& \sum_{\*s_{t-1}} P_{\alpha}(s_{i,t}|s_{k,t},\*{s}_{t-1}) 
    \\ & \cdot P_{\alpha}(s_{k,t}|\*{s}_{t-1}) P(\*{s}_{t-1}), 
\end{align}
with conditional probabilities given by
\begin{align}
	  P_\alpha(s_{i,t}| s_{k,t},\*{s}_{t-1}) =&  \frac{\mathrm{e}^{ s_{i,t} h_{i,t}(\alpha) }}{2\cosh h_{i,t}(\alpha)}, \label{eq:ising-alpha_C1}
	 \\ h_{i,t}(\alpha) =&    (1-\alpha) \theta_{i,t}(s_{k,t}) 
	 \nonumber\\ & + \alpha  (H_{i} + \sum_j J_{ij} s_{j,t-1}), \nonumber
	\\ P_{\alpha}(s_{k,t}|\*{s}_{t-1}) =& \frac{\mathrm{e}^{ s_{k,t} {h}_{k,t}(\alpha)}}{2 \cosh {h}_{k,t}(\alpha)}, 
	\label{eq:ising-alpha_C2}
	  \\ h_{k,t}(\alpha) = & (1-\alpha) \Theta_{k,t}
	  + \alpha (H_k + \sum_l J_{kl} s_{l,t-1}). \nonumber
\end{align}
Note that $\theta_{i,t}$ is a function of $s_{k,t}$ that accounts for equal-time correlations between $s_{i,t}$ and $s_{k,t}$. Computed similarly to delayed correlations, the second order approximation yields (see \SIPpairwise{}):
\begin{align}
    \theta_{i,t}^*(s_{k,t}) \approx& H_{i} + \sum_{j} J_{ij} m_{j,t-1} 
    \nonumber\\ & + ( \sum_{jl} J_{ij}J_{kl} C_{jl,t-1}) (s_{k,t} - m_{k,t})
     \nonumber\\ &-\tanh \theta_{i,t}^*(s_{k,t}) \sum_{jl} J_{ij}J_{il} C_{jl,t-1},
    \label{eq:PlefkaD_1C}
    \\ \Theta_{k,t}^* \approx& H_k + \sum_l J_{kl} m_{l,t-1}
    \nonumber\\  &- m_{k,t} \sum_{jl} J_{kj}J_{kl} C_{jl,t-1}.
    \label{eq:PlefkaD_2C}
\end{align}
Using these equations, approximate equal-time correlations are given as
\begin{align}
	C_{ik,t} \approx & \sum_{s_{k,t}} \tanh \theta_{i,t}^*(s_{k,t}) s_{k,t} Q^*(s_{k,t}) -m_{i,t}m_{k,t}.
\end{align}
Note that the approximation of equal-time correlations may not be symmetric for $C_{ik,t}$ and $C_{ki,t}$. In the results of this paper we use the average of the two.

\begin{figure*}[t]
\centering{
\includegraphics[width=17.9cm]{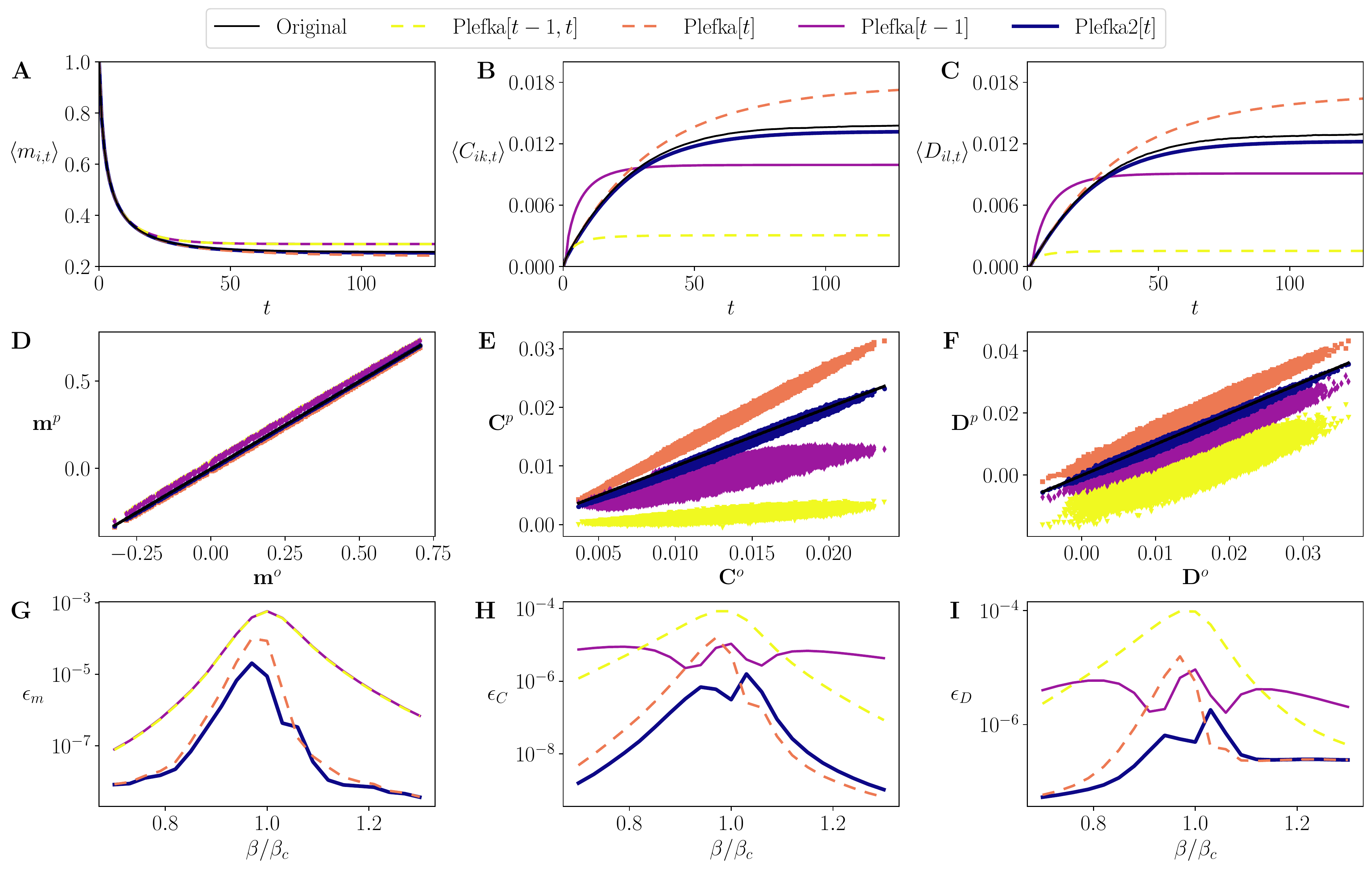}
}
\caption{\textbf{Forward Ising problem}. 
Top: Evolution of average activation rates (magnetizations) (A), equal-time correlations (B) and delayed correlations (C) found by different mean-field methods for $\beta=\beta_c$.
Middle: Comparison of the activation rates (D), equal-time correlations (E)  and delayed correlations (F) found by the different Plefka approximations (ordinate, $p$ superscript) with the original values (abscissa, $o$ superscript) for $\beta=\beta_c$ and $t=128$. Black lines represent the identity line.
Bottom: Average squared error of the magnetizations $\epsilon_{\*{m}}= \langle \langle (m^o_{i,t} - m^p_{i,t})^2 \rangle_{i} \rangle_t$ (G), equal-time correlations $\epsilon_{\*{C}}= \langle \langle (C^o_{ik,t} - C^p_{ik,t})^2 \rangle_{ik} \rangle_t $ (H), and delayed correlations $\epsilon_{\*{D}}= \langle \langle (D^o_{ik,t} - D^p_{ik,t})^2 \rangle_{il} \rangle_t$ (I) for 21 values of $\beta$ in the range $[0.7\beta_c, 1.3\beta_c]$.
} 
\label{fig:results-forward}
\end{figure*}

\subsection*{Comparison of the different approximations}
In the subsequent sections, we compare the family of Plefka approximation methods described above by testing their performance in the forward and inverse Ising problems. 
More specifically, we compare the second order approximations of Plefka[$t-1,t$] and Plefka[$t$], the first order approximation of Plefka[$t-1$], and the second order pairwise approximation of Plefka2[$t$].
We define an Ising model as an asymmetric version of the kinetic Sherrington-Kirkpatrick (SK) model, setting its parameters around the equivalent of a ferromagnetic phase transition in the equilibrium SK model. External fields $H_{i}$ are sampled from independent uniform distributions $\mathcal{U}(-\beta H_0, \beta H_0)$, $H_0=0.5$, whereas coupling terms $J_{ij}$ are sampled from independent Gaussian distributions $\mathcal{N}(\beta \frac{J_0}{N},\beta^2 \frac{J_\sigma^2}{N})$, $J_0=1, J_\sigma = 0.1$, where $\beta$ is a scaling parameter (i.e., an inverse temperature).

Generally, mean-field methods are suitable for approximating properties of systems with small fluctuations. However, there is evidence that many biological systems operate in critical, highly fluctuating regimes \cite{mora_dynamical_2015,tkacik_thermodynamics_2015}. In order to examine different approximations in such a biologically plausible yet challenging situation, we select the model parameters around a phase transition point displaying large fluctuations. 

To find such conditions, we employed path integral methods to solve the asymmetric SK model (\SISKmodel{}). 
We find that the stationary solution of the asymmetric model displays for our choice of parameters a non-equilibrium analogue of a critical point for a ferromagnetic phase transition, which takes place at $\beta_c \approx 1.1108$ in thermodynamic limit (see \SISKmodel{}, Supplementary Fig.~1). 
Note that the uniformly distributed bias terms $\*H$ shifts the phase transition point from $\beta=1$ obtained at $\*H=\*0$. By simulation of the finite size system, we confirmed that the maximum fluctuations in the model are found near the theoretical $\beta_c$, which shows maximal covariance values (see \SISKmodel{}, Supplementary Fig.~2).

Fluctuations of a system are generally expected to be maximized at a critical phase transition \cite{salinas_ising_2001}. In addition, entropy production (a signature of time irreversibility) has been suggested as an indicator of phase transitions. For example, it presents a peak at the transition point of a continuous phase transition in a non-equilibrium Curie-Weiss Ising model with oscillatory field \cite{zhang_critical_2016} and some instances of mean-field majority vote models \cite{crochik_entropy_2005, noa_entropy_2019}. We found that the entropy production of the kinetic Ising system is also maximized around $\beta_c$ (discussed later, see also Methods for its derivation). 

\subsection*{Forward Ising problem} 
We examine the performance of the different Plefka expansions in predicting the evolution of an asymmetric SK model of size $N=512$ with random $\*H$ and $\*J$.
To study the non-stationary transient dynamics of the model, we start from $\*s_0 = \*1$ (all elements set to 1 at $t=0$) and recursively update its state for $T=128$ steps. We repeated this stochastic simulation for $R=10^6$ trials for $21$ values of $\beta$ in the range $[0.7\beta_c, 1.3\beta_c]$, except for the reconstruction of the phase transition where we used $R=10^5$ and $201$ values of $\beta$ in the same range.
Using the $R$ samples, we computed the statistical moments and cumulants of the system, $\*m_{t}$, $\*C_{t}$, and $\*D_{t}$ at each time step. We then computed their averages over the system units, i.e., $\langle m_{i,t}\rangle_{i}$, $\langle C_{ik,t}\rangle_{ik}$ and $\langle D_{il,t}\rangle_{il}$, where the angle bracket denotes average over indices of its subscript. 

The black solid lines in Fig.~\ref{fig:results-forward}A,B,C display non-stationary dynamics of these averaged statistics from $t=0, \dots, 128$, simulated by the original model at $\beta = \beta_c$. In comparison, color lines display these statistics predicted by the family of Plefka approximations that are recursively computed using the obtained equations, starting from the initial state $\*m_0 = \*1$, $\*C_0 = \*0$ and $\*D_0 = \*0$.
We observe that although the recursive application of all the approximation methods provides good predictions for the transient dynamics of the mean activation rates $\*m_t$ until its convergence (Fig.~\ref{fig:results-forward}A), the predictions using Plefka[$t$] and especially the proposed Plefka2[$t$] approximations are closer to the true dynamics than the others. Evolution of the mean equal-time and time-delayed correlations $\*C_t, \*D_t$ is precisely captured only by our new method Plefka2[$t$]. In contrast, Plefka[$t$] overestimates correlations while Plefka[$t-1$] and Plefka[$t-1,t$] underestimate correlations. 

Performance of the methods in predicting individual activation rates and correlations are displayed in Fig.~\ref{fig:results-forward}D, E, F by comparing vectors $\*m_{t}$, $\*C_{t}$ and $\*D_{t}$ at the last time step ($t=128$) of the original model ($o$ superscript) and those of the Plefka approximations ($p$ superscript). For activation rates $\*m_t$, the proposed Plefka2[$t$] and Plefka[$t$] perform slightly better than the others (see also Fig.~\ref{fig:results-forward}A). While being overestimated by Plefka[$t$], underestimated moderately by Plefka[$t-1$] and significantly by Plefka[$t-1,t$], equal-time and time-delayed correlations $\*{C}_t, \*{D}_t$ are best predicted by Plefka2[$t$] (Fig.~\ref{fig:results-forward}E,F). 

The above results are obtained at the critical $\beta=\beta_c$, intuitively the most challenging point for mean-field approximations. In order to further show that our novel approximation Plefka2[$t$] systematically outperforms the others in a wider parameter range, we repeated the analysis for different inverse temperatures $\beta$ (the same random parameters are applied for all $\beta$). Fig.~\ref{fig:results-forward}G,H,U respectively show the averaged squared errors (averaged over time and units) of the activation rates $\epsilon_{\*{m}}$, equal-time correlations $\epsilon_{\*{C}}$ and delayed correlations $\epsilon_{\*{D}}$ between the original model and approximations, averaged over units and time for values of $\beta$ in the range $[0.7\beta_c, 1.3\beta_c]$. Figs.~\ref{fig:results-forward}G,H,I show that Plefka2[$t$] outperforms the other methods in computing $\*m_t,\*C_t,\*D_t$ (with the exception of a certain region of $\beta>\beta_c$ in which Plefka[$t$] is slightly better), yielding consistently a low error bound for all values of $\beta$. Errors of these approximations are smaller when the system is away from $\beta_c$. 

\begin{figure*}
\centering{
 \includegraphics[width=11.cm]{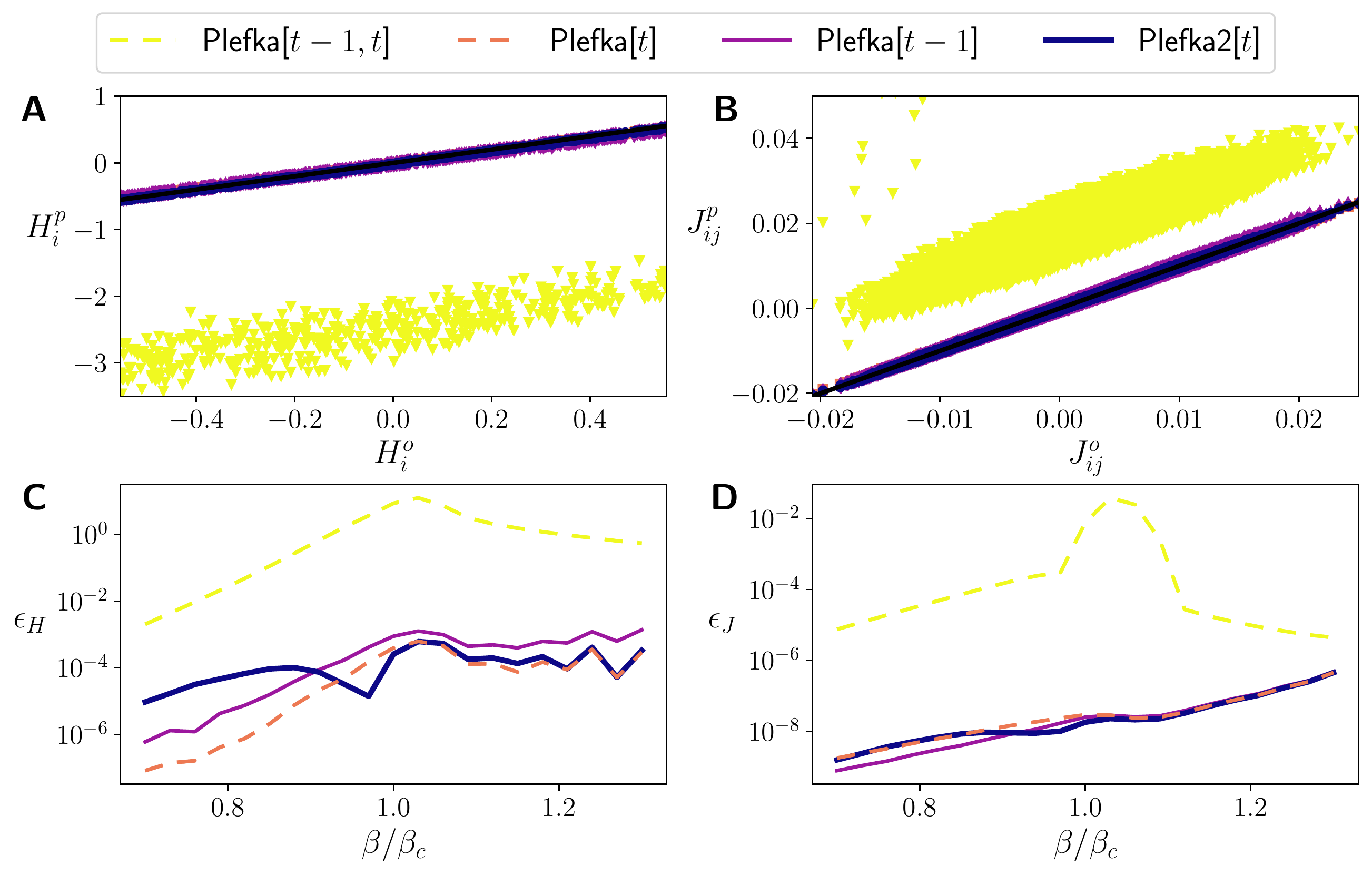}
}
\caption{\textbf{Inverse Ising problem}. Top: Inferred external fields (A)  and couplings (B) found by different mean-field models plotted versus the real ones for $\beta=\beta_c$. Black lines represent the identity line. Bottom: Average squared error of inferred external fields $\epsilon_{\*{H}}= \langle (H^o_{i} - H^p_{i})^2 \rangle_{i}$ (C) and couplings $\epsilon_{\*{J}}= \langle (J^o_{ij} - J^p_{ij})^2 \rangle_{ij}$ (D) for 21 values of $\beta$ in the range $[0.7\beta_c, 1.3\beta_c]$.
}
\label{fig:results-inverse}
\end{figure*}

\subsection*{Inverse Ising problem} 
We apply the approximation methods to the inverse Ising problem by using the data generated above for the trajectory of $T=128$ steps and $R=10^6$ trials to infer the parameters of the model, $\*H$ and $\*J$. The model parameters are estimated by the Boltzmann learning method under the maximum likelihood principle: $\*H$ and $\*J$ are updated to minimize the differences between the average rates $\*m_{t}$ or delayed correlations $\*D_t$ of the original data and the model approximations, which can significantly reduce computational time (see Methods).
While Boltzmann learning requires to compute the likelihood of every point in a trajectory and every trial ($RT$ calculations) each iteration, we can estimate the gradient at each iteration in a one-shot computation by applying the Plefka approximations (Methods).
At $\beta=\beta_c$  (Fig.~\ref{fig:results-inverse}A,B), we observe that the classical Plefka[$t-1,t$] approximation adds significant offset values to the fields $\*H$ and couplings $\*J$.
In contrast,  Plefka[$t$], Plefka[$t-1$] and  Plefka2[$t$] are all precise in estimating the values of $\*H$ and $\*J$.

Fig. \ref{fig:results-inverse}C and D show the mean squared error $\epsilon_{\*{H}}$, $\epsilon_{\*{J}}$ for bias terms and couplings between the original model and the inferred values for different $\beta$.
In this case, errors are large in the estimation of $\*J$ for Plefka[$t-1,t$].
In comparison, Plefka[$t$], Plefka[$t-1$] and Plefka2[$t$] work equally well even in the high fluctuation regime ($\beta \approx \beta_c$). Since the inverse Ising problem is solved by applying approximation one single time step (per iteration), it is not as challenging as the forward problem that can accumulate errors by recursively applying the approximations. Therefore, different approximations other than the classical mean-field Plefka[$t-1,t$] perform equally well in this case.

\subsection*{Phase transition reconstruction}

We have shown how different methods perform in computing the behavior of the system (forward problem) and inferring the parameters of a given network from its activation data (inverse problem). Combining the two, we can ask how well the methods explored here can reconstruct the behavior of a system from data, potentially exploring behaviors under different conditions than the recorded data.

\begin{figure*}[t]
\centering{
\includegraphics[width=17.9cm]{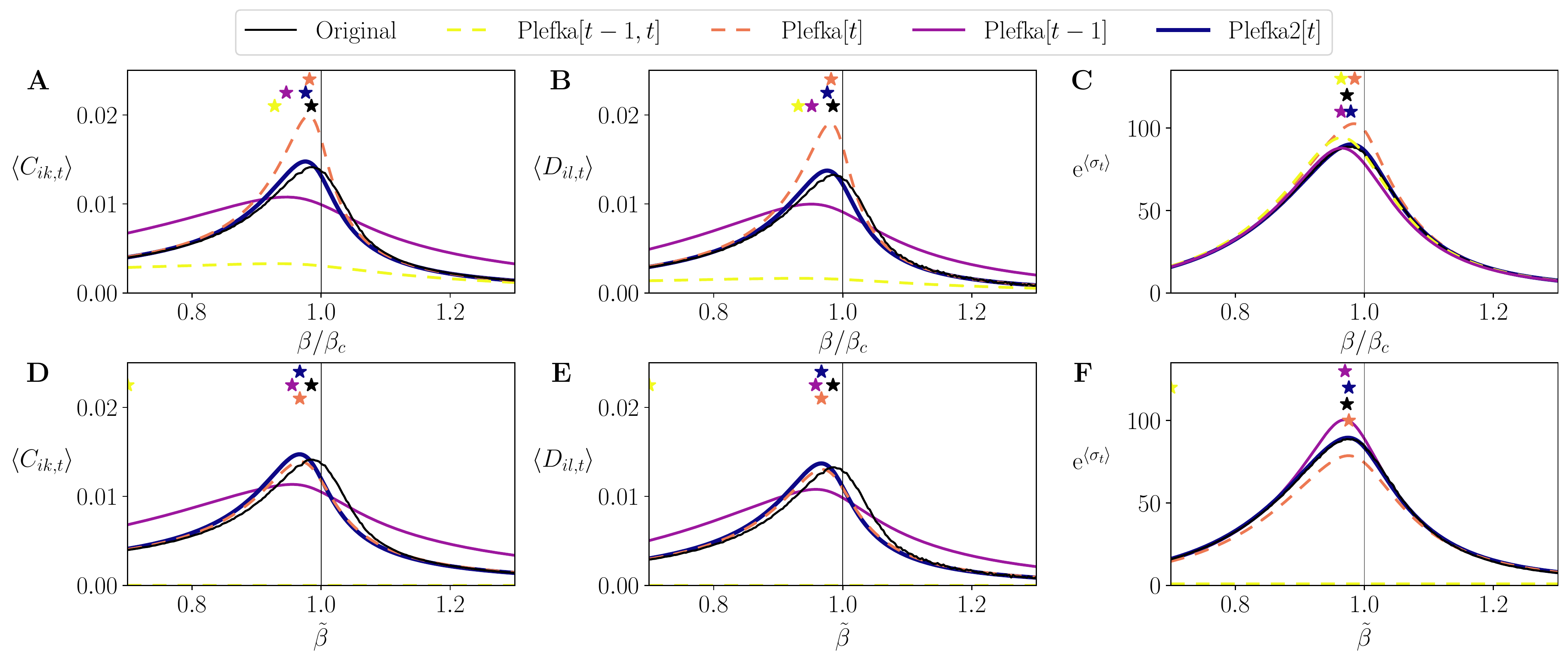}
}
\caption{\textbf{Reconstructing phase transition of kinetic Ising systems}. 
Top: Average of the Ising model's equal-time correlations (A),  delayed correlations (B), and entropy production (shown as an exponential for better presentation of its maximum) (C), at the last step $t=128$ found by different mean-field methods for $\beta=\beta_c$.
Bottom (D, E, F): The same as above using the reconstructed network $\*H, \*J$ by solving the inverse Ising problem at $\beta=\beta_c$ and multiplying a fictitious inverse temperature $\tilde \beta$ to the estimated parameters.
The stars are marked at the values of $\tilde \beta$ that yield maximum fluctuations or maximum entropy production.} 
\label{fig:results-reconstruction}
\end{figure*}

First, in Fig.~\ref{fig:results-reconstruction}.A-C we examine how the different approximation methods approximate fluctuations (equal-time and time-delayed covariances) and the entropy production (see Methods) at $t=128$ after solving the forward problem by recursively applying the approximations for the 128 steps. As we mentioned above, the asymmetric SK model explored here presents maximum fluctuations and maximum entropy production around $\beta=\beta_c$ (\SISKmodel{}, Supplementary Fig.~2).
However, we see that Plefka[$t-1,t$] and  Plefka[$t-1$] cannot reproduce the behavior  of correlations $\*C_t$ and $\*D_t$ of the original SK model around the transition point. Plefka[$t$] and Plefka2[$t$] show  much better  performance in capturing the behavior of $\*C_t$ and $\*D_t$ in the phase transition, although Plefka[$t$] overestimates both correlations. Additionally, all the methods capture the phase transition in entropy production, though Plefka[$t$] overestimates its value around $\beta_c$ and Plefka2[$t$] is more precise than the other methods. 

Next, we combine the forward and inverse Ising problem and try to reproduce the transition in the asymmetric SK model in the models inferred from the data.
We first obtain the values of $\*H, \*J$ by solving the inverse problem from the data sampled at $\beta=\beta_c$, and next we solve again the forward problem with those estimated parameters re-scaled by a new inverse temperature $\tilde\beta$. 
The results for the correlations  (Fig.~\ref{fig:results-reconstruction}.D-E) show that in this case Plefka[$t-1,t$] works badly, not being able to capture the transition. Plefka[$t-1$] shows similar performances as in the forward problem, and Plefka[$t$] and Plefka2[$t$] have a similar behavior, underestimating fluctuations slightly. 
When we analyze entropy production of the system (Fig.~\ref{fig:results-reconstruction}.F), we find that Plefka2[$t$] exhibits better performance with a high precision, with Plefka[$t-1$] slightly overestimating it, Plefka[$t$] underestimating it, and Plefka[$t-1,t$] not capturing the phase transition.
Overall, the results above suggest that Plefka2[$t$] is better suited to identify non-equilibrium phase transitions in models reconstructed from experimental data.

\section*{Discussion}

We have proposed a framework that unifies different mean-field approximations of the evolving statistical properties of non-equilibrium Ising models. This allows us to derive approximations premised on specific assumptions about the correlation structure of the system previously proposed in the literature. Furthermore, using our framework we derive a new approximation (Plefka2[$t$]) using atypical assumptions for mean-field methods, i.e. the maintenance of pairwise correlations in the system. This new pairwise approximation outperforms existing ones for approximating the behavior of an asymmetric SK model near the non-equilibrium equivalent of a ferromagnetic phase transition (see \SISKmodel{}), where classical mean-field approximations face problems. This shows that the proposed methods are useful tools to analyze large-scale, non-equilibrium dynamics near critical regimes expected for biological and social systems. However, we note that low-temperature spin phases (e.g. the spin-glass phase in symmetric models) also impose limitations on mean-field approximations \cite{plefka_convergence_1982, mezard_exact_2011}, which could be further explored with methods like the ones presented here.

The generality of this framework allows us to picture other approximations with atypical assumptions. For example, the Sessak-Monasson expansion 
\cite{sessak_small-correlation_2009} for an equilibrium Ising model assumes a linear relation between $\alpha$ and spin correlations. An  equivalent equilibrium expansion could use an effective field $h(\alpha)$ nonlinearly dependent on $\alpha$, satisfying linear $\*C_t(\alpha) = \alpha \*C_t$ or $\*D_t(\alpha) = \alpha \*D_t$ relations.
As another extension, Plefka2[$t$] could incorporate higher-order interactions. As Eq.~\ref{eq:PlefkaD_1}, \ref{eq:PlefkaD_1C}  are each equivalent to two mean-field approximations with $s_{l,t-1}=\pm 1$ respectively, a generalized PlefkaM[$t$] would involve $2^{M-1}$ equations, increasing accuracy but also computational costs.
In general, reference models $Q(\*s_t)$ set coupling parameters of the model to zero at some steps of its dynamics. Other parameters (e.g. fields) are either free parameters fitted as m-projection from $P(\*s_t)$, or preserved to their original value (see \SIcomparison{} for comparing free and fixed parameters of each model). Augmenting accuracy by increasing parameters often involves a computational cost. As a practical guideline for using each method, \SIcomparison{} compares their precision and computation time in the forward and inverse problems (see also Supplementary Fig.~3,4). 

Asides from its theoretical implications, our unified framework offers analysis tools for diverse data-driven research fields. In neuroscience, it has been popular to study the activity of ensembles of neurons by inferring an equilibrium Ising model with homogeneous (fixed) parameters \cite{schneidman_weak_2006} or inhomogeneous (time-dependent) parameters \cite{granot-atedgi_stimulus-dependent_2013, shimazaki_state-space_2012} from empirical data. 
Extended analyses based on the equilibrium model have reported that neurons operate near a critical regime \cite{mora_dynamical_2015, tkacik_thermodynamics_2015}. However, studies of non-equilibrium dynamics in neural spike trains are scarce \cite{tyrcha_effect_2013, hertz_ising_2013, cofre_introduction_2019}, partly due to the lack of systematic methods for analysing large-scale non-equilibrium data from neurons exhibiting large fluctuations. 
The proposed pairwise model Plefka2[$t$] is suitable for simulating such network activities, being more accurate than previous methods in predicting the network evolution at criticality (Fig.~\ref{fig:results-forward}) and in testing if the system is near the maximally fluctuating regime (Fig.~\ref{fig:results-reconstruction}). 
In particular, application of our methods for computing entropy production in non-equilibrium systems could provide tools for characterizing the non-equilibrium dynamics of neural systems \cite{lynn_non-equilibrium_2020}.

In summary, a unified framework of mean-field theories offers a systematic way to construct suitable mean-field methods in accordance with the statistical properties of the systems researchers wish to uncover. This is expected to foster a variety of tools to analyze large-scale non-equilibrium systems in physical, biological, and social systems.

\section*{Methods}

\begingroup
\allowdisplaybreaks
\subsection*{Boltzmann learning in the inverse Ising problem}
Let $\*{S}_t^r=\{S_{1,t}^r,S_{2,t}^r,\ldots,S_{N,t}^r\}$ for $t=1,\dots,T$ be observed states of a process described by Eq.~\ref{eq:Ising} at the $r$-th trial ($r=1,\dots,R$). We also define $\*S_{1:T}$ to represent the processes from all trials. 
The inverse Ising problem consists in inferring the external fields $\*{H}$ and couplings $\*{J}$ of the system. These parameters can be estimated by maximizing the log-likelihood $l(\*S_{1:T})$ of the observed states under the model:
\begin{align}
	l(\*S_{1:T}) = &  \log \prod_{t=1}^{T} \prod_{r=1}^{R} P(\*{S}_{t}^{r}|\*{S}_{t-1}^{r}) 
	\nonumber\\ = &  
	\sum_t \sum_{r} \sum_{i}\left( S_{i,t}^{r} h_{i,t}^{r} -  \log 2\cosh h_{i,t}^{r} \right),
\end{align}
with $h_{i,t}^{r} = H_{i} + \sum_j J_{ij} S_{j,t-1}^{r}$. The learning steps are obtained as:
\begin{align}
	\frac{\partial l(\*S_{1:T})}{\partial H_{i}} =& RT (\langle S_{i,t}^{r} \rangle_{r,t}  - \langle \tanh h_{i,t}^{r} \rangle_{r,t} ),
\label{eq:derivatives-projection-H} \\
	\frac{\partial l(\*S_{1:T})}{\partial J_{il}} = & RT( \langle S_{i,t}^{r} S_{l,t-1}^{r} \rangle_{r,t}  - \langle \tanh h_{i,t}^{r} S_{l,t-1}^{r,t} \rangle_{r,t}),  \label{eq:derivatives-projection-J}
\end{align}
where $\langle \cdot \rangle_{r}$ denotes average over trials. We solve the inverse Ising problem by applying these equations as a gradient ascent rule adjusting $\*{H}$ and $\*{J}$. The second terms of Eqs.~\ref{eq:derivatives-projection-H}, \ref{eq:derivatives-projection-J} need to be computed at every iteration, thus the computational cost grows linearly with $R \times T$. However, the use of mean-field approximations can significantly reduce the cost when a large number of samples $R$ and time bins $T$ are used to correctly estimate activation rates and correlations in large networks. Here the second terms can be written as
\begin{align}
     \langle \tanh h_{i,t}^{r}  \rangle_{r,t} = &  \sum_{\*s,\tilde{\*s}} s_{i} P(\*s|\tilde{\*s}) \overline{P}(\tilde{\*s})
     = m_{i}, 
    \\  \langle \tanh h_{i,t}^r  S_{l,t-1}^r\rangle_{r,t}   = &  \sum_{\*s,\tilde{\*s}}  s_{i}  \tilde s_{l}   P(\*s|\tilde{\*s})\overline{P}(\tilde{\*s})\nonumber \\
    = & D_{il} + m_{i} \tilde m_{l}, 
\end{align}
where $\overline{P}(\tilde{\*s}) = \frac{1}{RT}\sum_{r,t} \delta(\tilde{\*s},\*S_t^r)$ is the empirical distribution averaged over trials and trajectories (with $\delta$ being a Kronecker delta) and $\tilde m_{l}$ is the average activation rate computed from the empirical distribution. $P(\*s|\tilde{\*s})$ is defined as Eq.~\ref{eq:Ising}. We then approximate $m_{i}$ and $D_{il}$ using the mean-field equations. Note that when we apply the mean-field equations, we replaced all statistics related to the previous step with those computed by the empirical distribution.
By applying the mean-field methods, we reduced the computation of $R$ trials of trajectories of length $T$ into a single computation (instead of $RT$ calculations).
In our numerical tests, gradient ascent was executed using learning coefficients $\eta_H=0.1/RT, \eta_J=1/(RT\sqrt{N})$, starting from $\*H=\*0,\*J=\*0$.

\subsection*{Entropy production of the kinetic Ising model}
The entropy production is defined as the KL divergence between the forward and backward path, quantifying the irreversibility of the system \cite{schnakenberg_network_1976, seifert_stochastic_2012, cofre_introduction_2019}:
\begin{align}
	\langle \sigma_{t} \rangle = &\frac{1}{2} \sum_{\*s_{t},\*s_{t-1}} \big(P(\*s_{t-1}) P(\*s_{t}|\*s_{t-1}) -P(\*s_{t}) P_{\rm B}(\*s_{t-1}|\*s_{t})\big)
	\nonumber\\&\cdot\log \frac{P(\*s_{t}|\*s_{t-1}) P(\*s_{t-1})}{P_{\rm B}(\*s_{t-1}|\*s_{t})P(\*s_{t})}.
\end{align}
where $P_{\rm B}(\*s_{t-1}|\*s_{t})$ is a probability of the backward trajectory defined as in Eq.~\ref{eq:Ising} but switching $\*s_{t}$ and $\*s_{t-1}$.
Assuming a non-equilibrium steady state, where $P(\*s_{t}) = P(\*s_{t-1})$, the entropy production of the kinetic Ising system is computed as:
\begin{align}
	\langle \sigma_{t} \rangle =& \sum_{\*s_{t},\*s_{t-1}} P(\*s_{t},\*s_{t-1}) \Big( \sum_i H_i (s_{i,t} - s_{i,t-1}) 
	\nonumber\\ &+ \sum_{ij} J_{ij} (s_{i,t}s_{j,t-1} - s_{i,t-1}s_{j,t}) 
	\nonumber\\ & + \log P(\*s_{t-1})- \log P(\*s_{t})
	\nonumber\\ & -  \log (2 \cosh (H_i + \sum_j J_{ij} s_{j,t-1} ))
	\nonumber\\ & + \log ( 2 \cosh (H_i + \sum_j J_{ij} s_{j,t} ) )\Big)
	\nonumber\\ =& \sum_{ij} J_{ij} (D_{ij,t} - D_{ji,t}) 
	=  \sum_{ij} (J_{ij} - J_{ji}) D_{ij,t}. 
\end{align}
\endgroup

\begin{acknowledgments}
The authors thank Yasser Roudi and Masanao Igarashi for valuable comments and discussions on this manuscript. This work was supported in part by the Cooperative Intelligence Joint Research between Kyoto University and Honda Research Institute Japan, MEXT/JSPS KAKENHI Grant Number JP 20K11709, and the grant of Joint Research by the National Institutes of Natural Sciences (NINS Program No.~01112005). M.A. was funded by the European Union's Horizon 2020 research and innovation programme under the Marie Skłodowska-Curie grant agreement No \mbox{892715} and the University of the Basque Country \mbox{UPV/EHU} post-doctoral training program grant \mbox{ESPDOC17/17}, and supported in part by the Basque Government project \mbox{IT 1228-19} and project Outonomy (\mbox{PID2019-104576GB-I00}) by the Spanish Ministry of Science and Innovation.
\end{acknowledgments}

\section*{Code Availability}
The source code for implementing the results in this work is available under GPL license at GitHub \url{https://github.com/MiguelAguilera/kinetic-Plefka-expansions} \cite{aguilera_unifying_2020-1} (DOI:10.5281/zenodo.4357634). 

\section*{Data Availability}
The datasets generated and analysed during the current study are available under CC BY license at Zenodo \url{https://zenodo.org/record/4318983} \cite{aguilera_unifying_2020} (DOI:10.5281/zenodo.4318983).

\section*{Competing Interests}
The authors declare no competing interests.

\section*{Author contributions}
M.A., S.A.M. and H.S. designed and reviewed research; M.A. contributed analytical and numerical results; M.A., S.A.M. and H.S. wrote the paper.
%
\section*{Additional information}

\subsection*{Supplementary information}
The online version contains supplementary material
available at \url{https://doi.org/10.1038/s41467-021-20890-5}.


\bibliography{references}

\newpage
\onecolumngrid
\appendix
\renewcommand{\thesection}{\arabic{section}}
\renewcommand{\theequation}{SI.\thesection.\arabic{equation}}
\section*{Supplementary Notes}

\subsection*{List of changes respect to the published version}

This version of the Supplementary Information contains the following corrections respect to the published version
available at \url{https://doi.org/10.1038/s41467-021-20890-5}.
\begin{itemize}
    \item A $g_{i,u}$ term was removed from Eq. \ref{app-eq:trajectory-prob}.
    \item Missing $s_{j,v-1}$ and $\iu \hat\vartheta_{i,v}$ terms were added to Eq. \ref{app-eq:gaussian-integral-2}.
    \item Missing $\sum_i$ were added and extra $N$ terms removed in the set of equations starting at Eq. \ref{app-eq:saddle-point-extremization}
    \item $\*z$ was substituted by $\*{\tilde z}$ in Eq. \ref{app-eq:intermediate-solution}
    \item A missing $z$ term was added to Eq. \ref{app-eq:mean-field-solution-m}
    \item Missing limit operator has been added to Eq. \ref{app-eq:mean-activation-order-parameter} and \ref{app-eq:correlation-order-parameter_1}
    \item An intermediate step was added for deriving the critical exponents at Eq. \ref{eq:taylor-expansion-reduced-temperature}
\end{itemize}

\section{General approach for Plefka expansions}

Let $\*s_t = \{s_{i,t}\}_{i},~i=1,\dots,N$ be the state of the system at time $t$ and $\*s_{1:T} = \{\*s_{i,t} \}_{i,t},~i=1,\dots,N,~t=1,\dots,T$ a trajectory of the system. Given an initial state $\*s_{0}$, the probability of a trajectory $\*s_{1:T}$ of a kinetic Ising model is:
\begin{align}
    P(\*s_{1:T}) =& \prod_t P(\*s_{t}|\*s_{t-1}) =  \prod_t \exp\pr{\sum_{i} s_{i,t} h_{i,t} - \psi },
    \\ h_{i,t} =& H_{i} + \sum_j J_{ij} s_{j,t-1},
    \\ \psi =&  \sum_{t,i} \log 2\cosh h_{i,t}.
    \label{app-eq:psi}
\end{align}
Note that $\psi$ depends on the specific trajectory $\*s_{0:T}$.
The manifold $\mathcal{P} = \{ P(\*s_{1:T} | \*H, \*J) \}$ defines the family of probability distributions of the trajectories of all kinetic Ising models.
Within this manifold, we consider a submanifold $\mathcal{P}_0$ in which the probability distributions of a set of elements of the system $\{\{s_{i,t} \}_{i\in \mathcal{I}_t}\}_t$ is tractable and independent from the rest of the elements of the system. Here $\mathcal{I}_t$ denotes the indices of the tractable elements of the system at time $t$.
Trajectories in the submanifold are defined as:
\begin{align}
    P_0(\*s_{1:T}) =& \prod_t P_0(\*s_{t}|\*s_{1:t-1}) =  \prod_t\exp\pr{\sum_{i\in \mathcal{I}_{t}} s_{i,t} \theta_{i,t} + \sum_{i\in \overline{\mathcal{I}}_{t}} s_{i,t} h_{i,t} - \psi_0 } ,
    \\ \psi_0 =&  \sum_{t,i\in \mathcal{I}_t} \log 2\cosh \theta_{i,t}  +  \sum_{t,i\in \overline{\mathcal{I}}_t} \log 2\cosh h_{i,t}, 
\end{align}
where $\overline{\mathcal{I}}_t$ is the complement set of $\mathcal{I}_t$ for the elements at time $t$.
The mean-field definition of the new effective field is $\theta_{i,t}=\Theta_{i,t}$, although other definitions are possible.
We restrict the function $\theta_{i,t}$ to maximum caliber models composed of individual fields and pairwise couplings
\begin{equation}
    \theta_{i,t} = \sum_{I\in\mathcal{C}_{i,t}} \Theta_{i,t,I} s_{I},
\end{equation}
where $\mathcal{C}_{i,t}$ is a set of couplings $\{(j,\tau)\}, \tau< t, j\in \mathcal{I}_\tau$ between $s_{i,t}$ and other nodes $s_{j,\tau}$ from the set of nodes with tractable properties. We define the individual field of a unit with $\Theta_{i,t,\emptyset} \equiv \Theta_{i,t}$, by defining $s_\emptyset=1$. In general, the effect from the past spiking activities at $\tau<t$ can be modeled by this equation, which includes the generalized linear model for conditionally independent Bernoulli processes. In the approximations explored in this paper, however, we focus on the effect from the immediate past $\tau=t-1$. 

Different approximations are defined through different definitions of $\theta_{i,t}$ using a model connecting $P$ and $P_0$. This model is defined through a parameter $\alpha$:
\begin{align}
    P_{\alpha}(\*s_{1:T}) =&  \prod_t P_{\alpha}(\*s_{t}|\*s_{1:t-\tau-1}) 
    = \prod_t  \exp\pr{ \sum_{i\in \mathcal{I}_{t}} s_{i,t} ((1-\alpha) \theta_{i,t} + \alpha h_{i,t}) + \sum_{i\in \overline{\mathcal{I}}_{t}} s_{i,t} h_{i,t} - \psi_{\alpha} },
    \\ \psi_{\alpha} =& \sum_{t,i\in \mathcal{I}_t} \log 2\cosh ((1-\alpha) \theta_{i,t} + \alpha h_{i,t})) 
    +  \sum_{t,i\in \overline{\mathcal{I}}_t} \log 2\cosh h_{i,t}, 
\end{align}
such that $P_1(\*s_{1:T}) =P(\*s_{1:T})$.

The model $P_0$ that better approximates $P$ is the one that minimizes the Kullback Leibler divergence:
\begin{equation}
    D_{KL}(P||P_0) = \sum_{\*s_{1:T}} P(\*s_{1:T}) \log\frac{P(\*s_{1:T}) }{P_0(\*s_{1:T}) }.
\end{equation}
Thus its parameters $\Theta_{i,t,I}$ meet
\begin{align}
    \frac{\partial D_{KL}(P||P_0)}{\partial \Theta_{i,t,I}} =& -\sum_{\*s_{1:T}} P(\*s_{1:T})   s_{i,t} s_{I} +  \sum_{\*s_{1:T}} P(\*s_{1:T}) \sum_{\^\sigma_{t}} P_0(\^\sigma_t|\*s_{1:t-1})  \sigma_{i,t} s_{I} 
    \\ = & \langle s_{i,t} s_{I} \rangle_{0}^t - \langle s_{i,t} s_{I} \rangle_{1}^t = 0,
\end{align}
where $\langle \dots \rangle_{\alpha}^t = \sum_{\*s_{1:T}} \dots P_{\alpha}(\*s_t|\*s_{1:t-1}) P(\*s_{1:t-1})$.
Thus, the closest approximation $P_0$ to $P$ is the one in which $\langle s_{i,t} s_{I} \rangle_{0}^t = \langle s_{i,t} s_{I} \rangle_{1}^t, \forall i, I$.
Furthermore, models $P_0$ are tractable for indices $\{\mathcal{I}_t\}$, in the sense that knowing $\Theta_{i,t,I}$, it is easy to compute $\langle s_{i,t} s_{I} \rangle_{0}^t$.
The value of parameters $\Theta_{i,t,I}$ cannot be computed directly in general, but they can be approximated by computing a Plefka expansion. Approximating $\langle s_{i,t} s_{I} \rangle_{1}^t$ by the Taylor expansion of $\langle s_{i,t} s_{I} \rangle_{\alpha}^t$ from $\alpha=0$, we have
\begin{equation}
	 \langle s_{i,t} s_{I} \rangle_{\alpha}^t = \langle s_{i,t} s_{I} \rangle_{\alpha=0}^t + \sum_{k=1}^n  \frac{\alpha^k}{k!}  \frac{\partial^k  \langle s_{i,t} s_{I} \rangle_{\alpha=0}^t}{\partial \alpha^k} + \mathcal{O}(\alpha^{(n+1)}).
\end{equation}
Evaluating it at $\alpha=1$ knowing that $\langle s_{i,t} s_{I} \rangle_{0}^t = \langle s_{i,t} s_{I} \rangle_{1}^t, \forall i,I$, we have
\begin{equation}
    \sum_{k=1}^n  \left[ \frac{\alpha^k}{k!} \right]_{\alpha=1} \frac{\partial^k  \langle s_{i,t} s_{I} \rangle_{0}^t}{\partial \alpha^k}  =0 + \left[\mathcal{O}(\alpha^{(n+1)})\right]_{\alpha=1}.
    \label{app-eq:gen-plefka-solution}
\end{equation}

Let us define $\Delta h_{i,t} = -\theta_{i,t} +  h_{i,t}$, we have that
\begin{align}
    \frac{\partial P_{\alpha}(\*s)}{\partial \alpha} =& \sum_t \sum_{i\in \mathcal{I}_t}     \left( s_{i,t} \Delta  h_{i,t}-  \langle s_{i,t} \Delta  h_{i,t} \rangle_{t,\alpha} \right)   P_{\alpha}(\*s),
    \\ \frac{\partial^2 P_{\alpha}(\*s)}{\partial \alpha^2} =& 
    \left( \sum_t \sum_{i\in \mathcal{I}_t}   s_{i,t} \Delta  h_{i,t} - \langle s_{i,t} \Delta  h_{i,t} \rangle_{t,\alpha} \right)^2 P_{\alpha}(\*s)
    \\ \phantom{=}& - \sum_t  \left( \left\langle  \left(\sum_{i\in \mathcal{I}_t} s_{i,t} \Delta  h_{i,t}\right)^2 \right\rangle_{t,\alpha} - \left\langle \sum_{i\in \mathcal{I}_t} s_{i,t} \Delta  h_{i,t}  \right\rangle_{t,\alpha}^2 \right) P_{\alpha}(\*s)
    \\ =& \sum_t  \Bigg( \left( \left(\sum_{i\in \mathcal{I}_t} s_{i,t} \Delta  h_{i,t}\right)^2 - \left\langle  \left(\sum_{i\in \mathcal{I}_t} s_{i,t} \Delta  h_{i,t}\right)^2 \right\rangle_{t,\alpha}\right)
    \\ \phantom{=}& - 2 \left\langle \sum_{i\in \mathcal{I}_t} s_{i,t} \Delta  h_{i,t} \right\rangle_{t,\alpha} \left(\sum_{k\in \mathcal{I}_t} s_{k,t} \Delta  h_{k,t} - \left\langle \sum_{k\in \mathcal{I}_t} s_{k,t} \Delta  h_{k,t}  \right\rangle_{t,\alpha}\right) \Bigg) P_{\alpha}(\*s),
\end{align}
where  $\langle\dots\rangle_{t,\alpha} = \sum_{\*s_t} \dots P_{\alpha}(\*s_t|\*s_{1:t-1})$. From these equations we derive the first and second order approximations.

For the first order term, we have
\begin{align}
      \frac{\partial  \langle s_{i,t} s_{I} \rangle_{\alpha=0}^t}{\partial \alpha} =& \sum_{\*s_{1:T}} s_{i,t} s_{I} \frac{\partial P_{0}(\*s)}{\partial \alpha}
      \\ =& \sum_{\*s_{1:T}} \pr{\sum_t \sum_{k\in \mathcal{I}_t}     \left( s_{i,t} s_{I} (s_{k,t} \Delta  h_{k,t} -  \langle s_{k,t} \Delta  h_{k,t} \rangle_{t,0})   \right)} P_0(\*s).
\end{align}

The second order term is:
\begin{align}
      \frac{\partial^2  \langle s_{i,t} s_{I} \rangle_{\alpha=0}^t}{\partial \alpha^2} =&\sum_{\*s_{1:T}} s_{i,t} s_{I} \frac{\partial^2 P_{0}(\*s)}{\partial \alpha^2}
      \\ =&   \sum_{\*s_{1:T}}  s_{i,t} s_{I} \Bigg(   \Bigg(\sum_{k\in \mathcal{I}_t} s_{k,t} \Delta  h_{k,t}\Bigg)^2 - \left\langle  \Bigg(\sum_{k\in \mathcal{I}_t} s_{k,t} \Delta  h_{k,t}\Bigg)^2 \right\rangle_{t,0}
      \\ \phantom{=}& - 2 \left\langle \sum_{k\in \mathcal{I}_t} s_{k,t} \Delta  h_{k,t} \right\rangle_{t,0} \Bigg(\sum_{m\in \mathcal{I}_t} s_{m,t} \Delta  h_{m,t} - \left\langle \sum_{m\in \mathcal{I}_t} s_{m,t} \Delta  h_{m,t}  \right\rangle_{t,0}\Bigg) \Bigg)P_0(\*s).
\end{align}

Using the equations above to solve Supplementary Eq.~\ref{app-eq:gen-plefka-solution} for different orders and choices of $P_0$ will give us the different Plekfa approximations.
\newpage

\section{Plefka[$t-1,t$]}
\label{app:plefka_t-1_t}

This approximation uses the following approximated marginal probability distribution:
\begin{equation}
    P_\alpha^{[t-1:t]}(\*{s}_{t},\*{s}_{t-1})=\sum_{ \*{s}_{t-2}}  P_{\alpha}(\*{s}_{t}|\*{s}_{t-1})P_{\alpha}(\*{s}_{t-1}|\*{s}_{t-2})P(\*{s}_{t-2}),
\end{equation}
where
\begin{align}
	  P_{\alpha}(s_{i,t}|\*{s}_{t-1}) = & \frac{\mathrm{e}^{ s_{i,t} {h}_{i,t}(\alpha)}}{2 \cosh {h}_{i,t}(\alpha)},
	  \\ h_{i,t}(\alpha) 
	  =& (1-\alpha) \Theta_{i,t} + \alpha (H_i + \sum_j J_{ij} s_{j,t-1}).
\end{align}
Here, by increasing the value of $\alpha$ from $0$ to $1$, one can smoothly connect the independent and coupled models. Further, $h_{i,t}(\alpha)$ can be written as
\begin{equation}
     h_{i,t}(\alpha) 
	  = \Theta_{i,t} + \alpha \, \Delta h_{i,t},
\end{equation}
where $\Delta h_{i,t} = -\Theta_{i,t} + H_i + \sum_j J_{ij} s_{j,t-1}$ represents deviation from the independent model. We estimate $m_{i,t}$ by using its $\alpha$-dependent approximation, defined as
\begin{equation}
    m_{i,t}(\alpha) = \sum_{\*{s}_{t},\*{s}_{t-1}} s_{i,t} P_\alpha^{[t-1:t]}(\*{s}_{t},\*{s}_{t-1}) = \sum_{\*{s}_{t-1},\*{s}_{t-2}} \tanh h_{i,t}(\alpha)  P_{\alpha}(\*{s}_{t-1}|\*{s}_{t-2}) P(\*s_{t-2}).
\end{equation}
Approximating its value by expanding around $\alpha=0$ yields
\begin{equation}
	 m_{i,t}(\alpha) = m_{i,t}(\alpha=0) + \sum_{k=1}^n  \frac{\alpha^k}{k!}  \frac{\partial^k  m_{i,t}(\alpha=0)}{\partial \alpha^k} + \mathcal{O}(\alpha^{(n+1)}),
\end{equation}
By noting that $m_{i,t}(\alpha=0)=m_{i,t}(\alpha=1)$, we evaluate it at $\alpha=1$. This results in
\begin{equation}
    \left[ \sum_{k=1}^n  \frac{\alpha^k}{k!}  \frac{\partial^k  m_{i,t}(\alpha=0)}{\partial \alpha^k} \right]_{\alpha=1} =0 + \left[\mathcal{O}(\alpha^{(n+1)})\right]_{\alpha=1}.
    \label{app-eq:solution_Plefka_t_t-1}
\end{equation}

The approximation yields the nMF equations when we ignore quadratic and higher order terms in solving this equation, and the TAP equations when ignoring third and higher order terms. 

The first order derivative of $m_{i,t}(\alpha)$ is given as
\begin{align}
	\frac{\partial m_{i,t}(\alpha)}{\partial \alpha} &
		= \sum_{\*{s}_{t-1},\*{s}_{t-2}} \left[ \frac{\partial \tanh h_{i,t}(\alpha)}{\partial \alpha} P_\alpha(\*s_{t-1}|\*s_{t-2}) + \tanh h_{i,t}(\alpha) \frac{\partial P_\alpha(\*s_{t-1}|\*s_{t-2})}{\partial \alpha} \right] P(\*s_{t-2}).
\end{align}
Using the following equation, 
\begin{align}
    \frac{\partial \tanh h_{i,t}(\alpha)}{\partial \alpha} 
    &= (1-\tanh^2 h_{i,t}(\alpha))  \Delta h_{i,t}, 
\end{align}
the first order derivative is given as
\begin{align}
	\frac{\partial m_{i,t}(\alpha)}{\partial \alpha} 
	&= \sum_{\*{s}_{t-1},\*{s}_{t-2}} \Bigg[ (1-\tanh^2 h_{i,t}(\alpha)) \Delta h_{i,t}
	    + \tanh h_{i,t}(\alpha)\frac{\partial P_\alpha(\*s_{t-1}|\*s_{t-2})}{\partial \alpha}   \Bigg] P_{\alpha}(\*{s}_{t-1}|\*{s}_{t-2}) P(\*{s}_{t-2}).
\end{align}
Expectation of the first term at $\alpha=0$ is 
\begin{align}
    \sum_{\*{s}_{t-1},\*{s}_{t-2}} (1-\tanh^2 h_{i,t}(0)) \Delta h_{i,t} P_{0}(\*{s}_{t-1})
    = (1 - m_{i,t}^2)(-\Theta_{i,t}+H_{i}+\sum_{j}J_{ij}m_{j,t-1}).
\end{align}
The second term becomes zero at $\alpha=0$, since the derivative of $P_\alpha(\*{s}_{t-1}|\*{s}_{t-2})$ becomes independent of $\tanh h_{i,t}(\alpha)$ for $\alpha=0$ and we know that $\sum_{\*{s}_{t-1}} P_\alpha(\*{s}_{t-1}|\*{s}_{t-2})= 1$.
Thus we have
\begin{align} 
    \frac{\partial m_{i,t}(\alpha=0)}{\partial \alpha} & = (1 - m_{i,t}^2)(-\Theta_{i,t}+H_{i}+\sum_{j}J_{ij}m_{j,t-1}).
\end{align}
From here, we obtain the nMF equations, yielding $\left[\alpha (-\Theta_{i,t}+H_{i}+\sum_{j}J_{ij}m_{j,t-1})\right]_{\alpha=1} = 0 + \left[\mathcal{O}(\alpha^2)\right]_{\alpha=1} $ and
\begin{align}
	\Theta_{i,t} &= H_i + \sum_j J_{ij} m_{j,t-1}  + \left[\mathcal{O}(\alpha^1)\right]_{\alpha=1},
	\\ m_{i,t} &\approx \tanh[ H_i + \sum_j J_{ij} m_{j,t-1}].
\end{align}

The second order derivative of $m_{i,t}(\alpha)$ is given as
\begin{align}
	\frac{\partial^2 m_{i,t}(\alpha)}{\partial \alpha^2} &
		= \sum_{\*{s}_{t-1},\*{s}_{t-2}} \Bigg[ \frac{\partial^2 \tanh h_{i,t}(\alpha)}{\partial \alpha^2} P_\alpha(\*s_{t-1}|\*s_{t-2}) 
		+ 2 \frac{\partial \tanh h_{i,t}(\alpha)}{\partial \alpha} \frac{\partial P_\alpha(\*s_{t-1}|\*s_{t-2})}{\partial \alpha}
		\\ &\phantom{=} + \tanh h_{i,t}(\alpha) \frac{\partial^2 P_\alpha(\*s_{t-1}|\*s_{t-2})}{\partial \alpha^2} \Bigg] P(\*s_{t-2}).
\end{align}
Here we note that 
\begin{align}
    \frac{\partial^2 \tanh h_{i,t}(\alpha)}{\partial \alpha^2} 
    &= 
    -2 \tanh h_{i,t}(\alpha) (1-\tanh^2 h_{i,t}(\alpha))  \Delta h_{i,t}^2.
\end{align}
From these equations, the second derivative is computed as 
\begin{align}
	\frac{\partial^2 m_{i,t}(\alpha)}{\partial \alpha^2}
		= & \sum_{\*{s}_{t-1},\*{s}_{t-2}} \Big( - 2 \tanh h_{i,t}(\alpha) (1 - \tanh^2 h_{i,t}(\alpha)) \Delta h_{i,t}^2 P_{\alpha}(\*{s}_{t-1}|\*{s}_{t-2})
		\\ &\phantom{=} + 2 (1 - \tanh^2 h_{i,t}(\alpha)) \Delta h_{i,t}  \frac{\partial P_\alpha(\*s_{t-1}|\*s_{t-2}) }{\partial \alpha}
		\\ &\phantom{=} + \frac{\partial^2 P_\alpha(\*s_{t-1}|\*s_{t-2}) }{\partial \alpha^2}
		\Big)  P(\*{s}_{t-2}).
\end{align}
We evaluate the second derivative at $\alpha=0$. 
Here at $\alpha=0$ the third term is zero since $\sum_{\*{s}_{t-1}}P_{\alpha}(\*s_{t-1}|\*s_{t-2}) = 1$, so its derivatives are equal to zero. Thus we have
\begin{align}
	\frac{\partial^2 m_{i,t}(\alpha=0)}{\partial \alpha^2} = &-2 m_{i,t} (1 - m_{i,t}^2)((-\Theta_{i,t}+H_{i}+\sum_{j}J_{ij}m_{j,t-1})^2 + \sum_{j} J_{ij}^2 (1-m_{j,t-1}^2))
	\\ &\phantom{=} + 2 (1 -m_{i,t}^2)   \sum_j J_{ij} \frac{\partial m_{j,t-1}(\alpha)}{\partial \alpha},
	\label{app-eq:taylor-mean-alpha-1-00}
\end{align}
where the last term comes from the $s_{j,t-1}$ terms in $h_{i,t}(\alpha)$ multiplied by $ \frac{\partial P_\alpha(\*s_{t-1}|\*s_{t-2}) }{\partial \alpha}$.

Note that the second order term in Supplementary Eq.~\ref{app-eq:taylor-mean-alpha-1-00} contains the expression $\left[\alpha^2 (-\Theta_{i,t} + H_i + \sum_j J_{ij} m_{j,t-1})^2\right]_{\alpha=1}$ and $\left[\alpha^2\frac{\partial m_{k,t-1}(\alpha)}{\partial \alpha}\right]_{\alpha=1}$ that can be neglected as terms with order higher than quadratic. This is due to the fact that the second-order approximation is in the proximity of naive mean-field solution which is in the first order of $\alpha$. Therefore, we know that $\left[ \alpha (1-m_i^2) (-\Theta_{i,t} + H_i + \sum_j J_{ij} m_{j,t-1})\right]_{\alpha=1}  =       \left[\mathcal{O}(\alpha^{2})\right]_{\alpha=1}$ and thus {$\big[ \alpha^2 (-\Theta_{i,t} + H_i + \sum_j J_{ij} m_{j,t-1})^2\big]_{\alpha=1}  = \left[\mathcal{O}(\alpha^{4})\right]_{\alpha=1}$} and $\left[ \alpha^2 \frac{\partial m_{k,t-1}(\alpha)}{\partial \alpha}\right]_{\alpha=1}  = \left[\mathcal{O}(\alpha^{3})\right]_{\alpha=1}$ which can be neglected for the second order approximation.

The combination of the first and second order derivatives of $m_{i,t}$ evaluated at $\alpha=0$ allows to solve Supplementary Eq.~\ref{app-eq:solution_Plefka_t_t-1} for order $n=2$, yielding the TAP equations:
\begin{align}
	\Theta_{i,t} &= H_i + \sum_j J_{ij} m_{j,t-1} - m_{i,t}\sum_{jl} J_{ij}^2 (1-m_{j,t-1}^2) + \left[ \mathcal{O}(\alpha^2) \right]_{\alpha=1},
	\\ m_{i,t} &\approx \tanh[ H_i + \sum_j J_{ij} m_{j,t-1} - m_{i,t}\sum_{jl} J_{ij}^2 (1-m_{j,t-1}^2)].
\end{align}
These results have a form similar to the TAP equations obtained for symmetric and asymmetric networks \cite{kappen_mean_2000, roudi_dynamical_2011}.

\subsection*{Equal-time correlations}

Here we approximate $C_{ik,t}$ by evaluating the Plefka expansion of $C_{ik,t}(\alpha=1)$ around $\alpha=0$.

When $i=k$, we have $C_{ii,t}(\alpha) = 1 -m_{i,t}(\alpha)^2$.
When $i\neq k$, correlations in the system can be obtained by expanding the alpha-dependent correlations
\begin{equation}
    C_{ik,t}(\alpha) = \sum_{\*{s}_{t},\*{s}_{t-1}} (s_{i,t}-m_{i,t}(\alpha))(s_{k,t} - m_{k,t}(\alpha)) P_\alpha^{[t-1:t]}(\*{s}_{t},\*{s}_{t-1}),
\end{equation}
around $\alpha=0$: 
\begin{equation}
	 C_{ik,t}(\alpha) =  \sum_{n=0}^m  \frac{\alpha^n}{n!}  \frac{\partial^n C_{ik,t}(\alpha=0)}{\partial \alpha^n} + \mathcal{O}(\alpha^{(m+1)}).
\end{equation}
The derivatives with different orders, and their values evaluated at $\alpha=0$ are obtained as follows. 

The zeroth order term is:
\begin{align}
	C_{ik,t}(\alpha) = &\sum_{\*{s}_{t-1},\*{s}_{t-2}} (\tanh h_{i,t}(\alpha) -  m_{i,t}(\alpha))( \tanh h_{k,t}(\alpha) - m_{k,t}(\alpha)) P_{\alpha}(\*{s}_{t-1}|\*{s}_{t-2}) P(\*{s}_{t-2}), 
	\\ C_{ik,t}(\alpha=0) =& (m_{i,t}  - m_{i,t}) (m_{k,t} - m_{k,t}) = 0.
\end{align}

The first order term is:
\begin{align}
	\frac{\partial C_{ik,t}(\alpha)}{\partial \alpha} &
		= \sum_{\*{s}_{t-1},\*{s}_{t-2}} \Big( 
		\big( (1 - \tanh^2 h_{i,t}(\alpha)) \Delta h_{i,t} 
		- \frac{\partial m_{i,t}(\alpha)}{\partial \alpha}\big) 
		\big(\tanh h_{k,t}(\alpha) - m_{k,t}(\alpha) \big) P_{\alpha}(\*{s}_{t-1}|\*{s}_{t-2})
		\\ &\phantom{=} + \big( (1 - \tanh^2 h_{k,t}(\alpha)) \Delta h_{k,t} 
		- \frac{\partial m_{k,t}(\alpha)}{\partial \alpha}\big) 
		\big(\tanh h_{i,t}(\alpha) - m_{i,t}(\alpha) \big) P_{\alpha}(\*{s}_{t-1}|\*{s}_{t-2}) 
		\\ &\phantom{=} + \sum_m (\tanh h_{i,t}(\alpha) -  m_{i,t}(\alpha))( \tanh h_{k,t}(\alpha)- m_{k,t}(\alpha)) \frac{\partial P_\alpha(\*s_{t-1}|\*s_{t-2})}{\partial \alpha}
		\Big)  P(\*{s}_{t-2}).
\end{align}
Here, all terms are equal to zero for $\alpha=0$ because the terms $\tanh h_{i,t}(\alpha=0) = m_{i,t}$ cancel out. Then we have
\begin{align}
	\frac{\partial C_{ik,t}(\alpha=0)}{\partial \alpha}& = 0.
\end{align}
Therefore the nMF equation is obtained as
\begin{align}
	C_{ik,t} &\approx 0. 
\end{align}

The second order term is:
\begin{align}
	\frac{\partial^2 C_{ik,t}(\alpha)}{\partial \alpha^2} 
		=& \sum_{\*{s}_{t-1},\*{s}_{t-2}} \Big(  
		2 \big( (1 - \tanh^2  h_{i,t}(\alpha)) \Delta h_{i,t}
		- \frac{\partial  m_{i,t}(\alpha)}{\partial \alpha}\big) 
		\big( (1 - \tanh^2  h_{k,t}(\alpha))\Delta h_{k,t} 
		- \frac{\partial  m_{k,t}(\alpha)}{\partial \alpha}\big) P_{\alpha}(\*{s}_{t-1}|\*{s}_{t-2})
		\\ &+  \big( -2 \tanh h_{i,t}(\alpha)(1 - \tanh^2 h_{i,t}(\alpha))\Delta h_{i,t}^2 - \frac{\partial^2  m_{i,t}(\alpha)}{\partial \alpha^2}\big) 
		\big(\tanh h_{k,t}(\alpha) -  m_{k,t}(\alpha) \big) P_{\alpha}(\*{s}_{t-1}|\*{s}_{t-2})
		\\ &+  \big( -2 \tanh h_{k,t}(\alpha)(1 - \tanh^2 h_{k,t}(\alpha))\Delta h_{k,t}^2 - \frac{\partial^2  m_{k,t}(\alpha)}{\partial \alpha^2}\big) 
		\big(\tanh h_{i,t}(\alpha) - m_{i,t}(\alpha) \big)  P_{\alpha}(\*{s}_{t-1}|\*{s}_{t-2})
		\\ &  + (\tanh h_{i,t}(\alpha) -  m_{i,t}(\alpha))( \tanh h_{k,t}(\alpha) - m_{k,t}(\alpha)) \frac{\partial^2 P_\alpha(\*s_{t-1}|\*s_{t-2})}{\partial \alpha^2}
		\\ &  +  \big( (1 - \tanh^2  h_{i,t}(\alpha)) \Delta h_{i,t}
		- \frac{\partial  m_{i,t}(\alpha)}{\partial \alpha}\big) ( \tanh h_{k,t}(\alpha)  - m_{k,t}(\alpha)) \frac{\partial P_\alpha(\*s_{t-1}|\*s_{t-2})}{\partial \alpha}
		\\ &  + (\tanh h_{i,t}(\alpha) -  m_{i,t}(\alpha))\big( (1 - \tanh^2  h_{k,t}(\alpha))\Delta h_{k,t} - \frac{\partial  m_{k,t}(\alpha)}{\partial \alpha}\big) \frac{\partial P_\alpha(\*s_{t-1}|\*s_{t-2})}{\partial \alpha}
		\Big)  P(\*{s}_{t-2}).
\end{align}
Except for the term in the first line, for $\alpha=0$ all terms are equal to zero because $\tanh h_{i,t}(\alpha=0) = m_{i,t}$ cancel out. This gives
\begin{align}
	\frac{\partial^2 C_{ik,t}(\alpha=0)}{\partial \alpha^2} =& 2  (1 - m_{i,t}^2)  (1 - m_{k,t}^2)
	\sum\limits_j J_{ij} J_{kj}(1-m_{j,t-1}^2).
\end{align}
Hence the correlations expanded up to the second order can be described as:
\begin{align}
C_{ik,t}(\alpha)
	&=  C_{ik,t}(\alpha=0) + \alpha \frac{\partial C_{ik,t}(\alpha=0)}{\partial \alpha} 
	+ \frac{\alpha^2}{2}  \frac{\partial^2 C_{ik,t}(\alpha=0)}{\partial \alpha^2} + \mathcal{O}(\alpha^3)
	\\ &= \alpha^2 (1 - m_{i,t}^2)  (1 - m_{k,t}^2) \sum\limits_j J_{ij} J_{kj}(1-m_{j,t-1}^2)
	+ \mathcal{O}(\alpha^3).
\end{align}
Hence, the corresponding TAP approximation is obtained as
\begin{align}
C_{ik,t}
	&\approx  (1 - m_{i,t}^2)  (1 - m_{k,t}^2) \sum\limits_j J_{ij} J_{kj}(1-m_{j,t-1}^2).
\label{app-eq:update_correlations_0}
\end{align}

\subsection*{Time-delayed correlations}

Similarly to the equal-time correlations, we approximate $D_{il,t}$ by evaluating the Plefka expansion of $D_{il,t}(\alpha=1)$ around $\alpha=0$.

We describe time-delayed correlations of the system 
\begin{equation}
    D_{il,t}(\alpha) = \sum_{\*{s}_{t},\*{s}_{t-1}}  (s_{i,t} - m_{i,t}(\alpha)) (s_{l,t-1} - m_{l,t-1}(\alpha)) P_\alpha^{[t-1:t]}(\*{s}_{t},\*{s}_{t-1}),
\end{equation}
using an expansion:
\begin{equation}
	 D_{il,t}(\alpha) =  \sum_{k=0}^n  \frac{\alpha^k}{k!}  \frac{\partial^k D_{il,t}(\alpha=0)}{\partial \alpha^k} + \mathcal{O}(\alpha^{(n+1)}).
\end{equation}

Likewise, the zeroth order term yields:
\begin{align}
	D_{il,t}(\alpha)&=  \sum_{\*{s}_{t-1},\*{s}_{t-2}}   (\tanh h_{i,t}(\alpha)- m_{i,t}(\alpha))(s_{l,t-1} - m_{l,t-1}(\alpha))  P_\alpha(\*{s}_{t-1}|\*{s}_{t-2})  P(\*{s}_{t-2}).
\end{align}
\begin{align}
	D_{il,t}(\alpha=0) &= 0.
\end{align}

The first order term is:
\begin{align}
	\frac{\partial D_{il,t}(\alpha)}{\partial \alpha}& 
		= \sum_{\*{s}_{t-1},\*{s}_{t-2}}   
		\Big( 
		\big((1 - \tanh^2 h_{i,t}(\alpha)) 
		\Delta h_{i,t}
		- \frac{\partial m_{i,t}(\alpha)}{\partial \alpha}\big) 
		 (s_{l,t-1} - m_{l,t-1}(\alpha)) P_\alpha(\*{s}_{t-1}|\*{s}_{t-2}) 
		 \\ &\phantom{=}+(\tanh h_{i,t}(\alpha)- m_{i,t}(\alpha))  \big( (s_{l,t-1} - m_{l,t-1}(\alpha)) \frac{\partial P_\alpha(\*s_{t-1}|\*s_{t-2})}{\partial \alpha}  - \frac{\partial m_{l,t-1}(\alpha)}{\partial \alpha} P_\alpha(\*{s}_{t-1}|\*{s}_{t-2})  \big)
		 \Big)  P(\*{s}_{t-2}).
\end{align}
When evaluated at $\alpha=0$, the terms in the second line disappear because the terms $\tanh h_{i,t}(\alpha=0) = m_{i,t}$ cancel out. Note that the term in the first line is computed as
\begin{equation}
\big((1 - \tanh^2 h_{i,t}(0)) \Delta h_{i,t}
		- \frac{\partial m_{i,t}(\alpha)}{\partial \alpha}
	= (1-m_{i,t}^2)\sum_{j} J_{ij} (s_{j,t-1} - m_{j,t-1}),
\end{equation}
which is multiplied by $s_{l,t-1} - m_{l,t-1}(\alpha)$. Because we take expectation over $\*s_{t-1}$ with the independent model, only expectation of the term $(s_{l,t-1}-m_{l,t-1})^2$ is preserved. Thus we have 
\begin{align}
    \frac{\partial D_{il,t}(\alpha=0)}{\partial \alpha}& = (1 - m_{i,t}^2) J_{il} (1-m_{l,t-1}^2).
\end{align}
Therefore the nMF equation is obtained as:
\begin{equation}
D_{il,t} \approx J_{il} (1 - m_{i,t}^2) (1-m_{l,t-1}^2).
\end{equation}

The second order term is:
\begin{align}
	\frac{\partial^2 D_{il,t}(\alpha)}{\partial \alpha^2}
		=& \sum_{\*{s}_{t-1},\*{s}_{t-2}}   
		\Big( - 2 \tanh h_{i,t}(\alpha) (1 - \tanh^2 h_{i,t}(\alpha)) 
		\Delta h_{i,t}^2 - \frac{\partial^2 m_{i,t}(\alpha)}{\partial \alpha^2}\big) 
		 (s_{l,t-1} - m_{l,t-1}(\alpha)) P_\alpha(\*{s}_{t-1}|\*{s}_{t-2}) 
         \\ & + 2 \big((1 - \tanh^2 h_{i,t}(\alpha)) 
		\Delta h_{i,t}
		- \frac{\partial m_{i,t}(\alpha)}{\partial \alpha}\big)  \big( (s_{l,t-1} - m_{l,t-1}(\alpha))\frac{\partial P_\alpha(\*s_{t-1}|\*s_{t-2})}{\partial \alpha}
		 \\ &-\frac{\partial m_{l,t-1}(\alpha)}{\partial \alpha}  P_\alpha(\*s_{t-1}|\*s_{t-2}) \big)
		 \\ &+ (\tanh h_{i,t}(\alpha)- m_{i,t}(\alpha)) \Big( 
		 \big( (s_{l,t-1} - m_{l,t-1}(\alpha)) \frac{\partial^2 P_\alpha(\*s_{t-1}|\*s_{t-2})}{\partial \alpha^2}
		 \\ & - 2\frac{\partial m_{l,t-1}(\alpha)}{\partial \alpha} \frac{\partial P_\alpha(\*s_{t-1}|\*s_{t-2})}{\partial \alpha} -  \frac{\partial^2 m_{l,t-1}(\alpha)}{\partial \alpha^2} P_\alpha(\*s_{t-1}|\*s_{t-2})
		 \Big)  P(\*{s}_{t-2}).
\end{align}
Here, in the first line, only the  $\Delta h_{i,t}^2 (s_{l,t-1} - m_{l,t-1})$ survives as
\begin{align}
    \Delta h_{i,t}^{2}(s_{l,t-1}-m_{l,t-1})
    	&= \Big(-\Theta_{i,t}+H_{i}+\sum_{j}J_{ij}m_{j,t-1}+\sum_{j}J_{ij}(s_{j,t-1}-m_{j,t-1}) \Big)^{2}(s_{l,t-1}-m_{l,t-1})
	\\&=\Big(-\Theta_{i,t}+H_{i}+\sum_{j}J_{ij}m_{j,t-1}\Big)^{2}(s_{l,t-1}-m_{l,t-1})
	\\&+2\Big(-\Theta_{i,t}+H_{i}+\sum_{j}J_{ij}m_{j,t-1}\Big)\sum_{j}J_{ij}(s_{j,t-1}-m_{j,t-1})(s_{l,t-1}-m_{l,t-1})
	\\&+\Big(\sum_{j}J_{ij}(s_{j,t-1}-m_{j,t-1})\Big)^{2}(s_{l,t-1}-m_{l,t-1}),
\end{align}
which results in $2(-\Theta_{i,t}+H_{i}+\sum_{j}J_{ij}m_{j,t-1})(1-m_{l,t-1}^2) -2 J_{il}^2 m_{l,t-1}(1-m_{l,t-1}^2)$ when evaluated for the mean field model.

The term in the second and third lines of the previous equation can be decomposed in two terms. The second part results in $\sum_j J_{ij} (s_{j,t-1} - m_{j,t-1}) \frac{\partial m_{l,t-1}(\alpha)}{\partial \alpha}$, which results in zero. From the first term, only survives the part containing $\sum_j J_{ij} (s_{j,t-1} - m_{j,t-1})(s_{l,t-1} - m_{l,t-1}) \frac{\partial P_\alpha(\*s_{t-1}|\*s_{t-2})}{\partial \alpha}$ which  for $\alpha=0$ results in
\begin{align}
    &\sum_{\*s_{t-1},\*s_{t-2}} \sum_j  J_{ij}(s_{j,t-1}-m_{j,t-1})(s_{l,t-1}-m_{l,t-1})\pd{P_{0}(s_{t-1}|s_{t-2})}{\alpha}P(s_{t-2})\\
    &=\sum_{\*s_{t-2}}  \sum_{jk}  J_{ij}(s_{j,t-1}-m_{j,t-1})(s_{l,t-1}-m_{l,t-1})(s_{k,t-1}-m_{k,t-1})\Delta h_{k,t-1} Q(s_{t-1})P(s_{t-2})\\
    &=\sum_{\*s_{t-1}}  J_{il}(s_{l,t-1}-m_{l,t-1})^3 Q(s_{t-1})\sum_{\*s_{t-2}} \Delta h_{k,t-1}P(s_{t-2})\\
    &= -2J_{il}m_{l,t-1}(1-m_{l,t-1} ^2) (-\Theta_{i,t-1}+H_{i}+\sum_j J_{ij}m_{j,t-2}).
\end{align}


The term in the last two lines disappear because the terms $\tanh h_{i,t}(\alpha=0) = m_{i,t}$ cancel out.
Thus, evaluated at $\alpha=0$ we have
\begin{align}
	\frac{\partial^2 D_{il,t}(\alpha=0)}{\partial \alpha^2} =& 4 m_{i,t} (1 - m_{i,t}^2) J_{il}^2  m_{l,t-1} (1-m_{l,t-1}^2) 
	\\ & - 4 m_{i,t} (1 - m_{i,t}^2) J_{il} (1-m_{l,t-1}^2) (-\Theta_{i,t} + H_i + \sum_j m_{j,t-1}) 
	\\& -  4 (1 - m_{i,t}^2)J_{il}m_{l,t-1}(1-m_{l,t-1} ^2) (-\Theta_{i,t-1}+H_{i}+\sum_j J_{ij}m_{j,t-2}).
\end{align}
The second and third term above are equal to  $0 + \mathcal{O}(\alpha)$, which makes them negligible (i.e. order larger than quadratic) when computing $\alpha^2 \frac{\partial^2 D_{il,t}(\alpha=0)}{\partial \alpha^2}$.

Hence the second order expansion is
\begin{align}
D_{il,t}(\alpha)
	&= D_{il,t}(\alpha=0) + \alpha \frac{\partial D_{il,t}(\alpha=0)}{\partial \alpha} 
	+ \frac{\alpha^2}{2}  \frac{\partial^2 D_{il,t}(\alpha=0)}{\partial \alpha^2} + \mathcal{O}(\alpha^3)
	\\ &= \alpha (1 - m_{i,t}^2) J_{il} (1-m_{l,t-1}^2) + \alpha^2  2 m_{i,t} (1 - m_{i,t}^2) J_{il}^2  m_{l,t-1} (1-m_{l,t-1}^2) 
	+ \mathcal{O}(\alpha^3).
\end{align}
Thus the TAP equation for the time-delayed correlations can be described as:
\begin{equation}
D_{il,t} \approx J_{il} (1 - m_{i,t}^2)  (1-m_{l,t-1}^2)(1 + 2 J_{il} m_{i,t} m_{l,t-1}).
\end{equation}

\newpage
\section{Plefka[$t$]}
\label{app:plefka_t}

This approximation uses the following approximated marginal probability distribution:
\begin{equation}
    P_\alpha^{[t]}(\*{s}_{t})= \sum_{\*{s}_{t-1}} P_{\alpha}(\*{s}_{t}|\*{s}_{t-1}) P(\*{s}_{t-1}),
\end{equation}
where
\begin{align}
	  P_{\alpha}(s_{i,t}|\*{s}_{t-1}) = & \frac{\mathrm{e}^{ s_{i,t} {h}_{i,t}(\alpha)}}{2 \cosh {h}_{i,t}(\alpha)},
	  \\ h_{i,t}(\alpha) 
	  =& (1-\alpha) \Theta_{i,t} + \alpha (H_{i} + \sum_j J_{ij} s_{j,t-1}).
\end{align}
We define $h_{i,t}(\alpha) = \Theta_{i,t} + \alpha \, \Delta h_{i,t}$, where $\Delta h_{i,t} = -\Theta_{i,t} + H_i + \sum_j J_{ij} s_{j,t-1}$ represents deviation from the independent model.
We estimate $m_{i,t}$ by using its $\alpha$-dependent approximation, whose element is defined as
\begin{equation}
    m_{i,t}(\alpha) = \sum_{\*{s}_{t},\*{s}_{t-1}} s_{i,t} P_\alpha^{[t]}(\*{s}_{t},\*{s}_{t-1}) = \sum_{\*{s}_{t-1}} \tanh h_{i,t}(\alpha) P(\*s_{t-1}).
\end{equation}
Approximating its value by expanding around $\alpha=0$ yields
\begin{equation}
	 m_{i,t}(\alpha) = m_{i,t}(\alpha=0) + \sum_{k=1}^n  \frac{\alpha^k}{k!}  \frac{\partial^k  m_{i,t}(\alpha=0)}{\partial \alpha^k} + \mathcal{O}(\alpha^{(n+1)}).
\end{equation}
By noting that $m_{i,t}(\alpha=0)=m_{i,t}(\alpha=1)$, we evaluate it at $\alpha=1$. This results in
\begin{equation}
    \left[ \sum_{k=1}^n  \frac{\alpha^k}{k!}  \frac{\partial^k  m_{i,t}(\alpha=0)}{\partial \alpha^k} \right]_{\alpha=1} =0 + \left[\mathcal{O}(\alpha^{(n+1)})\right]_{\alpha=1}.
\end{equation}

The approximation yields the nMF equations when we ignore the quadratic term and higher, and the TAP equations when ignoring the third and higher order terms. The terms at each order and its evaluation at $\alpha=0$ are obtained as follows.

The first order term is:
\begin{align}
   \frac{\partial m_{i,t}(\alpha)}{\partial \alpha} &= \sum_{\*{s}_{t-1}}
   \Big(1-\tanh^{2}{h_{i,t}(\alpha)}\Big) \Delta h_{i,t} P(\*{s}_{t-1})  ,
    \\ \frac{\partial m_{i,t}(\alpha=0)}{\partial \alpha}
    &=  (1-m_{i,t}^2) \Big(-\Theta_{i,t} + H_{i}+\sum_{j}J_{ij}m_{j,t-1}\Big).
\end{align}

The first order or naive mean-field approximation is then
\begin{align}
     \Theta_{i,t} &= H_{i} + \sum_j J_{ij} m_{j,t-1}  + \left[\mathcal{O}(\alpha^1)\right]_{\alpha=1},
     \\ m_{i}(t) &\approx \tanh{[H_{i}+\sum_{j}J_{ij}m_{j,t-1}]}.
\end{align}

For the second order approximation, we have
\begin{align}
   \frac{\partial^{2} m_{i,t}(\alpha)}{\partial \alpha^{2}_{t}}
   &= -2 \sum_{\*{s}_{t-1}} 
   \tanh{h_{i,t}(\alpha)} \Big(1-\tanh^{2}{h_{i,t}(\alpha)}\Big)
   \Delta h_{i,t}^{2}  P(\*{s}_{t-1}),
    \\\frac{\partial^{2} m_{i,t}(\alpha=0)}{\partial \alpha^{2}_{t}} &= -2  m_{i,t} \Big(1 - m_{i,t}^2\Big) \bigg[\Big(-\Theta_{i,t} + H_{i}+\sum_{j}J_{ij}m_{j,t-1}\Big)^{2}
     \\&\phantom{=}  + \sum_{j,k} J_{ij}J_{ik}C_{jk,t-1} \bigg].
\end{align}

Then we solve the following second order equation:
\begin{equation}
	 \frac{\partial   m_{i,t}(\alpha=0)}{\partial \alpha}
	 + \frac{1}{2} \frac{\partial^2   m_{i,t}(\alpha=0)}{\partial \alpha^2}
	= 0 +\mathcal{O}(\alpha^3)\rvert_{\alpha=1}.
\end{equation}

To compute the second order approximation, we can take advantage of the fact that the second order term contains the expression $\left[\alpha^2 (-\Theta_{i,t} + H_{i} + \sum_j J_{ij} m_j)^2\right]_{\alpha=1}$. We know that $\left[ \alpha (1-m_i^2) (-\Theta_{i,t} + H_{i} + \sum_j J_{ij} m_j)\right]_{\alpha=1}  =       \left[\mathcal{O}(\alpha^{2})\right]_{\alpha=1}$ and thus $\left[ \alpha^2 (-\Theta_{i,t} + H_{i} + \sum_j J_{ij} m_j)^2\right]_{\alpha=1}  = \left[\mathcal{O}(\alpha^{4})\right]_{\alpha=1}$. This yields the TAP equations:
\begin{align}
     \Theta_{i,t} 
     &= H_{i}+\sum_{j}J_{ij}m_{j,t-1} - m_{i,t}\sum_{j,k} J_{ij}J_{ik}C_{jk,t-1} + \left[\mathcal{O}(\alpha^2)\right]_{\alpha=1},
     \label{app-eq:P01-Theta-TAP}
     \\ m_{i}(t) &\approx \tanh{[H_{i}+\sum_{j}J_{ij} m_{j,t-1}- m_{i,t} \sum_{j,k} J_{ij} J_{ik} C_{jk,t-1}]}.
     \label{app-eq:P01-m-TAP}
\end{align}

This result is a reminiscence of the TAP equations obtained for symmetric networks and asymmetric networks, which generally take the form $\Theta_{i,t} \approx H_{i} + \sum_j J_{ij} m_{j,t-1} - m_{i,t}\sum_{j} J_{ij}^2 (1-m_j^2)$ \cite{kappen_mean_2000, roudi_dynamical_2011}. The only difference is that previous results approximated either the stationary state of the network, or the probability of trajectories over several updates of the network dynamics. The consequence of these is that previous results ignored correlations at previous states (since they were also expanded from the independent model, thus $C_{jl}$ terms become zero when $i\neq j$ and $1-m_i^2$ otherwise).

\subsection*{Equal-time correlations}

When $i=j$, $C_{ii,t} = 1 -m_{i,t}(\alpha)^2$.
When $i\neq j$, correlations in the system can be obtained by expanding
\begin{equation}
    C_{ik,t}(\alpha) = \sum_{\*{s}_{t},\*{s}_{t-1}} (s_{i,t}-m_{i,t}(\alpha)) (s_{k,t}-m_{k,t}(\alpha)) P_\alpha(\*{s}_{t}|\*{s}_{t-1}) P(\*{s}_{t-1}),
\end{equation}
over $\alpha=0$:
\begin{equation}
	 C_{ik,t}(\alpha) =  \sum_{n=0}^m  \frac{\alpha^n}{n!}  \frac{\partial^n C_{ik,t}(\alpha=0)}{\partial \alpha^n} + \mathcal{O}(\alpha^{(m+1)}).
\end{equation}

The expanded terms of each order and its evaluation at $\alpha=0$ are obtained as follows.

The zeroth order term is:
\begin{align}
	C_{ik,t}(\alpha) =& \sum_{\*{s}_{t-1}}   (\tanh h_{i,t}(\alpha)-  m_{i,t}(\alpha))( \tanh h_{k,t}(\alpha)- m_{k,t}(\alpha))   P(\*{s}_{t-1}), 
	\\ C_{ik,t}(\alpha=0) =&(m_{i,t}  - m_{i,t}) (m_{k,t} - m_{k,t}) = 0.
\end{align}

The first order term yields
\begin{align}
	\frac{\partial C_{ik,t}(\alpha)}{\partial \alpha} &= \sum_{\*{s}_{t-1}}   \Big( 
		\big( (1 - \tanh^2 h_{i,t}(\alpha))\Delta h_{i,t}
		\\ &\phantom{=} - \frac{\partial m_{i,t}(\alpha)}{\partial \alpha}\big) 
		\big(\tanh h_{i,t}(\alpha) -m_{k,t}(\alpha) \big)
		\\ &\phantom{=} + \big( (1 - \tanh^2 h_{k,t}(\alpha))\Delta h_{k,t}
		\\ &\phantom{=} - \frac{\partial m_{k,t}(\alpha)}{\partial \alpha}\big) 
		\big(\tanh h_{i,t}(\alpha) - m_{i,t}(\alpha) \big) 
		\Big)  P(\*{s}_{t-1}),
	\\ \frac{\partial C_{ik,t}(\alpha=0)}{\partial \alpha} &= 0.
\end{align}
Therefore the nMF equation is obtained as
\begin{equation}
    C_{ik,t}  \approx 0.
    \label{app-eq:P01-C-nMF}
\end{equation} 

The second order term is:
\begin{align}
	\frac{\partial^2 C_{ik,t}(\alpha)}{\partial \alpha^2}&= \sum_{\*{s}_{t-1}}   \Big(  
		2 \big( (1 - \tanh^2 h_{i,t}(\alpha))\Delta h_{i,t}
		\\ &\phantom{=}- \frac{\partial m_{i,t}(\alpha)}{\partial \alpha}\big) 
		\big( (1 - \tanh^2 h_{k,t}(\alpha))\Delta h_{k,t}
		- \frac{\partial m_{k,t}(\alpha)}{\partial \alpha}\big) 
		\\ &\phantom{=}+  \big( -2 \tanh h_{i,t}(\alpha)(1 - \tanh^2 h_{i,t}(\alpha))\Delta h_{i,t}^2 
		\\ &\phantom{=}- \frac{\partial^2 m_{i,t}(\alpha)}{\partial \alpha^2}\big) 
		\big(\tanh h_{k,t}(\alpha) - m_{k,t}(\alpha) \big)
		\\ &\phantom{=}+  \big( -2 \tanh h_{k,t}(\alpha)(1 - \tanh^2 h_{k,t}(\alpha))\Delta h_{k,t}^2 
		\\ &\phantom{=}- \frac{\partial^2 m_{k,t}(\alpha)}{\partial \alpha^2}\big) 
		\big(\tanh h_{i,t}(\alpha) - m_{i,t}(\alpha) \big) 
		\Big)  P(\*{s}_{t-1}),
	\\ \frac{\partial^2 C_{ik,t}(\alpha=0)}{\partial \alpha^2} &= 2  (1 - m_{i,t}^2)  (1 - m_{k,t}^2) ( \sum\limits_{ln} J_{il} J_{jn} C_{ln,t-1}).
\end{align}
So the TAP equation for the correlations can be described as:
\begin{align}
C_{ik,t}
	&\approx (1 - m_{i,t}^2)  (1 - m_{k,t-1}^2) \sum\limits_{ln} J_{il} J_{jn} C_{ln,t-1}.
\label{app-eq:P01-C-TAP}
\end{align}
To obtain this result, correlations of order $3$ (and more) $C_{ijk,t} = \sum_{\*{s}_t} (s_{i,t} - m_{i,t}) (s_{j,t}-m_{j,t}) (s_{k,t}-m_{k,t}) P(\*{s}_t) = 0+\left[\mathcal{O}(\alpha^3)\right]_{\alpha=1}$ are ignored in the TAP equations if $i\neq j \neq k$. Otherwise we used $C_{iik,t} = -2 m_{i,t} C_{ik,t}$ and $C_{iii,t} = -2 m_{i,t} (1-m_{i,t}^2)$.

The obtained expression has a form that is similar to the equations in \cite{roudi_mean_2011} but presents some differences since is computed from the same expansion as the TAP equations, instead of performing a new expansion from the approximation of $\Theta_{i,t}$.

\subsection*{Time-delayed correlations}
We introduce the $\alpha$-dependent time-delayed correlations as \begin{equation}
    D_{il,t}(\alpha)= \sum_{\*{s}_{t},\*{s}_{t-1}} (s_{i,t}- m_{i,t}(\alpha))  (s_{l,t-1}-m_{l,t-1}) P_\alpha(\*{s}_{t}|\*{s}_{t-1}) P(\*{s}_{t-1}).
\end{equation}
We approximate $D_{il,t}$ by expanding this equation around $\alpha=0$.

The zeroth order term yields:
\begin{align}
	D_{il,t}(\alpha)=& \sum_{\*{s}_{t-1}}   (\tanh h_{i,t}(\alpha)- m_{i,t}(\alpha))(s_{l,t-1}-m_{l,t-1})   P(\*{s}_{t-1}), 
	\\ D_{il,t}(\alpha=0) =& 0.
\end{align}

The first order term is:
\begin{align}
	\frac{\partial D_{il,t}(\alpha)}{\partial \alpha}&= \sum_{\*{s}_{t-1}}  \sum_{\*{s}_{t-1}}   \Big( 
		\big( (1 - \tanh^2 h_{i,t}(\alpha))\Delta h_{i,t}
		\\ &\phantom{=}- \frac{\partial m_{i,t}(\alpha)}{\partial \alpha}\big) 
		 (s_{l,t-1}-m_{l,t-1}) \big)
		 P(\*{s}_{t-1}),
	\\ \frac{\partial D_{il,t}(\alpha=0)}{\partial \alpha}&=   (1 - m_{i,t}^2) \sum\limits_{j} J_{ij} C_{jl,t-1},
\end{align}
therefore obtaining that for the nMF equation:
\begin{equation}
D_{il,t} \approx (1 - m_{i,t}^2) \sum\limits_{j} J_{ij} C_{jl,t-1}.
\label{app-eq:update_delayed-correlations}
\end{equation}

The second order term is:
\begin{align}
	\frac{\partial^2 D_{il,t}(\alpha)}{\partial \alpha^2} 
	&= \sum_{\*{s}_{t-1}}   \Big( 
		\big( -2 \tanh h_{i,t}(\alpha) (1 - \tanh^2 h_{i,t}(\alpha))\Delta h_{i,t}^2 
		\\ &\phantom{=} - \frac{\partial^2 m_{i,t}(\alpha)}{\partial \alpha^2} \big) 
		 (s_{l,t-1}-m_{l,t-1}) \Big)
		 P(\*{s}_{t-1}) \\
	&=  -2 
		\tanh h_{i,t}(\alpha) (1 - \tanh^2 h_{i,t}(\alpha)) \sum_{\*{s}_{t-1}} \Delta h_{i,t}^2 
		 (s_{l,t-1}-m_{l,t-1}) 
		 P(\*{s}_{t-1}).	 
\end{align}
Here we note that
\begin{align}
    & \sum_{\*{s}_{t-1}}  \Delta h_{i,t}^{2}(s_{l,t-1}-m_{l,t-1}) P(\*{s}_{t-1}) 
    \\	&=\sum_{\*s_{t-1}} \Big(-\Theta_{i,t}+H_{i}+\sum_{j}J_{ij}m_{j,t-1}+\sum_{j}J_{ij}(s_{j,t-1}-m_{j,t-1}) \Big)^{2}(s_{l,t-1}-m_{l,t-1})P(\*s_{t-1})
	\\&=\Big(-\Theta_{i,t}+H_{i}+\sum_{j}J_{ij}m_{j,t-1}\Big)^{2}\sum_{\*s_{t-1}}(s_{l,t-1}-m_{l,t-1})P(\*s_{t-1})
	\\&+2\Big(-\Theta_{i,t}+H_{i}+\sum_{j}J_{ij}m_{j,t-1}\Big)\sum_{\*s_{t-1}}\sum_{j}J_{ij}(s_{j,t-1}-m_{j,t-1})(s_{l,t-1}-m_{l,t-1})P(\*s_{t-1})
	\\&+\sum_{\*s_{t-1}}\Big(\sum_{j}J_{ij}(s_{j,t-1}-m_{j,t-1})\Big)^{2}(s_{l,t-1}-m_{l,t-1})P(\*s_{t-1})
	\\&=2\Big(-\Theta_{i,t}+H_{i}+\sum_{j}J_{ij}m_{j,t-1}\Big)\sum_{j}J_{ij}C_{jl,t-1}+\sum_{j,k}J_{ij}J_{ik}C_{jkl,t-1}.
\end{align}
where $C_{jkl,t-1} \equiv \sum_{\*s_{t-1}}(s_{j,t-1}-m_{j,t-1})(s_{k,t-1}-m_{k,t-1})(s_{l,t-1}-m_{l,t-1})P(\*s_{t-1})$. 
We then obtain 
\begin{align}
	\frac{\partial^2 D_{il,t}(\alpha=0)}{\partial \alpha^2} &= -2 m_{i,t} (1 - m_{i,t}^2) \sum\limits_{j,k} J_{ij} J_{ik} C_{jkl,t-1}
	-4 m_{i,t} (1 - m_{i,t}^2) (-\Theta_{i,t} + H_i + \sum_j J_{ij} m_j) \sum\limits_{j} J_{ij} C_{jl,t-1}.
\end{align}
Here, the second term above is equal to  $0 + \mathcal{O}(\alpha)$, which makes it negligible (i.e. order larger than quadratic) when computing $\alpha^2 \frac{\partial^2 D_{il,t}(\alpha=0)}{\partial \alpha^2}$.
Thus the second order expansion for delayed correlations can be described as:
\begin{equation}
D_{il,t} =  (1 - m_{i,t}^2) \big( \sum\limits_{j} J_{ij} C_{jl,t-1} -m_{i,t} \sum\limits_{j,k} J_{ij} J_{ik} C_{jkl,t-1}  \big) + \left[\mathcal{O}(\alpha^3)\right]_{\alpha=1}.
\end{equation}
Moreover, if we consider that $C_{jkl,t-1}$ terms are going to be equal to $0+\left[\mathcal{O}(\alpha^3)\right]_{\alpha=1}$ when $j\neq l \neq n$, we have the TAP approximation
\begin{equation}
    D_{il,t} \approx (1 - m_{i,t}^2) \sum\limits_{j} J_{ij} C_{jl,t-1} (1 
    + 2 J_{il} m_{i,t}  m_{l,t-1} )
	+ \left[\mathcal{O}(\alpha^3)\right]_{\alpha=1}.
\end{equation}

The nMF Equation for time-delayed correlations is similar to the first order approximation obtained by \cite{roudi_mean_2011}. The second order approximation differs since they do a new expansion over one time step of the obtained TAP expression obtained for the whole trajectory. In our case, since we apply the TAP expansion for one-step updates in all cases, we can derive an expression from the same expansion that obtains the TAP equation for updating the mean fields of the system.
\newpage
\section{Plefka[$t-1$]}
\label{app:plefka_t-1}

This approximation uses the following approximated marginal probability distribution.
\begin{equation}
    P_\alpha^{[t-1]}(\*{s}_{t},\*{s}_{t-1})=\sum_{\*{s}_{t-2}}  P(\*{s}_{t}|\*{s}_{t-1}) P_{\alpha}(\*{s}_{t-1}|\*{s}_{t-2}) P(\*{s}_{t-2}),
\end{equation}
where activity at time $t$ is normally defined by
\begin{align}
	  P(s_{i,t}|\*{s}_{t-1}) = & \frac{\mathrm{e}^{ s_{i,t} {h}_{i,t}}}{2 \cosh {h}_{i,t}},
	  \\ h_{i,t}
	  =& H_{i} + \sum_j J_{ij} s_{j,t-1},
\end{align}
whereas activity at time $t-1$ is mediated by the parameter $\alpha$
\begin{align}
	  P_{\alpha}(s_{i,t-1}|\*{s}_{t-2}) = & \frac{\mathrm{e}^{ s_{i,t-1} {h}_{i,t-1}(\alpha)}}{2 \cosh {h}_{i,t-1}(\alpha)},
	  \\ h_{i,t-1}(\alpha) 
	  =& (1-\alpha) \Theta_{i,t-1} + \alpha (H_{i} + \sum_j J_{ij} s_{j,t-2}).
\end{align}

Computing the values of $\*\Theta_{t-1}$ is equivalent to the calculations in \ref{app:plefka_t}. 
However, in the calculations below we will see that we will need only $ m_{i,t-1}$ to compute the statistics at time $t$. Namely, $ m_{i,t-1}$ is the only value we need to know to make the approximation at $t-1$.

Now, we calculate $m_{i,t}$ by using its $\alpha$-dependent approximation, defined as
\begin{equation}
    m_{i,t}(\alpha) = \sum_{\*{s}_{t},\*{s}_{t-1}} s_{i,t} P_\alpha^{[t-1]}(\*{s}_{t},\*{s}_{t-1}) = \sum_{\*{s}_{t-1},\*{s}_{t-2}} \tanh h_{i,t} P_{\alpha}(\*{s}_{t-1}|\*{s}_{t-2}) P(\*{s}_{t-2}).
\end{equation}
Approximating its value by expanding around $\alpha=0$ yields
\begin{equation}
	 m_{i,t}(\alpha) = m_{i,t}(\alpha=0) + \sum_{k=1}^n  \frac{\alpha^k}{k!}  \frac{\partial^k  m_{i,t}(\alpha=0)}{\partial \alpha^k} + \mathcal{O}(\alpha^{(n+1)}).
\end{equation}
We solve this equation at $\alpha=1$. The approximation yields the nMF equations when we ignore quadratic terms and higher, and the TAP equations when ignoring third and higher order terms. However, in this case we will only compute the terms in the nMF equation, since the second order yields marginals that are complicated to evaluate. In the case of the first order term, we will show how to estimate the corresponding marginals at the end of this Supplementary Note.

The derivatives of this distribution with respect to $\alpha$ is given by
\begin{align}
	  \frac{\partial P_{\alpha}(s_{i,t-1}|\*{s}_{t-2})}{\partial \alpha} = &  (s_{i,t-1}- \tanh {h}_{i,t-1}(\alpha))\Delta h_{i,t-1}P_{\alpha}(s_{i,t-1}|\*{s}_{t-2}).
\end{align}
Again, we define $h_{i,t-1}(\alpha) = \Theta_{i,t-1} + \alpha \, \Delta h_{i,t-1}$, where $\Delta h_{i,t-1} = -\Theta_{i,t-1} + H_i + \sum_j J_{ij} s_{j,t-2}$ represents deviation from the independent model. The derivatives of each order are derived as follows. 

The zeroth order term is:
\begin{align}
	m_{i,t}(\alpha)&= \sum_{\*{s}_{t-1},\*{s}_{t-2}} \tanh h_{i,t}P_\alpha(\*{s}_{t-1}|\*{s}_{t-2}) P(\*{s}_{t-2}),
	\\ m_{i,t}(\alpha=0) & = \sum_{\*{s}_{t-1}} \tanh h_{i,t}Q(\*{s}_{t-1}).
\end{align}

The first order terms are:
\begin{align}
	\frac{\partial m_{i,t}(\alpha)}{\partial \alpha} &
		= \sum_{\*{s}_{t-1},\*{s}_{t-2}} \tanh h_{i,t}\sum_k  (s_{k,t-1} - \tanh h_{k,t-1}(\alpha) )\Delta h_{k,t-1}	P_\alpha(\*{s}_{t-1}|\*{s}_{t-2}) P(\*{s}_{t-2}), 
	\\ \frac{\partial m_{i,t}(\alpha=0)}{\partial \alpha}  & = \sum_k (- \Theta_{k,t-1} + H_k + \sum_l J_{kl} m_{l,t-2})  \sum_{\*{s}_{t-1}} \tanh h_{i,t} (s_{k,t-1}-m_{k,t-1}) Q(\*{s}_{t-1}).
\end{align}

Since  $\left[\alpha(H_k - \Theta_{k,t-1} + \sum_l J_{kl} m_{l,t-2})\right]_{\alpha=1} = 0 + \left[ \mathcal{O}(\alpha^2) \right]_{\alpha=1}$ (\ref{app:plefka_t}) we can ignore the first order terms, leading to the first order expansion
\begin{align}
	m_{i,t}& \approx \Gamma_{i,t}+ \left[ \mathcal{O}(\alpha^2) \right]_{\alpha=1},
\end{align}
where $\Gamma_{i,t}= \sum_{\*{s}_{t-1}} \tanh h_{i,t}Q(\*{s}_{t-1})$.
Moreover, since $Q(\*{s}_{t-1})$ is an independent distribution, for a large number of units, applying the central limit theorem, it can be approximated by a Gaussian distribution (see last section of this Supplementary Note). 

%

\subsection*{Equal-time correlations}
When $i\neq k$, correlations in the system are calculated as 
\begin{equation}
    C_{ik,t}(\alpha) = \sum_{\*{s}_{t},\*{s}_{t-1}} (s_{i,t}-m_{i,t}(\alpha))(s_{k,t} - m_{k,t}(\alpha)) P_\alpha(\*{s}_{t-1}|\*{s}_{t-2}) \*{s}_{t},\*{s}_{t-1}(\*{s}_{t},\*{s}_{t-1}).
\end{equation}{}
Again, we compute this using a Plefka expansion:
\begin{equation}
	 C_{ik,t}(\alpha) =  \sum_{n=0}^m  \frac{\alpha^n}{n!}  \frac{\partial^n C_{ik,t}(\alpha=0)}{\partial \alpha^n} + \mathcal{O}(\alpha^{(m+1)}).
\end{equation}
Otherwise, when $i=j$, we have $C_{ii,t} = 1 -m_{i,t}^2$. The derivatives of each order are derived as follows. 

The zeroth order term:
\begin{align}
	C_{ik,t}(\alpha)& = \sum_{\*{s}_{t-1},\*{s}_{t-2}} (\tanh h_{i,t}-  m_{i,t}(\alpha))( \tanh h_{k,t}- m_{k,t}(\alpha))  P_\alpha(\*{s}_{t-1}|\*{s}_{t-2}) P(\*{s}_{t-2}), 
	\\ C_{ik,t}(\alpha=0) & = \sum_{\*{s}_{t-1}} \tanh h_{i,t}\tanh h_{k,t} Q(\*{s}_{t-1}) - m_{i,t}m_{k,t} \equiv  \Gamma_{ik,t} .
\end{align}

The first order term yields:
\begin{align}
	\frac{\partial C_{ik,t}(\alpha)}{\partial \alpha} = &
		\sum_{s(t-1)} \Big((\tanh h_{i,t}-  m_{i,t}(\alpha))( \tanh h_{k,t}- m_{k,t}(\alpha))  \sum_m (s_{m,t-1} - \tanh h_{m,t-1}(\alpha)) \Delta h_{m,t-1}
		\\  & - \frac{\partial m_{i,t}(\alpha)}{\partial \alpha} ( \tanh h_{k,t}- m_{k,t}(\alpha)) - ( \tanh h_{i,t}- m_{i,t}(\alpha))  \frac{\partial m_{k,t}(\alpha)}{\partial \alpha} 
	 \Big) P_\alpha(\*{s}_{t-1}|\*{s}_{t-2}) P(\*{s}_{t-2}),
	\\ \frac{\partial C_{ik,t}(\alpha=0)}{\partial \alpha} = & - \sum_m (H_m - \Theta_{m,t-1} + \sum_n J_{mn} m_{n,t-2}) ( \sum_{\*s_{t-1}} ( \tanh h_{k,t}- m_{k,t} 
	\\&+  \tanh h_{i,t}- m_{i,t}) (s_{m,t-1}-m_{m,t-1}) Q(\*s_{t-1}).
\end{align}

Since  $\left[\alpha(H_m - \Theta_{m,t-1} + \sum_n J_{mn} m_{n,t-2})\right]_{\alpha=1} = 0 + \left[ \mathcal{O}(\alpha^2) \right]_{\alpha=1}$ (\ref{app:plefka_t}), the nMF equation is:
\begin{equation}
	C_{ik,t} \approx  \Gamma_{ik,t}.
\end{equation} 

\subsection*{Time-delayed correlations}

Time-delayed correlations in the system are calculated as 
\begin{equation}
    D_{il,t}(\alpha) = \sum_{\*{s}_{t},\*{s}_{t-1}} (s_{i,t}-m_{i,t}(\alpha))(s_{l,t-1} - m_{l,t-1}(\alpha)) P_\alpha^{[t-1]}(\*{s}_{t},\*{s}_{t-1}).
    \label{app-eq:gamma_i_t_l}
\end{equation}
We compute this using a Plefka expansion:
\begin{equation}
	 D_{il,t}(\alpha) =  \sum_{n=0}^m  \frac{\alpha^n}{n!}  \frac{\partial^n D_{il,t}(\alpha=0)}{\partial \alpha^n} + \mathcal{O}(\alpha^{(m+1)}).
\end{equation}

The zeroth order terms are:
\begin{align}
	D_{il,t}(\alpha)& = \sum_{\*{s}_{t-1},\*{s}_{t-2}} (\tanh h_{i,t}-  m_{i,t}(\alpha))( s_{l,t-1}- m_{l,t-1}(\alpha))  P_\alpha(\*{s}_{t-1}|\*{s}_{t-2}) P(\*{s}_{t-2}), 
	\\ D_{il,t}(\alpha=0) &= \sum_{\*{s}_{t-1}} (\tanh h_{i,t}-  m_{i,t})(s_{l,t-1}- m_{l,t-1})  Q(\*{s}_{t-1}) \equiv \Gamma_{i,t}^{(l)}.
\end{align}

The first order terms are:
\begin{align}
	\frac{\partial D_{il,t}(\alpha)}{\partial\alpha}& = \sum_{\*{s}_{t-1},\*{s}_{t-2}} \Big(
	(\tanh h_{i,t}-  m_{i,t}(\alpha))( s_{l,t-1}- m_{l,t-1}(\alpha)) \sum_k (s_{k,t-1} - \tanh h_{k,t-1}(\alpha))\Delta h_{k,t-1}
	\\ &\phantom{=}   -  \frac{\partial m_{i,t}(\alpha))}{\partial\alpha}( s_{l,t-1}- m_{l,t-1}(\alpha)) - (\tanh h_{i,t}-  m_{i,t}(\alpha))\frac{\partial m_{l,t-1}(\alpha)}{\partial\alpha}
	\Big) P_\alpha(\*{s}_{t-1}|\*{s}_{t-2}) P(\*{s}_{t-2}), 
	\\ \frac{\partial D_{il,t}(\alpha=0)}{\partial\alpha} &=  \sum_k \sum_{\*{s}_{t-1}}
	(\tanh h_{i,t}-  m_{i,t})( s_{l,t-1}- m_{l,t-1}(\alpha))  (s_{k,t-1} 
	\\ &\phantom{=} - m_{k,t-1}(\alpha)) (- \Theta_{k,t-1} + H_k  + \sum_n J_{kn} m_{n,t-2}) Q(\*{s}_{t-1}).
\end{align}

Since  $\left[\alpha(H_k - \Theta_{k,t-1} + \sum_n J_{kn} m_{n,t-2})\right]_{\alpha=1} = 0 + \left[ \mathcal{O}(\alpha^2) \right]_{\alpha=1}$ (\ref{app:plefka_t}), the nMF equation is:
\begin{equation}
	D_{il,t} \approx  \Gamma_{i,t}^{(l)}.
	\label{app-eq:D_nMF}
\end{equation}

\subsection*{Gaussian approximations}
\label{sec:gaussian-approximations}
The integrals for computing the first order means and correlations can be directly obtained applying the central limit theorem to approximate a set of independent binary signals to a Gaussian distribution.

Thus we obtain
\begin{equation}
 \Gamma_{i,t}= \sum_{\*{s}_{t-1}} \tanh h_{i,t}Q(\*{s}_{t-1}) \approx \int \mathrm{D}_{x} \tanh [g_{i,t}+ x \sqrt{\Delta_{i,t}} ],
\end{equation}
where $\mathrm{D}_{x}$ denotes an integral using a Gaussian distribution:
\begin{equation}
    \mathrm{D}_{x} = \frac{\mathrm{d}x}{\sqrt{2 \pi}} \exp\pr{-\frac{1}{2}x^2}.
\end{equation}
The other parameters are $g_{i,t}= H_i + \sum_j J_{ij} m_{j,t-1}$ and  $\Delta_{i,t} = \mathrm{Var} [h_{i,t}]_{\alpha=0}= \sum_j J_{ij}^2 (1- m_{j,t-1}^2)$. 

Similarly, we have
\begin{align}
 \Gamma_{ik,t} &= \sum_{\*{s}_{t-1}} \tanh h_{i,t}\tanh h_{k,t} Q(\*{s}_{t-1}) 
 \\ & \approx \int \mathrm{D}_{xy}^{\rho_{ik}} \tanh [g_{i,t}+ x \sqrt{\Delta_{i,t}} ]  \tanh [g_{k,t} + y \sqrt{\Delta_{k,t}} ]
 \\ & = \int \mathrm{D}_{x} \mathrm{D}_{y} \tanh [g_{i,t}+ (x \sqrt{\frac{1+\rho_{ik}}{2}} + y\sqrt{\frac{1-\rho_{ik}}{2}}) \sqrt{\Delta_{i,t}} ]  \tanh [g_{k,t} + (x \sqrt{\frac{1+\rho_{ik}}{2}} - y\sqrt{\frac{1-\rho_{ik}}{2}}) \sqrt{\Delta_{k,t}} ],
\end{align}
where $\mathrm{D}_{xy}^{\rho_{ik}}$ denotes an integral using a bivariate Gaussian distribution:
\begin{equation}
    \mathrm{D}_{xy}^{\rho_{ik}} = \frac{\mathrm{d}x \mathrm{d}y}{2 \pi \sqrt{1-\rho_{ik}^2}} \exp\pr{-\frac{1}{2}\frac{(x^2+y^2) - 2\rho_{ik}  x y}{1-\rho_{ik}^2} }.
\end{equation}
The other parameters are $\rho_{ij} = \frac{\Delta_{ij,t}}{\sqrt{\Delta_{ii,t}\Delta_{kk,t}}}$, with $\Delta_{ik,t}=\mathrm{Cov} (h_{i,t}, h_{k,t})_{\alpha=0} = \sum_{j} J_{ij} J_{kj} (1-m_{j,t-1}^2)$.
In the following step, we applied an orthogonal transformation to the previous bivariate distribution, involving a change of variables $x' = x \sqrt{\frac{1+\rho_{ik}}{2}} + y\sqrt{\frac{1-\rho_{ik}}{2}}$ and $y' = x \sqrt{\frac{1+\rho_{ik}}{2}} - y\sqrt{\frac{1-\rho_{ik}}{2}}$ that removes the coupling terms in the bivariate normal distribution.

The integrals of the Gaussian approximations of $\Gamma_{i,t}$ and $\Gamma_{ik,t}$ are easy to compute. The problem arises when we deal with the more complex terms $\Gamma_{i,t}^{(l)}$ in Supplementary Eq.~\ref{app-eq:D_nMF}, which can be computed by similar approximations but multiplies the number of integrals to be solved. 

For large system sizes, terms obtained from the first and second order expressions like $\Gamma_{i,t}^{(l)}$ (Supplementary Eq.~\ref{app-eq:gamma_i_t_l}) can be obtained by assuming that the values of individual weights are small (e.g. $J_{ij} = \mathcal{O}(1/N)$). 
We can  compute 
\begin{align}
    D_{il,t} \approx \Gamma_{i,t}^{(l)} = \sum_{\*{s}_{t-1}} (s_{l,t}- m_{l,t})  \tanh h_{i,t} Q(\*{s}_{t-1}),
\end{align}
where the constant term $m_{i,t}$ was removed from Supplementary Eq.~\ref{app-eq:gamma_i_t_l} as it yields zero when averaging its product with $(s_{l,t}- m_{l,t})$.

With independent spins at $t-1$, we can approximate this quantity with the aid of a bivariate Gaussin distribution of fields $h_{k,t}, h_{i,t}$. Knowing that $h_{k,t}- \langle h_{k,t}\rangle = \sum_l J_{kl} (s_{l,t-1}- m_{l,t-1}) $, we can approximate the following quantity 
\begin{align}
    \sum_l J_{kl} D_{il,t} \approx &  \sum_{\*{s}_{t-1}} \sum_l J_{kl} (s_{l,t-1}- m_{l,t-1})  \tanh h_{i,t} Q(\*{s}_{t-1})
    \\ = & \sum_{\*{s}_{t-1}} ( h_{k,t}- \langle h_{k,t}\rangle)  \tanh h_{i,t} Q(\*{s}_{t-1})
    \\ \approx &   \int \mathrm{D}_{xy}^{\rho_{ik}} y \sqrt{\Delta_{k,t}} \tanh [g_{i,t}+ x \sqrt{\Delta_{i,t}} ]. 
\end{align}
where $\langle h_{k,t} \rangle = H_k + \sum_l m_{l,t-1} $.

As in \cite{mezard_exact_2011}, if we assume that $\rho_{ik}$ is small,$\mathrm{D}_{xy}^{\rho_{ik}} $ can be approximated as:
\begin{equation}
	\mathrm{D}_{xy}^{\rho_{ik}} \approx \mathrm{D}_{x} \mathrm{D}_{y}(1 +  \rho_{ik} x y).
\end{equation}

Then 
\begin{align}
    \sum_l J_{kl} D_{il,t} \approx &   \int \mathrm{D}_{xy}^{\rho_{ik}} y \sqrt{\Delta_{j,t}} \tanh [g_{i,t}+ x \sqrt{\Delta_{i,t}} ] 
    \\ \approx &   \int \mathrm{D}_{x} \mathrm{D}_{y} (y + \rho_{ik} x y^2) \sqrt{\Delta_{k,t}} \tanh [g_{i,t}+ x \sqrt{\Delta_{i,t}} ] 
    \\ = & \rho_{ik}\sqrt{\Delta_{k,t}}  \int D_{x} x  \tanh [g_{i,t}+ x \sqrt{\Delta_{i,t}} ] 
    \\ = & \rho_{ik}\sqrt{\Delta_{i,t} \Delta_{k,t}} \int D_{x} \left(1 - \tanh^2 [g_{i,t}+ x \sqrt{\Delta_{i,t}} ]\right),
\end{align}
where the last step was obtained by partial integration.

As $\rho_{ik}\sqrt{\Delta_{i,t}\Delta_{k,t}} = \sum_{jl} J_{ij} J_{kl} C_{jl,t-1}$
we have that
\begin{equation}
	\sum_l J_{kl} D_{il,t} \approx a_{i,t} \sum_{jl} J_{ij} J_{kl} C_{jl,t-1},
\end{equation}
where 
\begin{equation}
	a_{i,t} = \int D_{x} \left(1 - \tanh^2 [g_{i,t}+ x \sqrt{\Delta_{i,t}} ]\right).
\end{equation}

Therefore, we have the following approximation for the delayed correlations,
\begin{equation}
	 D_{il,t} \approx a_{i,t} \sum_{j} J_{ij} C_{jl,t-1}.
\end{equation}

\newpage

\section{Pairwise Plefka expansions, Plefka2[$t$]}
\label{app:plefka_D}

\subsection*{Pairwise delayed spin correlations}
\label{app:plefka_D1}
Instead of a manifold of independent distributions, in this approximation we consider a manifold $\mathcal{Q}$ with a pairwise probability distribution:
\begin{equation}
	  Q(s_{i,t},s_{l,t-1}) = Q(s_{i,t}|s_{l,t-1}) Q(s_{l,t-1}) =  \frac{\mathrm{e}^{s_{i,t} \theta_{i,t}(s_{l,t-1})} }{2 \cosh \theta_{i,t}(s_{l,t-1})}
	  \frac{\mathrm{e}^{s_{l,t-1} \Theta_{l,t-1}}}{2 \cosh \Theta_{l,t-1}},
\end{equation}
where $\theta_{i,t}(s_{l,t-1}) = \Theta_{i,t} + \Delta_{il,t} s_{l,t-1}$.
Here $Q(s_{l,t-1})$ is the independent probability distribution for $s_{l,t-1}$ computed as in \ref{app:plefka_t}, and $Q(s_{i,t}|s_{l,t-1})$ is a conditional probability distribution we use to construct the pairwise probability distribution $Q(s_{i,t},s_{l,t-1})$ using the chain rule.

We use this pairwise model to approximate the distribution
\begin{equation}
    P(s_{i,t},s_{l,t-1}) = \sum_{\*s_{\setminus l,t-1}} P(s_{i,t},\*{s}_{t-1})
     = \sum_{\*s_{\setminus l,t-1}} P(s_{i,t}|\*{s}_{t-1}) P(\*{s}_{t-1}).
\end{equation}
with $s_{\setminus l,t-1}$ containing all elements of $\*s_{t-1}$ except $s_{l,t-1}$.

As in previous cases, we want to find an approximation of the probability distribution at time $t$ that minimizes the relative entropy
\begin{align}
	D(P(s_{i,t},s_{l,t-1})||Q(s_{i,t},s_{l,t-1})) &= \sum_{\substack{s_{i,t}\\s_{l,t-1}}}
	 P(s_{i,t},s_{l,t-1}) \log \frac{P(s_{i,t},s_{l,t-1})}{Q(s_{i,t},s_{l,t-1})}.
\end{align}

Specifically, the mean-field approximation that minimizes the relative entropy is the one that satisfies
\begin{align}
	\frac{\partial D(P(s_{i,t},s_{l,t-1})||Q(s_{i,t},s_{l,t-1}))}{\partial \Theta_{i,t}} &= - \sum_{\substack{s_{i,t}\\s_{l,t-1}}} \Big( s_{i,t} 
	- \tanh[\theta_{i,t}(s_{l,t-1})] \Big) P(s_{i,t},s_{l,t-1})
	\\ &=  \langle s_{i,t} \rangle_Q - \langle s_{i,t} \rangle_P = 0,
	\\ \frac{\partial D(P(s_{i,t},s_{l,t-1})||Q(s_{i,t},s_{l,t-1}))}{\partial \Delta_{il,t}} &= - \sum_{\substack{s_{i,t}\\s_{l,t-1}}} \Big( s_{i,t} s_{l,t-1}  - \tanh[\theta_{i,t}(s_{l,t-1})] s_{l,t-1} \Big) P(s_{i,t},s_{l,t-1})
	\\ &=  \langle \tanh[\theta_{i,t}(s_{l,t-1})] s_{l,t-1} \rangle_{P} - \langle s_{i,t} s_{l,t-1} \rangle_P 
	\\ &=  \langle s_{i,t} s_{l,t-1} \rangle_{(Q\cdot P)} - \langle s_{i,t} s_{l,t-1} \rangle_P = 0,
\end{align}
where $\langle s_{i,t} s_{l,t-1} \rangle_{(Q\cdot P)} = \sum_{s_{i,t},s_{l,t-1}}  s_{i,t} s_{l,t-1} Q(s_{i,t}|s_{l,t-1}) P(s_{l,t-1})$, and we used the equivalence $\tanh[\theta_{i,t}(s_{l,t-1})] = \sum_{s_{i,t}} s_{i,t} \frac{\mathrm{e}^{s_{i,t} \theta_{i,t}(s_{l,t-1})} }{2 \cosh \theta_{i,t}(s_{l,t-1})}$.

This equation states that the closest factorized model has its first and second order moments equal to the first moments of the target distribution $P$. That is, $\langle s_{i,t} \rangle_Q = \langle s_{i,t} \rangle_P = m_{i,t}$ and  $\langle s_{i,t} s_{l,t-1} \rangle_{(Q\cdot P)} = \langle s_{i,t} s_{l,t-1} \rangle_P = D_{il,t} + m_{i,t} m_{l,t-1}$. This is equivalent to having the marginalized distribution for spins $i$ and $l$ equal to the model $Q$, i.e., $P(s_{i,t},s_{l,t-1}) =Q(s_{i,t},s_{l,t-1})$.
If we assume that the distribution $P$ is close to $Q$, we can compute $P(s_{i,t},s_{l,t-1})$ as an expansion with respect to $\alpha$ of the probability distribution:
\begin{equation}
P_{\alpha}(s_{i,t}, s_{l,t-1}) 
= \sum_{\substack{\*s_{\setminus l,t-1}\\ \*{s}_{t-2}}} P_{\alpha}(s_{i,t}|\*{s}_{t-1}) 
P_{\alpha}(s_{l,t-1}|\*{s}_{t-2}) P(\*s_{\setminus l,t-1}|\*{s}_{t-2}) P(\*{s}_{t-2}),
\end{equation}
with
\begin{align}
	  P_\alpha(s_{i,t}| \*{s}_{t-1}) &= \frac{\mathrm{e}^{ s_{i,t} h_{i,t}(\alpha) }}{2\cosh h_{i,t}(\alpha)} 
	  \label{app-eq:pairwise-Dil-ising-alpha-1}
	  \\   h_{i,t}(\alpha) &=    (1 -\alpha) \theta_{i,t}(s_{l,t-1})  +\alpha  (H_{i} + \sum_j J_{ij} s_{j,t-1}),
\end{align}
and
\begin{align}
	  P_{\alpha}(s_{l,t-1}|\*{s}_{t-2}) &= \frac{\mathrm{e}^{ s_{l,t-1} {h}_{l,t-1}(\alpha)}}{2 \cosh {h}_{l,t-1}(\alpha)},
	  \label{app-eq:pairwise-Dil-ising-alpha-2}
	   \\ h_{l,t-1}(\alpha) 
	  &= (1-\alpha) \Theta_{l,t-1} + \alpha (H_l + \sum_n J_{ln} s_{n,t-2}).
\end{align}
When $\alpha$ is set to zero, $P_{\alpha=0}(s_{i,t}, s_{l,t-1}) = Q(s_{i,t}, s_{l,t-1})$, whereas when $\alpha=1$, $P_{\alpha =1}(s_{i,t}, s_{l,t-1}) = P(s_{i,t}, s_{l,t-1})$.
We approximate the values of $\Theta_{i,t},\Theta_{l,t-1},\Delta_{il,t}$ as follows:
\begin{align}
	\frac{\partial P_\alpha(s_{i,t},s_{l,t-1})}{\partial \alpha} =& \sum_{\substack{\*s_{\setminus l,t-1}\\\*s_{t-1}}} 
	\Big( (s_{i,t}-\tanh h_{i,t}(\alpha)) (-\theta_{i,t}(s_{l,t-1}) 
	+ H_{i} + \sum_j J_{ij} s_{j,t-1} )
	\\ &+ (s_{l,t-1}-\tanh h_{l,t-1}(\alpha)) ( - \Theta_{l,t-1}
	+ H_l + \sum_n J_{ln} s_{n,t-2} )
	\Big) 
	\\ &\cdot P_{\alpha}(s_{i,t}|\*{s}_{t-1}) 
P_{\alpha}(s_{l,t-1}|\*{s}_{t-2}) P(\*s_{\setminus l,t-1}|\*{s}_{t-2}) P(\*{s}_{t-2}),
	\\ \frac{\partial P_\alpha(s_{i,t},s_{l,t-1})}{\partial \alpha}\Big \rvert_{\alpha=0} =& \Big( (s_{i,t}-\tanh \theta_{i,t}(s_{l,t-1})) ( -\theta_{i,t}(s_{l,t-1}) + J_{il}(s_{l,t-1}-m_{l,t-1})  + H_{i} + \sum_{j} J_{ij} m_{j,t-1} )
	\\ &+  (s_{l,t-1}-m_{l,t-1}) ( - \Theta_{l,t-1}  + H_l + \sum_n J_{ln} m_{n,t-2} )
	\Big)  Q(s_{i,t}, s_{l,t-1}).
\end{align}

From here, we obtain the nMF equations:
\begin{align}
    \theta_{i,t}(s_{l,t-1}) &\approx H_{i} + \sum_{j} J_{ij} m_{j,t-1} + J_{il}(s_{l,t-1}-m_{l,t-1}),
    \\ \Theta_{l,t-1} &\approx H_l + \sum_n J_{ln} m_{n,t-1}.
\end{align}
The second order expressions are obtained by expanding:
\begin{align}
	\frac{\partial^2 P_\alpha(s_{i,t},s_{l,t-1})}{\partial \alpha^2} 
		= &\sum_{\substack{\*s_{\setminus l,t-1}\\\*s_{t-1}}} 
	\Big( - (1-\tanh^2 h_{i,t}(\alpha)) (-\theta_{i,t}(s_{l,t-1}) + H_{i} + \sum_j J_{ij} s_{j,t-1} )^2  
	\\ &- (1-\tanh^2 h_{l,t-1}(\alpha)) ( - \Theta_{l,t-1} + H_l + \sum_n J_{ln} s_{n,t-2} )^2
	\\ & + \big( (s_{i,t}-\tanh h_{i,t}(\alpha)) (-\theta_{i,t}(s_{l,t-1}) + H_{i} + \sum_j J_{ij} s_{j,t-1} )
	\\ & + (s_{l,t-1}-\tanh h_{l,t-1}(\alpha)) ( - \Theta_{l,t-1}  + H_l + \sum_n J_{ln} s_{n,t-2} )
	\big)^2 \Big) 
	\\ &
	\cdot P_{\alpha}(s_{i,t}|\*{s}_{t-1}) 
P_{\alpha}(s_{l,t-1}|\*{s}_{t-2}) P(\*s_{\setminus l,t-1}|\*{s}_{t-2}) P(\*{s}_{t-2}), \\
	\frac{\partial^2 P_\alpha(s_{i,t},s_{l,t-1})}{\partial \alpha^2}\Big \rvert_{\alpha=0}
	     =& \Big( -2 \tanh \theta_{i,t}(s_{l,t-1}) (s_{i,t}-\tanh \theta_{i,t}(s_{l,t-1})) \big( W_{i,t}^2 + V_{ii,t}\big)
	     \\ & -2 m_{l,t-1} (s_{l,t-1}-m_{l,t-1}) \big( W_{l,t-1}^2 + V_{ll,t-1}\big)
	     \\ &+ 2 (s_{i,t}-\tanh \theta_{i,t}(s_{l,t-1})) (s_{l,t-1}-m_{l,t-1}) \big( W_{i,t} W_{l,t-1} 
	     \\ &+\sum_{j \neq l, n} J_{ij}J_{ln} D_{jn,t-1} \big)
	     \Big) Q(s_{i,t},s_{l,t-1}),
\end{align}
where $V_{ii,t} = \sum_{j\neq l, n\neq l} J_{ij}J_{in} C_{jn,t-1}$, $V_{ll,t-1} = \sum_{mn} J_{lm}J_{ln} C_{mn,t-2}$, $W_{i,t}=-\theta_{i,t}(s_{l,t-1}) + H_{i} + \sum_{j} J_{ij} m_{j,t-1} + J_{il}(s_{l,t-1}-m_{l,t-1})$ and $W_{l,t-1}=-\Theta_{l,t-1} + H_l + \sum_n J_{ln} m_{n,t-1}$.
Note that the first term in the second equation comes from the combination of the term in the first line with the squared term in the third line of the first equation. Similarly, the second term in the second equation derives from the combination of the second line and the squared term in the fourth line of the first equation.

We know that $\big[ \alpha^2 W_{i,t}^2 \big]_{\alpha=1}= 0 + \big[\mathcal{O}(\alpha^{4})\big]_{\alpha=1}$, that $\big[\alpha^2 W_{l,t-1}^2 \big]_{\alpha=1}= 0 + \big[\mathcal{O}(\alpha^{4})\big]_{\alpha=1}$ and
$\big[ \alpha^2 W_{i,t} W_{l,t-1}  \big]_{\alpha=1}= 0 + \big[\mathcal{O}(\alpha^{4})\big]_{\alpha=1}$. This allows us to dismiss terms of the equations above. Then, grouping together the terms that multiply $(s_{i,t}-\tanh \theta_{i,t}(s_{l,t-1}))$ and those that do not, we obtain the TAP equations:
\begin{align}
    \theta_{i,t}(s_{l,t-1}) =& H_{i} + \sum_{j} J_{ij} m_{j,t-1} 
    +\left(J_{il} + \sum_{j \neq l, n} J_{ij}J_{ln} D_{jn,t-1} \right)(s_{l,t-1} - m_{l,t-1}) 
    \label{app-eq:TAP-D-1}
    \\ &-\tanh\left[\theta_{i,t}(s_{l,t-1}) \right] \sum_{jn\neq l} J_{ij}J_{in} C_{jn,t-1}
    + \big[\mathcal{O}(\alpha^{2})\big]_{\alpha=1},
    \\ \Theta_{l,t-1} =& H_l + \sum_n J_{ln} m_{n,t-2} 
    - m_{l,t-1} \sum_{mn} J_{lm}J_{ln} C_{mn,t-2}
    + \big[\mathcal{O}(\alpha^{2})\big]_{\alpha=1}.
    \label{app-eq:TAP-D-2}
\end{align}

With these parameters we can compute the TAP approximations of the system's sufficient statistics as follows: 
\begin{align}
	m_{i,t} \approx&  \sum_{s_{l,t-1}} \tanh [\theta_{i,t}(s_{l,t-1})] Q(s_{l,t-1}),
	\\ m_{l,t-1} \approx&  \tanh \Theta_{l,t-1},
	\\ D_{il,t} \approx&  \sum_{s_{l,t-1}} (\tanh [\theta_{i,t}(s_{l,t-1})]-m_{i,t} ) (s_{l,t-1}-m_{l,t-1}) Q(s_{l,t-1}) ,
\end{align}
where $Q(s_{l,t-1})=P(s_{l,t-1})= \frac{1+s_{l,t-1} m_{l,t-1}}{2}$ is the factorised distribution for $s_{l,t-1}$.

\subsection*{Pairwise equal-time spin correlations}
\label{app:plefka_D2}

We can use a similar approximation in the case above to compute equal-time spin correlations. We do this by substituting $s_{l,t-1}$ in the approximation above by $s_{k,t}$. As $s_{i,t}$, $s_{k,t}$ are conditionally independent, we can assume that $s_{k,t}$ is computed first and that then $s_{i,t}$ is computed conditioned on $s_{k,t}$, (i.e., consider $P(s_{i,t}|s_{k,t},\*s_{t-1}) = P(s_{i,t}|\*s_{t-1})$).
In this approximation we consider a manifold $\mathcal{Q}$ with a pairwise probability distribution:
\begin{equation}
	  Q(s_{i,t},s_{k,t}) = Q(s_{i,t}|s_{k,t}) Q(s_{k,t}) =  \frac{\mathrm{e}^{s_{i,t} \theta_{i,t}(s_{k,t})} }{2 \cosh \theta_{i,t}(s_{k,t})}
	  \frac{\mathrm{e}^{s_{k,t} \Theta_{k,t}}}{2 \cosh \Theta_{k,t}},
\end{equation}
where $\theta_{i,t}(s_{k,t}) = \Theta_{i,t} + \Delta_{ik,t} s_{k,t}$.
Here $Q(s_{k,t})$ is the independent probability distribution for $s_{k,t}$ computed as in \ref{app:plefka_t}, and $Q(s_{i,t}|s_{k,t})$ is a conditional probability distribution we use to construct the pairwise probability distribution $Q(s_{i,t},s_{k,t})$ using the chain rule.

We use this pairwise model to approximate the distribution
\begin{equation}
    P(s_{i,t},s_{k,t}) = \sum_{\*s_{t-1}} P(s_{i,t},s_{k,t},\*{s}_{t-1})
     = \sum_{\*s_{t-1}} P(s_{i,t}|\*{s}_{t-1}) P(s_{k,t},\*{s}_{t-1}).
\end{equation}

As in previous cases, we want to find an approximation of the probability distribution at time $t$ that minimizes the relative entropy
\begin{align}
	D(P(s_{i,t},s_{k,t})||Q(s_{i,t},s_{k,t})) &= \sum_{\substack{s_{i,t}\\s_{k,t}}}
	 P(s_{i,t},s_{k,t}) \log \frac{P(s_{i,t},s_{k,t})}{Q(s_{i,t},s_{k,t})}.
\end{align}

Specifically, the mean-field approximation that minimizes the relative entropy is the one that satisfies
\begin{align}
	\frac{\partial D(P(s_{i,t},s_{k,t})||Q(s_{i,t},s_{k,t}))}{\partial \Theta_{i,t}} &= - \sum_{\substack{s_{i,t}\\s_{k,t}}} \Big( s_{i,t} 
	- \tanh[\theta_{i,t}(s_{k,t})]) \Big) P(s_{i,t},s_{k,t})
	\\ &=  \langle s_{i,t} \rangle_Q - \langle s_{i,t} \rangle_P = 0,
	\\ \frac{\partial D(P(s_{i,t},s_{k,t})||Q(s_{i,t},s_{k,t}))}{\partial \Delta_{ik,t}} &= - \sum_{\substack{s_{i,t}\\s_{k,t}}} \Big( s_{i,t} s_{k,t}  - \tanh[\theta_{i,t}(s_{k,t})])  s_{k,t} \Big) P(s_{i,t},s_{k,t})
	\\ &=  \langle s_{i,t} s_{k,t} \rangle_{(Q\cdot P)} - \langle s_{i,t} s_{k,t} \rangle_P = 0,
\end{align}
where $\langle s_{i,t} s_{k,t} \rangle_{(Q\cdot P)} = \sum_{s_{i,t},s_{k,t}}  Q(s_{i,t}|s_{k,t}) P(s_{k,t})$.

This equation states that the closest factorized model has its first moments equal to the first and second order moments of the target distribution $P$. That is, $\langle s_{i,t} \rangle_Q = \langle s_{i,t} \rangle_P = m_{i,t}$ and  $\langle s_{i,t} s_{k,t} \rangle_{(Q\cdot P)} = \langle s_{i,t} s_{k,t} \rangle_P = C_{ik,t} + m_{i,t} m_{k,t}$. This is equivalent to having the marginalized distribution for spins $i$ and $l$ equal to the model $Q$, i.e. $P(s_{i,t},s_{k,t}) =Q(s_{i,t},s_{k,t})$.
If we assume that the distribution $P$ is close to $Q$, we can compute $P(s_{i,t},s_{k,t})$ as an expansion with respect to $\alpha$ of the probability distribution:
\begin{equation}
P_{\alpha}(s_{i,t}, s_{k,t}) 
= \sum_{\*{s}_{t-1}} P_{\alpha}(s_{i,t}|s_{k,t},\*{s}_{t-1}) 
P_{\alpha}(s_{k,t}|\*{s}_{t-1}) P(\*s_{t-1}),
\end{equation}
with
\begin{align}
	  P_\alpha(s_{i,t}|s_{k,t}, \*{s}_{t-1}) &= \frac{\mathrm{e}^{ s_{i,t} h_{i,t}(\alpha) }}{2\cosh h_{i,t}(\alpha)} 
	  \label{app-eq:pairwise-Dik-ising-alpha-1}
	  \\   h_{i,t}(\alpha) &=    (1 -\alpha) \theta_{i,t}(s_{k,t})  +\alpha  (H_{i} + \sum_j J_{ij} s_{j,t-1}),
\end{align}
and
\begin{align}
	  P_{\alpha}(s_{k,t}|\*{s}_{t-1}) &= \frac{\mathrm{e}^{ s_{k,t} {h}_{k,t}(\alpha)}}{2 \cosh {h}_{k,t}(\alpha)},
	  \label{app-eq:pairwise-Dik-ising-alpha-2}
	   \\ h_{k,t}(\alpha) 
	  &= (1-\alpha) \Theta_{k,t} + \alpha (H_k + \sum_l J_{kl} s_{l,t-1}).
\end{align}
When $\alpha$ is set to zero, $P_{\alpha=0}(s_{i,t}, s_{k,t}) = Q(s_{i,t}, s_{k,t})$, whereas when $\alpha=1$, $P_{\alpha =1}(s_{i,t}, s_{k,t}) = P(s_{i,t}, s_{k,t})$.
We approximate the values of $\Theta_{i,t},\Theta_{k,t},\Delta_{ik,t}$ as follows:
\begin{align}
	\frac{\partial P_\alpha(s_{i,t},s_{k,t})}{\partial \alpha} =& \sum_{\*{s}_{t-1}} 
	\Big( (s_{i,t}-\tanh h_{i,t}(\alpha)) (-\theta_{i,t}(s_{k,t}) 
	+ H_{i} + \sum_j J_{ij} s_{j,t-1} ) 
	\\ &+ (s_{k,t}-\tanh h_{k,t}(\alpha)) ( - \Theta_{k,t}
	+ H_k + \sum_l J_{kl} s_{l,t-1} )
	\Big) 
	\\ &\cdot P_{\alpha}(s_{i,t}|s_{k,t},\*{s}_{t-1}) 
P_{\alpha}(s_{k,t}|\*{s}_{t-1}) P(\*{s}_{t-1}),
	\\ \frac{\partial P_\alpha(s_{i,t},s_{k,t})}{\partial \alpha}\Big \rvert_{\alpha=0} =& \Big( (s_{i,t}-\tanh \theta_{i,t}(s_{k,t})) ( -\theta_{i,t}(s_{k,t})  + H_{i} + \sum_{j} J_{ij} m_{j,t-1} )
	\\ &+  (s_{k,t}-m_{k,t}) ( - \Theta_{k,t}  + H_k + \sum_l J_{kl} m_{l,t-1} )
	\Big)  Q(s_{i,t}, s_{k,t}).
\end{align}

From here, we obtain the nMF expansions:
\begin{align}
    \theta_{i,t}(s_{k,t}) &= H_{i} + \sum_{j} J_{ij} m_{j,t-1} + \big[\mathcal{O}(\alpha^{1})\big]_{\alpha=1},
    \\ \Theta_{k,t} &= H_k + \sum_l J_{kl} m_{l,t-1} + \big[\mathcal{O}(\alpha^{1})\big]_{\alpha=1}.
\end{align}
The second order expressions are obtained by expanding:
\begin{align}
	\frac{\partial^2 P_\alpha(s_{i,t},s_{k,t})}{\partial \alpha^2} 
		=& \sum_{\*{s}_{t-1}} 
	\Big( - (1-\tanh^2 h_{i,t}(\alpha)) (-\theta_{i,t}(s_{k,t}) + H_{i} + \sum_j J_{ij} s_{j,t-1} )^2  
	\\ &- (1-\tanh^2 h_{k,t}(\alpha)) ( - \Theta_{k,t} + H_k + \sum_l J_{kl} s_{l,t-1} )^2
	\\ & + \big( (s_{i,t}-\tanh h_{i,t}(\alpha)) (-\theta_{i,t}(s_{k,t}) + H_{i} + \sum_j J_{ij} s_{j,t-1} )
	\\ & + (s_{k,t}-\tanh h_{k,t}(\alpha)) ( - \Theta_{k,t}  + H_k + \sum_l J_{kl} s_{n,t-1} )
	\big)^2 \Big) 
	\\ &
	\cdot P_{\alpha}(s_{i,t}|s_{k,t},\*{s}_{t-1}) 
P_{\alpha}(s_{k,t}|\*{s}_{t-1})  P(\*{s}_{t-1}),
\end{align}
\begin{align}
	\frac{\partial^2 P_\alpha(s_{i,t},s_{k,t})}{\partial \alpha^2}\Big \rvert_{\alpha=0}
	     =& \Big( -2 \tanh \theta_{i,t}(s_{k,t}) (s_{i,t}-\tanh \theta_{i,t}(s_{k,t})) \big( W_{i,t}^2 + V_{ii,t}\big)
	     \\ & -2 m_{k,t} (s_{k,t}-m_{k,t}) \big( W_{k,t}^2 + V_{kk,t}\big)
	     \\ &+ 2 (s_{i,t}-\tanh \theta_{i,t}(s_{k,t})) (s_{k,t}-m_{k,t}) 
	     \big( W_{i,t} W_{k,t}  + V_{ik,t} \big)
	     \Big) Q(s_{i,t},s_{k,t}),
\end{align}
where $V_{ik,t} = \sum_{jl} J_{ij}J_{kl} C_{jl,t-1}$ and $W_{i,t}=-\theta_{i,t} + H_{i} + \sum_{j} J_{ij} m_{j,t-1}$.
Again, note that the first term in the second equation comes from the combination of the term in the first line with the squared term in the third line of the first equation. Similarly, the second term in the second equation derives from the combination of the second line and the squared term in the fourth line of the first equation.

We know that $\big[\alpha^2 W_{i,t}^2 \big]_{\alpha=1}= 0 + \big[\mathcal{O}(\alpha^{4})\big]_{\alpha=1}$, that $\big[\alpha^2 W_{k,t}^2 \big]_{\alpha=1}= 0 + \big[\mathcal{O}(\alpha^{4})\big]_{\alpha=1}$ and
$\big[ \alpha^2 W_{i,t} W_{k,t}  \big]_{\alpha=1}= 0 + \big[\mathcal{O}(\alpha^{4})\big]_{\alpha=1}$. This allows us to dismiss terms of the equations above. Again, grouping terms that contain $(s_{i,t}-\tanh \theta_{i,t}(s_{l,t-1}))$ and those that not we obtain the follwoing Plefka expansions for the TAP approximations:
\begin{align}
    \theta_{i,t}(s_{k,t}) =& H_{i} + \sum_{j} J_{ij} m_{j,t-1} 
    + (s_{k,t} - m_{k,t})  \sum_{jl} J_{ij}J_{kl} C_{jl,t-1}
    \label{app-eq:TAP-DC-1}
    \\ &-\tanh\left[\theta_{i,t}(s_{k,t}) \right] \sum_{jl} J_{ij}J_{il} C_{jl,t-1}
    + \big[\mathcal{O}(\alpha^{2})\big]_{\alpha=1},
    \\ \Theta_{k,t} =& H_k + \sum_l J_{kl} m_{l,t-1}
    - m_{k,t} \sum_{jl} J_{kj}J_{kl} C_{jl,t-1}
    + \big[\mathcal{O}(\alpha^{2})\big]_{\alpha=1}.
    \label{app-eq:TAP-DC-2}
\end{align}

With these parameters we can the TAP approximations of the system's statistics as follows: 
\begin{align}
	m_{i,t} \approx&  \sum_{s_{k,t}} \tanh [\theta_{i,t}(s_{k,t})] Q(s_{k,t}),
	\\ m_{k,t} \approx&  \tanh \Theta_{k,t},
	\\ C_{ik,t} \approx&  \sum_{s_{k,t}} (\tanh [\theta_{i,t}(s_{k,t})]-m_{i,t} ) (s_{k,t}-m_{k,t}) Q(s_{k,t}),
\end{align}
where $Q(s_{k,t})=P(s_{k,t})= \frac{1+s_{k,t} m_{k,t}}{2}$ is the factorised probability of unit $s_{k,t}$.

\newpage

\section{Solution of the asymmetric kinetic Sherrington-Kirkpatrick model}

The infinite kinetic Ising model with Gaussian couplings used in this article is generally referred in its symmetric version as the Sherrington-Kirkpatrick (SK) model. The SK model behaviour is well studied in statistical mechanics, and its solution can be obtained using the replica trick \cite{nishimori_statistical_2001}. In the symmetric case, dynamics can be represented as a bipartite network that can also be solved using the replica trick \cite{brunetti_asymmetric_1992}.
Here, we extend the solution to the kinetic, asymmetric version of the model.
As this model does not have an equilibrium distribution or a free energy defined in classical terms as the SK model, we need to recur to an dynamical equivalent in the form of a generating functional.

\subsection*{Generating functional}

We start with a kinetic Ising model
\begin{align}
    P\left(\*s_{t}\,|\,\*{s}_{t-1}\right)=& \prod_i \frac{\mathrm{e}^{ \beta s_{i,t} h_{i,t}}}{2 \cosh \pr{\beta h_{i,t}}},
 \label{app-eq:ing}
    \\ h_{i,t}=& H_{i,t}+\sum_j J_{ij} s_{j,t-1},
 \label{app-eq:ing-field}
\end{align}
where $\beta$ is the inverse temperature, $H_{i,t}$ are time-varying fields and $J_{ij}$ couplings. The results apply similarly to fixed fields.

The probability of a specific trajectory $\*s_{0:t}$, is defined as
\begin{align}
    P(\*s_{0:t}) =& \prod_{u=1}^{t} P(\*s_u|\*s_{u-1}) P(\*s_0)
    \\ =& \exp \Bigg( \sum_u \pr{ \sum_i s_{i,u} \pr{\beta H_{i,u} + \beta \sum_j J_{ij} s_{j,u-1} }}
    \\ & - \sum_{i,u} \log 2 \cosh \pr{\beta \pr{ H_{i,u} + \sum_j J_{ij} s_{j,u-1} }}  \Bigg) P(\*s_0).
    \label{app-eq:trajectory-prob}
\end{align}

Instead of a partition function, the distribution of the trajectories of the asymmetric SK model can be described  by defining a generating functional or a dynamical partition function:
\begin{align}
    Z_t(\*g) =& \sum_{\*s_{0:t}} \exp{\pr{ \sum_{iu} g_{i,u} s_{i,u}}} P(\*s_{0:t})
    \\ =&
    \sum_{\*s_{0:t}} \exp \Bigg( \sum_u \pr{ \sum_i s_{i,u} \pr{\beta H_{i,u} + g_{i,u} + \beta \sum_j J_{ij} s_{j,u-1} }}
    \\ & - \sum_{i,u} \log 2 \cosh \pr{\beta \pr{ H_{i,u} + \sum_j J_{ij} s_{j,u-1} }}  \Bigg) P(\*s_0).
    \label{app-eq:SK-partition-function}
\end{align}
Note that $Z_t(\*0)=1$. 
For simplicity we will assume an initial distribution with a single possible state $\*s_0$, i.e. $P(\*s_0^{\prime}) = \delta_{\*s_0^{\prime}, \*s_0}$, which allows us to drop the $P(\*s_0)$ term in the equation above.
In the $t\to\infty$ limit, the logarithmic dynamical partition function converges to the 
large deviation function
\begin{equation}
	\lim_{t\to\infty} \frac{1}{t} \log Z_t(\*g) = \varphi(\*g),
\end{equation}
which plays the role of a free-energy function for trajectories \cite{lecomte_thermodynamic_2007}.

The generating functional can be used to compute moments of the system as:
\begin{align}
	m_{i,u} =& \lim_{\*g\to\*0} \frac{\partial Z_t(\*g)}{\partial g_{i,u}} = \lim_{\*g\to\*0} \ang{ s_{i,u}}_{\*g} = \ang{ s_{i,u}},
	\\  R_{ij,uv} =& \lim_{\*g\to\*0} \frac{\partial^2 Z_t(\*g)}{\partial g_{i,u} \partial g_{j,v}} =  \lim_{\*g\to\*0} \ang{s_{i,u} s_{j,v}}_{\*g} = \ang{s_{i,u} s_{j,v}},
\end{align}
where the brackets are defined as
\begin{align}
    \ang{f(\*s)}_{\*g} =& \sum_{\*s_{0:t}} f(\*s) \exp{\pr{ \sum_{iu} g_{i,u} s_{i,u}}} P(\*s_{0:t}),
    \\  \ang{f(\*s)} =& \sum_{\*s_{0:t}} f(\*s)  P(\*s_{0:t}).
\end{align}
It also allows us to derive identities that will be helpful in eliminating spurious solutions
\begin{align}
	\lim_{\*g\to\*0}  \frac{\partial Z_t(\*g)}{\partial H_{i,u}} = \beta\pr{m_{i,u} - m_{i,u}} =& 0,
	\\ \lim_{\*g\to\*0} \frac{\partial^2 Z_t(\*g)}{\partial H_{i,u}\partial H_{j,v}} = \beta\pr{ \frac{\partial m_{i,u}}{\partial H_{j,v}} - \frac{\partial m_{i,u}}{\partial H_{j,v}}}   =& 0.
\end{align}

\subsection*{Path integral}

In the asymmetric SK model, the couplings $J_{ij}$ are quenched variables with a Gaussian distribution function
\begin{equation}
    P(J_{ij}) = \frac{1}{\sqrt{2\pi J_\sigma^2/N}} \exp{\pr{ -\frac{1}{2J_\sigma^2/N} \left(J_{ij} - \frac{J_0}{N}\right)^2 }},
\end{equation}
where the mean and the variance are proportional to $1/N$.

In the thermodynamic limit, one can study the system by computing the configurational average 
\begin{equation}
    \br{Z_t(\*g)} = \int \prod_{ij} \mathrm{d}J_{ij} P(J_{ij}) Z_T(\*g).
\end{equation}

The quenched average can be solved using path integral methods, by inserting an appropriate delta integral for the effective fields of each unit
\begin{equation}
	1 = \int  \mathrm{d}\^\theta \prod_{i,u} \delta(\theta_{i,u} - \beta H_{i,u} - \beta \sum_j J_{ij} s_{j,u-1}) = \frac{1}{\pr{2\pi}^{Nt}}\int  \mathrm{d}\^\theta \mathrm{d}\^{\hat\theta} \exp \pr{ \sum_{iu} \iu \hat\theta_{i,t} (\theta_{i,t} - \beta H_{i,u} - \beta \sum_j J_{ij} s_{j.t-1}) }.
\end{equation}
where $\^\theta$ represent the effective fields of units and $\^{\hat\theta}$ are conjugates of these fields.

By inserting the above equation, the configurational average is written as 
\begin{align}
    \br{Z_t(\*g)} =& \frac{1}{\pr{2\pi}^{Nt}}\int \mathrm{d}\^\theta \mathrm{d}\^{\hat\theta} \prod_{ij} \mathrm{d}J_{ij} P(J_{ij}) \sum_{\*s_{1:t}} \exp \Bigg( \sum_{iu}  s_{i,u} (g_{i,u}+ \theta_{i,u})  - \sum_{iu}  \log 2 \cosh \pr{ \theta_{i,u}} 
    \\ &+  \sum_{iu} \iu  \hat\theta_{i,u} (\theta_{i,u} - \beta H_{i,u} - \beta \sum_j J_{ij} s_{j,u-1})  \Bigg)
    \\ =&  \frac{1}{\pr{2\pi}^{Nt}} \int \mathrm{d}\^\theta \mathrm{d}\^{\hat\theta}  \sum_{\*s_{1:t}} \exp \Bigg( \sum_{iu}  s_{i,u} (g_{i,u}+ \theta_{i,u}) - \sum_{iu} \log 2 \cosh \pr{\theta_{i,u}}   + \sum_{iu} \iu \hat\theta_{i,u} (\theta_{i,u} - \beta H_{i,u})
    \\ &  - \sum_{iu} \iu \frac{\beta J_0}{N} \hat\theta_{i,u}  \sum_j s_{j,u-1} + \sum_{ij} \frac{\beta^2 J_\sigma^2}{2N}  \pr{ \iu  \sum_{u} \hat\theta_{i,u} s_{j,u-1}}^2 \Bigg)
    \\ =&  \frac{1}{\pr{2\pi}^{Nt}} \int \mathrm{d}\^\theta \mathrm{d}\^{\hat\theta}  \sum_{\*s_{1:t}} \exp \Bigg(\sum_{iu}  s_{i,u} (g_{i,u}+ \theta_{i,u}) - \sum_{iu} \log 2 \cosh \pr{\theta_{i,u}}  +  \sum_{iu} \iu \hat\theta_{i,u} (\theta_{i,u} - \beta H_{i,u})
    \\ & - N \beta J_0  \sum_{u} \frac{1}{N}\sum_i \iu \hat\theta_{i,u} \frac{1}{N} \sum_j s_{j,u-1} 
     +  N \frac{\beta^2 J_\sigma^2}{2} \sum_{uv} \frac{1}{N}\sum_i  \iu \hat\theta_{i,u} \iu \hat\theta_{i,v} \frac{1}{N}\sum_j s_{j,u-1} s_{j,v-1} \Bigg).
\end{align}

\subsubsection*{Gaussian integral}

We want to simplify the expression above by introducing new variables that will become order parameters. We do so by introducing a double Gaussian integral with the form:
\begin{align}
    \exp\pr{C xy} &=  \exp\pr{\frac{C}{2}\pr{\frac{1}{2}(x+y)^2 + \frac{1}{2}(\iu(x-y))^2} }
    \\ &=  \frac{C}{\pi} \int  \mathrm{d}z_R \mathrm{d}z_I \exp\pr{\frac{C}{2}\pr{-\frac{1}{2}z_R^2 -\frac{1}{2}z_I^2 + (x+y)z_R + \iu(x-y)z_I}}
    \\ &=  \frac{C}{\pi} \int  \mathrm{d}z_R \mathrm{d}z_I \exp\pr{\frac{C}{2}\pr{-\frac{1}{2}z_R^2  -\frac{1}{2}z_I^2  +x  (z_R + \iu z_I) + y  (z_R - \iu z_I)}}
    \\ &=  \frac{C}{\pi} \int  \mathrm{d}z_1 \mathrm{d}z_2 \exp\pr{C\pr{- z_1 z_2 + x z_1 + y z_2}},
\end{align}
where in the last step we applied a change of variables $z_1 =  \frac{1}{2}(z_R + \iu z_I), z_2 = \frac{1}{2}(z_R - \iu z_I)$.

We apply the Gaussian integral above: 
\begin{align}
    \exp\pr{N\beta J_0 \frac{1}{N} \sum_j s_{j,u-1} \frac{1}{N}\sum_{i} (-\iu \hat{\theta}_{i,u}) } =& \frac{N\beta J_0 }{\pi} \int  \mathrm{d}\mu_{u} \mathrm{d}m_{u-1} \exp\Bigg( -N\beta J_0 \mu_u  m_{u-1}
    \\ & + \beta J_0\mu_u \sum_j s_{j,u-1} - \beta J_0 m_{u-1} \sum_{i} \iu \hat{\theta}_{i,u} \Bigg),
    \label{app-eq:gaussian-integral-1}
\end{align}
by using $z_1 = \mu_u$, $z_2 = m_{u-1}$, $x= \frac{1}{N} \sum_j s_{j,u-1}$, $y=\frac{1}{N}\sum_{i} (-\iu \hat{\theta}_{i,u})$, and $C=N \beta J_0$. Similarly, we have
\begin{align}
    \exp\pr{N\beta^2 J_\sigma^2  \frac{1}{N} \sum_j s_{j,u-1}s_{j,v-1} \frac{1}{N}\sum_{i} \iu \hat{\theta}_{i,u}\iu \hat{\theta}_{i,v}} =& \frac{N\beta^2 J_\sigma^2 }{\pi} \int \mathrm{d}\rho_{u,v}\mathrm{d}q_{u-1,v-1} \exp\Bigg( -N\beta^2 J_\sigma^2 \rho_{u,v} q_{u-1,v-1} 
    \\ &+ \beta^2 J_\sigma^2 \mu_u \sum_j s_{j,u-1}s_{j,v-1} + \beta^2 J_\sigma^2 q_{u-1,v-1} \sum_{i} \iu \hat{\theta}_{i,u} \iu \hat{\theta}_{i,v} \Bigg).
    \label{app-eq:gaussian-integral-2}
\end{align}
With these transformations, we can rewrite
\begin{align}
    \br{Z_t(\*g)} =&  \frac{(N\beta J_0)^{t} (N\beta^2 J_\sigma^2)^{t}}{ \pi^{2t}} \int  \mathrm{d}\*m \mathrm{d}\^\mu \mathrm{d}\*q \mathrm{d}\^\rho 
    \exp  \Bigg( - N \beta J_0\sum_{u}  \mu_{u} m_{u-1}
    - N \beta^2 J_\sigma^2  \sum_{u>v} \rho_{u,v} q_{u-1,v-1}  
    \\ & + \log \sum_{\*s_{1:t}}\int \mathrm{d}\^\theta \mathrm{d}\^{\hat \theta} \mathrm{e}^{ \Phi(\*s, \^\theta) + \Omega(\^{\hat \theta}, \^\theta) } \Bigg),
\end{align}
where the rest of the terms from the Gaussian integral are contained in the terms
\begin{align}
	\Phi(\*s, \^\theta) =&  \sum_{iu}  \pr{g_{i,u} +  \theta_{i,u}} s_{i,u}
	+ \sum_{iu} \beta J_0 \mu_{u}  s_{i,u-1}  + \sum_{i,u>v} \beta^2 J_\sigma^2 \rho_{u,v} s_{i,u-1} s_{i,v-1}  - \sum_{iu} \log 2 \cosh \pr{\theta_{i,u}},  
\\ \Omega(\^{\hat \theta}, \^\theta)  =&  \sum_{iu}  (\theta_{i,u} - \beta H_{i,u} - \beta J_0 m_{u-1}) \iu  \hat\theta_{i,u} 
	+  \frac{\beta^2 J_\sigma^2 }{2}  \sum_{i,u}\pr{ \iu \hat\theta_{i,u} }^2 
	+ \beta^2 J_\sigma^2  \sum_{i,u>v} q_{u-1,v-1} \iu \hat\theta_{i,u}  \iu \hat\theta_{i,v} - Nt \log 2\pi. 
\end{align}
We will see now that the values of $\*m, \^\mu, \*q, \^\rho$ will act as order parameters of the system.

\subsubsection*{Order parameters}
The exponent of the above integrand is proportional to $N$, being it possible to evaluate the integral by steepest descent, giving the saddle-point solution as
\begin{align}
    \br{Z_t(\*g)} =& 
    \exp  \Bigg\{  - N \beta J_0\sum_{u}  \mu_{u} m_{u-1}
    - N \beta^2 J_\sigma^2  \sum_{u>v} \rho_{u,v} q_{u-1,v-1}  
      + \log \sum_{\*s_{1:t}}\int \mathrm{d}\^\theta \mathrm{d}\^{\hat \theta} \mathrm{e}^{ \Phi(\*s, \^\theta) + \Omega(\^{\hat \theta}, \^\theta) } \Bigg\},
\end{align}
where the values of $\*m, \^\mu, \*q, \^\rho$ are chosen to extremize (maximize or minimize) the quantity between the braces $\{\}$. 
Notice that integration over disordered connections has removed coupling between units and replaced it with same-unit temporal couplings $\*\rho$ and varying effective fields, which are also independent between units, resulting in a mean-field solution where the activity of different spins is independent.

From here, knowing that $\lim_{\*g\to\*0} \frac{ \partial \br{Z_t(\*g)} }{\partial g_{i,u}} = \br{\ang{s_{i,u}}}$, $\lim_{\*g\to\*0} \frac{ \partial^2 \br{Z_t(\*g)} }{\partial g_{i,u} \partial g_{j,v}} =\br{\ang{s_{i,u} s_{j,v}}}$, and taking into account that $\lim_{\*g\to \*0}  \br{Z_t(\*g)} = 1$ we can compute the order parameters of the system as
\begin{align}
	\lim_{\*g\to\*0} \frac{ \partial \br{Z_t(\*g)} }{\partial g_{i,u}} 
	=&  \lim_{\*g\to\*0} \frac{ \sum_{\*s_{1:t}}\int \mathrm{d}\^\theta \mathrm{d}\^{\hat \theta} s_{i,u}\mathrm{e}^{ \Phi(\*s, \^\theta) + \Omega(\^{\hat \theta}, \^\theta) } }{ \sum_{\*s_{1:t}}\int \mathrm{d}\^\theta \mathrm{d}\^{\hat \theta} \mathrm{e}^{ \Phi(\*s, \^\theta) + \Omega(\^{\hat \theta}, \^\theta) } } \br{Z_t(\*g)} 
	=  \ang{s_{i,u}}_*  	=  \br{\ang{s_{i,u}}},
	\\ \lim_{\*g\to\*0} \frac{ \partial^2 \br{Z_t(\*g)} }{\partial g_{i,u} \partial g_{j,v}} 
	=&  \lim_{\*g\to\*0} \frac{ \sum_{\*s_{1:t}}\int \mathrm{d}\^\theta \mathrm{d}\^{\hat \theta} s_{i,u} s_{j,v}\mathrm{e}^{ \Phi(\*s, \^\theta) + \Omega(\^{\hat \theta}, \^\theta) } }{ \sum_{\*s_{1:t}}\int \mathrm{d}\^\theta \mathrm{d}\^{\hat \theta} \mathrm{e}^{ \Phi(\*s, \^\theta) + \Omega(\^{\hat \theta}, \^\theta) } } 
	\br{Z_t(\*g)}
	=  \ang{s_{i,u} s_{j,v}}_* = \br{\ang{s_{i,u} s_{j,v}}},
	\\ \lim_{\*g\to\*0} \frac{ \partial \br{Z_t(\*g)} }{\partial H_{i,u}} 
	=&  \lim_{\*g\to\*0} \frac{ \sum_{\*s_{1:t}}\int \mathrm{d}\^\theta \mathrm{d}\^{\hat \theta} -  \beta\iu \hat \theta_{i,u} \mathrm{e}^{ \Phi(\*s, \^\theta) + \Omega(\^{\hat \theta}, \^\theta) } }{ \sum_{\*s_{1:t}}\int \mathrm{d}\^\theta \mathrm{d}\^{\hat \theta} \mathrm{e}^{ \Phi(\*s, \^\theta) + \Omega(\^{\hat \theta}, \^\theta) } }  
	\br{Z_t(\*g)}
	= - \beta \ang{ \iu \hat\theta_{i,u} }_* = 0,
	\\ \lim_{\*g\to\*0} \frac{ \partial^2 \br{Z_t(\*g)} }{\partial H_{i,u} \partial H_{j,v}} =&  
	\lim_{\*g\to\*0} \frac{ \sum_{\*s_{1:t}}\int \mathrm{d}\^\theta \mathrm{d}\^{\hat \theta}  \beta^2 \iu \hat\theta_{i,u} \iu\hat\theta_{j,v}\mathrm{e}^{ \Phi(\*s, \^\theta) + \Omega(\^{\hat \theta}, \^\theta) } }{ \sum_{\*s_{1:t}}\int \mathrm{d}\^\theta \mathrm{d}\^{\hat \theta} \mathrm{e}^{ \Phi(\*s, \^\theta) + \Omega(\^{\hat \theta}, \^\theta) } }
	\br{Z_t(\*g)}
	= \beta^2 \ang{ \iu \hat\theta_{i,u}  \iu \hat\theta_{j,v}}_* = 0,
\end{align}
where
\begin{align}
    \ang{f(\*s,\^{\hat \theta})}_* =  \frac{ \sum_{\*s_{1:t}}\int \mathrm{d}\^\theta \mathrm{d}\^{\hat \theta}  f(\*s,\^{\hat \theta}) \mathrm{e}^{ \Phi(\*s, \^\theta) + \Omega(\^{\hat \theta}, \^\theta) }}{\sum_{\*s_{1:t}}\int \mathrm{d}\^\theta \mathrm{d}\^{\hat \theta} \mathrm{e}^{ \Phi(\*s, \^\theta) + \Omega(\^{\hat \theta}, \^\theta) }}.
\end{align}

Here we should note that, as there is no coupling between units, for $i\neq j$ we have a factorised solution $\br{\ang{s_{i,u} s_{j,v}}} = \ang{s_{i,u} s_{j,v}}_* =  \ang{s_{i,u}}\ang{ s_{j,v}}_*$. 

To obtain the values of the order parameters, we extremize the contents of the brackets, finding
\begin{align}
    \label{app-eq:saddle-point-extremization}
	\lim_{\*g\to\*0} \frac{ \partial \log \br{Z_t(\*g)} }{\partial \mu_{u+1}} 
	=&  - N \beta J_0 m_u + \lim_{\*g\to\*0} \frac{ \sum_{\*s_{1:t}}\int \mathrm{d}\^\theta \mathrm{d}\^{\hat \theta} \beta J_0 \sum_i s_{i,u}\mathrm{e}^{ \Phi(\*s, \^\theta) + \Omega(\^{\hat \theta}, \^\theta) } }{ \sum_{\*s_{1:t}}\int \mathrm{d}\^\theta \mathrm{d}\^{\hat \theta} \mathrm{e}^{ \Phi(\*s, \^\theta) + \Omega(\^{\hat \theta}, \^\theta) } }
	\\ =&  \beta J_0 \pr{ \sum_i \ang{s_{i,u}}_* - N m_u }=0; \qquad  m_u  =  \frac{1}{N}\sum_i \br{\ang{s_{i,u}}},
	\\ 	\lim_{\*g\to\*0} \frac{ \partial \log \br{Z_t(\*g)} }{\partial m_{u-1}} 
	=&  -  N\beta J_0 \mu_u + \lim_{\*g\to\*0} \frac{ \sum_{\*s_{1:t}}\int \mathrm{d}\^\theta \mathrm{d}\^{\hat \theta} - \beta J_0 \sum_i \iu\hat\theta_{i,u}\mathrm{e}^{ \Phi(\*s, \^\theta) + \Omega(\^{\hat \theta}, \^\theta) } }{ \sum_{\*s_{1:t}}\int \mathrm{d}\^\theta \mathrm{d}\^{\hat \theta} \mathrm{e}^{ \Phi(\*s, \^\theta) + \Omega(\^{\hat \theta}, \^\theta) } }
	\\ =&  -\beta J_0 \pr{ \sum_i \ang{\iu \hat\theta_{i,u}}_* + N \mu_u }=0; \qquad  \mu_u  = 0,
	\\ \lim_{\*g\to\*0} \frac{ \partial \log \br{Z_t(\*g)} }{\partial \rho_{u+1,v+1}} 
	 =& - N \beta^2 J_\sigma^2 q_{u,v} + \lim_{\*g\to\*0} \frac{ \sum_{\*s_{1:t}}\int \mathrm{d}\^\theta \mathrm{d}\^{\hat \theta} \beta^2 J_\sigma^2 \sum_i s_{i,u} s_{i,v}\mathrm{e}^{ \Phi(\*s, \^\theta) + \Omega(\^{\hat \theta}, \^\theta) } }{ \sum_{\*s_{1:t}}\int \mathrm{d}\^\theta \mathrm{d}\^{\hat \theta} \mathrm{e}^{ \Phi(\*s, \^\theta) + \Omega(\^{\hat \theta}, \^\theta) } }
	\\ =&  \beta^2 J_\sigma^2 \pr{ \sum_i \ang{s_{i,u} s_{i,v}}_* - N q_{uv} }=0; \qquad q_{uv} = \frac{1}{N}  \sum_i \br{\ang{s_{i,u}s_{i,v}}},
	\\ \lim_{\*g\to\*0} \frac{ \partial \log \br{Z_t(\*g)} }{\partial q_{u-1,v-1}} 
	=&  -  N\beta^2 J_\sigma^2 \rho_{u,v} + \lim_{\*g\to\*0} \frac{ \sum_{\*s_{1:t}}\int \mathrm{d}\^\theta \mathrm{d}\^{\hat \theta} \beta^2 J_\sigma^2 \sum_i \iu\hat\theta_{i,u} \iu\hat\theta_{j,v} \mathrm{e}^{ \Phi(\*s, \^\theta) + \Omega(\^{\hat \theta}, \^\theta) } }{ \sum_{\*s_{1:t}}\int \mathrm{d}\^\theta \mathrm{d}\^{\hat \theta} \mathrm{e}^{ \Phi(\*s, \^\theta) + \Omega(\^{\hat \theta}, \^\theta) } }
	\\ =& \beta^2 J_\sigma^2 \pr{ \sum_i \ang{\iu \theta_{i,u} \iu\theta_{i,v}}_* - N \rho_{uv} }=0; \qquad\rho_{uv} = 0.
\end{align}

\subsection*{Saddle-point solution}

After solving the saddle-point integral, we have the following dynamical partition function
\begin{align}
	\br{Z_t(\*g)} &= \sum_{\*s_{1:t}}\int \mathrm{d}\^\theta \mathrm{d}\^{\hat \theta} \mathrm{e}^{ \Phi(\*s, \^\theta) + \Omega(\^{\hat \theta}, \^\theta) }.
\end{align}


At this point, we want to remove the effective conjugate fields $\^{\hat \theta}$ by recovering a delta function. We first rewrite
\begin{align}
	\mathrm{e}^{\Omega(\^{\hat \theta}, \^\theta)} =& \frac{1}{(2\pi)^{Nt}}\exp\pr{ \sum_{iu}(\theta_{i,u} - \beta H_{i,u} - \beta J_0 m_{u-1})   \iu  \hat\theta_{i,u} 
	 +   \frac{\beta^2 J_\sigma^2 }{2} \sum_{i,uv}q_{u-1,v-1}  \iu \hat\theta_{i,u} \iu \hat\theta_{i,v} },
\end{align}
where we defined $q_{u-1,u-1}=1$ and $q_{u-1,v-1}=q_{v-1,u-1}$.
We can remove the quadratic terms of $\^{\hat \theta}$ by applying a multivariate Gaussian integral of the form
\begin{align}
	\mathrm{e}^{\frac{1}{2}\sum_{uv} K_{uv} x_u x_v} = \frac{1}{\sqrt{(2\pi)^t |K^{-1}|}}\int \mathrm{d} \*z \mathrm{e}^{-\frac{1}{2}\sum_{uv} K_{uv} z_u z_v + \sum_{uv} K_{uv} x_u z_v  }.
\end{align}
For $x_u = -\beta J_\sigma  \iu  \hat\theta_{i,u}$ and $K_{uv} = q_{u-1,v-1}$, we get
\begin{align}
	  \int \mathrm{d}\^{\hat\theta} \mathrm{e}^{\Omega(\^{\hat \theta}, \^\theta)} 
	  =&  \frac{1}{(2\pi)^{Nt}} \int \mathrm{d}\^{\hat\theta}  \mathrm{d} \*z  P(\*z) 
	  \exp\Bigg(
  \sum_{iu}  \iu  \hat\theta_{i,u} (\theta_{i,u} - \beta H_{i,u} - \beta J_0 m_{u-1})  
	-    \beta J_\sigma \sum_{i,uv} q_{u-1,v-1} \iu \hat\theta_{i,u}   z_{i,v}  
	    \Bigg)
	\\ =&   \int \mathrm{d} \*z  P(\*z) \prod_{iu} \delta\Bigg( \theta_{i,u} - \hat H_{i,u}(\*z)\Bigg),
\end{align}
where $\hat H_{i,u}(\*z)$ is 
\begin{align}
	\hat H_{i,u}(\*z) =& 
	\beta H_{i,u} + \beta J_0 m_{u-1} 
	 + \beta J_\sigma  \sum_v z_{i,v} q_{u-1,v-1}.
\end{align}
Here the distribution $P(\*z)$ is a multivariate Gaussian $\mathcal{N}(0,\*\Sigma)$, with $(\*\Sigma^{-1})_{uv}= q_{u-1,v-1}=  \frac{1}{N}  \sum_i \br{\ang{s_{i,u-1}s_{i,v-1}}}$.

To simplify calculations, we can perform a change of variables
\begin{align}
	\tilde z_{i,u} = \sum_v z_{i,v} q_{u-1,v-1}.
\end{align}
Here, we can find that the covariance matrix of $\^{\tilde z}_i$ is the inverse of the covariance matrix of $\^{z}_i$. We can show this which we can describe using the characteristic function of the multivariate Gaussian $P(\*z)$
\begin{align}
	\phi_{\*z_i}(\^\tau) = \ang{\exp \pr{\iu \sum_u \tau_u z_{i,u}}}_{\*z_i}  =\exp\pr{-\frac{1}{2}\sum_{uv} \tau_u \tau_v \Sigma_{uv} },
\end{align}
and computing then the characteristic function of $P(\^{\tilde z}_i )$
\begin{align}
	\phi_{\^{\tilde z}_i}(\^\tau) =& \ang{\exp \pr{\iu \sum_u \tau_u \tilde z_{i,u}}}_{\^{\tilde z}_i} 
	 = \ang{\exp \pr{\iu \sum_u \tau_u  \sum_v z_{i,v} q_{u-1,v-1}}}_{\*z_i} 
	 \\ =& \exp\pr{ -\frac{1}{2} \sum_{uu'}\tau_u \tau_{u'} \sum_{vv'}q_{u-1,v-1} \Sigma_{vv'}q_{u'-1,v'-1} }
	 \\ =& \exp\pr{ -\frac{1}{2} \sum_{uv} \tau_u \tau_{v}  q_{u-1,v-1}  },
\end{align}
where in the last step we took into account that $ \^\tau^T \*q^{T} \*\Sigma \*q\^\tau = \^\tau^T \*q\^\tau$ as $\*\Sigma^{-1}=\*q$.
This result indicates that $P(\^{\tilde z}_i )$  is a multivariate Gaussian $\mathcal{N}(0,\*q)$.


After solving the saddle-point equation we have
\begin{align}
	\Phi(\*s, \^\theta) =&  \sum_{iu}  s_{i,u} \pr{g_{i,u}+  \theta_{i,u}}   -  \sum_{iu} \log 2 \cosh \pr{\theta_{i,u}},
\end{align}
which leads us to
\begin{align}
     \br{Z_t(\*g)}
	=&   \int \mathrm{d}\^\theta  \exp \pr{ \log \sum_{\*s_{1:t}} \mathrm{e}^{\Phi(\*s, \^\theta)} } \int \mathrm{d}\*{\tilde z} P(\*{\tilde z}) \prod_{iu} \delta\Bigg( \theta_{i,u} - \hat H_{i,u}(\*{\tilde z})\Bigg)
	\\ =&  \int  \mathrm{d}\*{\tilde z} P(\*{\tilde z})
	\exp\Bigg( \log \Bigg(  \sum_{\*s_{1:t}} \exp\pr{ \sum_{iu} s_{i,u} \pr{g_{i,u} +  \hat H_{i,u}(\*{\tilde z})} }
		 -  \sum_{iu} \log  2 \cosh \pr{ \hat H_{i,u}(\*{\tilde z})}  \Bigg)
	\\ =&  \int  \mathrm{d}\*{\tilde z} P(\*{\tilde z}) \prod_{iu} \frac{\cosh\pr{g_{i,u} + \hat H_{i,u}(\*{\tilde z})}}{ \cosh \pr{ \hat H_{i,u}(\*{\tilde z})}}.
	\label{app-eq:intermediate-solution}
\end{align}

And, as the diagonal of the covariance matrix of $\*{\tilde z}$ is equal to 1, we can derive
\begin{align}
	m_{i,u} =& \lim_{\*g\to\*0} \frac{\partial Z_t(\*g)}{\partial g_{i,u}} =  \int   \mathrm{D}z \tanh(\hat H_{i,u}(z)),
	\\  R_{ii,uv} =& \lim_{\*g\to\*0} \frac{\partial^2 Z_t(\*g)}{\partial g_{i,u} \partial g_{i,v}} =  \int \mathrm{D}xy(q_{u-1,v-1}) \tanh(\hat H_{i,u}(x)) \tanh(\hat H_{i,v}(y)),
\end{align}
where 
\begin{align}
	\mathrm{D}z =&  \frac{1}{\sqrt{2\pi}} \mathrm{e}^{-\frac{1}{2}z^2},
	\\ \mathrm{D}xy(q) =& \frac{1}{2\pi\sqrt{1-q^2}}\mathrm{e}^{-\frac{x^{2}+y^{2}-2q xy}{2(1-q^2)}},
	\\\hat H_{i,u}(z) =& 
	\beta H_{i,u} + \beta J_0 m_{u-1} 
	 + \beta J_\sigma  z.
\end{align}
The resulting equations are similar to the symmetric SK model, but the same-spin  coupling parameter vanishes for the computation of $m_{u}$. 
\begin{align}
	m_{u} =& \frac{1}{N}\sum_i \int   \mathrm{D}z \tanh\pr{\beta H_{i,u} + \beta J_0 m_{u-1} 
	 + \beta J_\sigma  z},
	 \label{app-eq:mean-field-solution-m}
	\\  q_{uv} =& \frac{1}{N}\sum_i \int \mathrm{D}xy(q_{u-1,v-1}) \tanh\pr{\beta H_{i,u} + \beta J_0 m_{u-1} 
	 + \beta J_\sigma  x} \tanh\pr{H_{i,v} + \beta J_0 m_{v-1} 
	 + \beta J_\sigma  y}.
\end{align}
This is consistent with findings of the asymmetric SK model lacking a spin-glass phase as $m_u$ is independent of $q_{u,v}$. Note also that all $q_{u,v}$ only depends on the previous  $q_{u-1,v-1}$, this in the $t\to\infty$ limit the value of $q_{t,t-d}$ tends to the same value for any finite $d$.

\subsection*{Ferromagnetic critical phase transition in the infinite kinetic Ising model with Gaussian couplings and uniform weights}

We define an Ising network of infinite size, with randomly defined bias $H_{i,u}=H_i$, where $H_i$ has a distribution $\mathcal{U}(-H_0 ,H_0)$ and couplings $J_{ij}$ with a Gaussian distribution $\mathcal{N}(\frac{1}{N}, \frac{J_\sigma^2}{N})$.
We choose $H_0=0.5$, $J_\sigma=0.1$.

As we have found that the asymmetric SK model with arbitrary fields follows a mean-field solution, calculating the effects of disorder in the fields becomes much easier, as we can approximate the update equations of the order parameters in the thermodynamic limit $N\to\infty$ as:
\begin{align}
    m_t = \lim_{N\to\infty} \frac{1}{N}\sum_{i=1}^{N}  m_{i,t} =& \frac{1}{2H_0}\int_{-H_0}^{H_0} \mathrm{d}h \int \mathrm{D}z \tanh \pr{\beta\pr{h  + J_0  m_{t-1} + J_\sigma z }} 
    \\ =& \frac{1}{2\beta H_0} \int \mathrm{D}z \log\pr{ \frac{\cosh \pr{\beta\pr{H_0 + J_0  m_{t-1} + J_\sigma z} }}{\cosh\pr{\beta\pr{-H_0 + J_0  m_{t-1} + J_\sigma z} }}}.
    \label{app-eq:mean-activation-order-parameter}
\end{align}

Similarly, the delayed self-correlation parameter is given as
\begin{align}
    q_{t,t'} =\lim_{N\to\infty} \frac{1}{N}\sum_{i=1}^{N}  R_{ii,t,t'} =&  \frac{1}{2H_0} \int_{-H_0}^{H_0} \mathrm{d}h  \int  \mathrm{D}xy(q_{t-1,t'-1}) \tanh \pr{\beta \pr{h + J_0  m_{t-1} + J_\sigma x} }\tanh \pr{\beta\pr{h + J_0  m_{t'-1} + J_\sigma y}}
    \\ =&  \frac{1}{2\beta H_0}   \int  \mathrm{D}xy(q_{t-1,t'-1})
    \Bigg( 
     \frac{1}{2}\pr{ \log\pr{\frac{1 + \tanh \pr{\beta\pr{H_0 + H_{t,x}}}}{1- \tanh \pr{\beta\pr{H_0 + H_{t,x}}}} \frac{1 - \tanh \pr{\beta\pr{-H_0 + H_{t,x}}}}{1 + \tanh \pr{\beta\pr{-H_0 + H_{t,x} }}}}}
     \label{app-eq:correlation-order-parameter_1}
    \\&- \frac{1}{\tanh \pr{ \beta \pr{ H_{t',y}-H_{t,x}}}}
    \log\pr{\frac{1+\tanh \pr{\beta\pr{H_{t',y}-H_{t,x}}}\tanh \pr{\beta\pr{H_0 + H_{t,x}}}}{1+\tanh \pr{\beta\pr{H_{t',y}-H_{t,x}}}\tanh \pr{\beta\pr{-H_0 + H_{t,x}}}}}\Bigg)
     .
    \label{app-eq:correlation-order-parameter_2}
\end{align}
with $H_{t,x} = J_0  m_{t-1} + J_\sigma x$ and $H_{t',y}  = J_0  m_{t'-1} + J_\sigma y$.

By recursively updating Supplementary Eqs.~\ref{app-eq:mean-activation-order-parameter} and \ref{app-eq:correlation-order-parameter_1}-\ref{app-eq:correlation-order-parameter_2}, we can obtain the order parameters of the system. In Supplementary Fig.~\ref{fig:infty-ising}, we show the order parameters for $J_0=1$, $J_\sigma=0.1$ and $H_0=0.5$, which resulted in a ferromagnetic transition around $\beta_c \approx 1.1108$.

\begin{figure}[t]
\begin{center}
\begin{tabular}{ll}
 \textbf{A} & \textbf{B}\\
  \includegraphics[width=5.0cm]{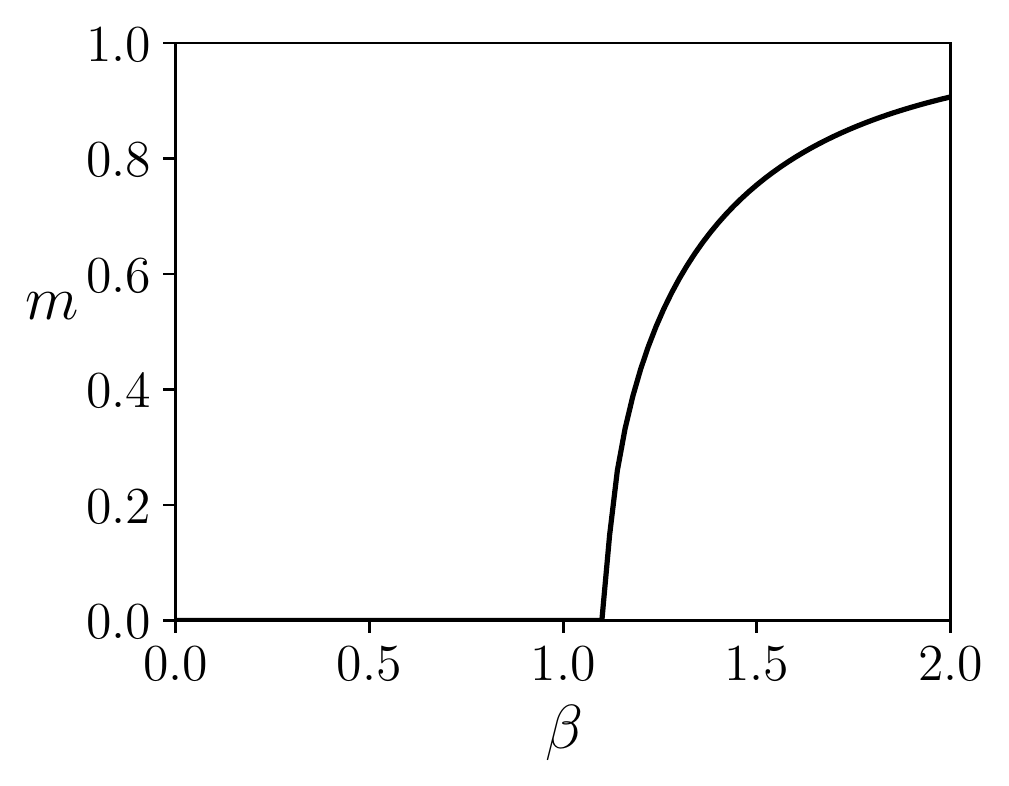}&
  \includegraphics[width=5.0cm]{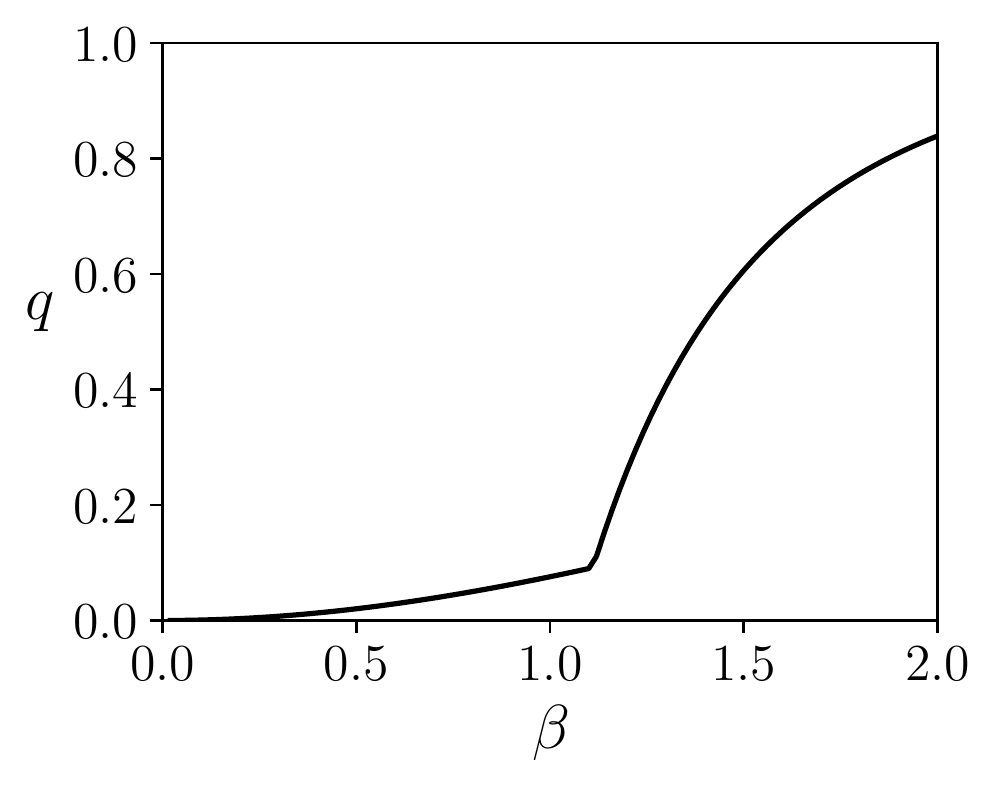} \\
 \textbf{C} & \textbf{D}\\
  \includegraphics[width=5.0cm]{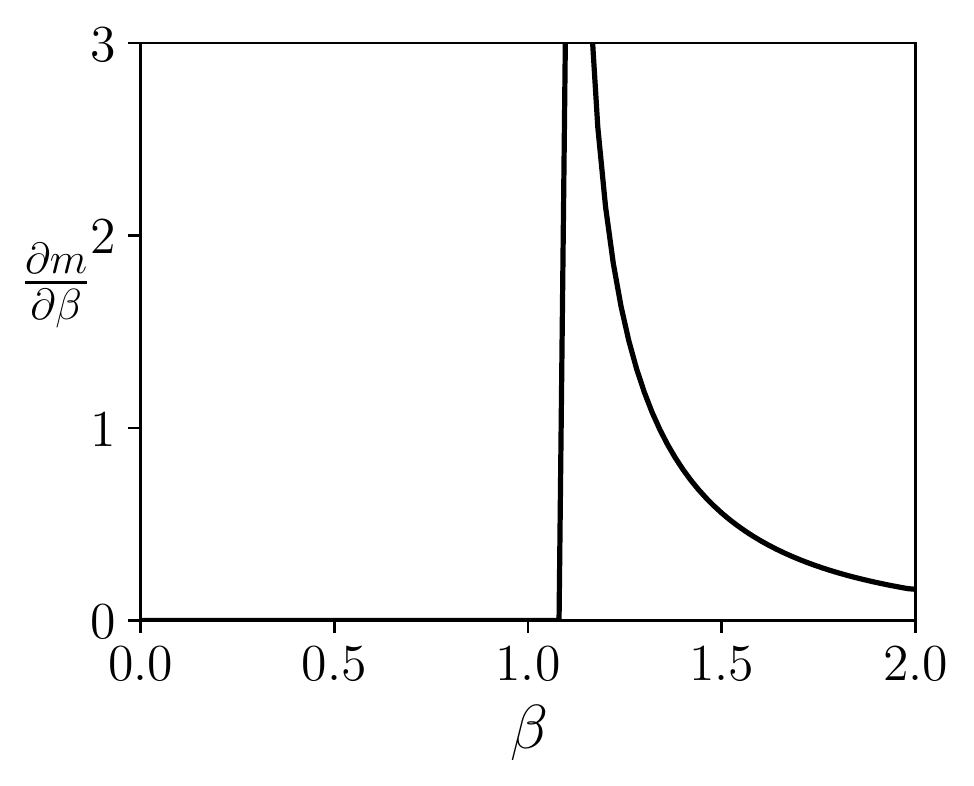}&
  \includegraphics[width=5.0cm]{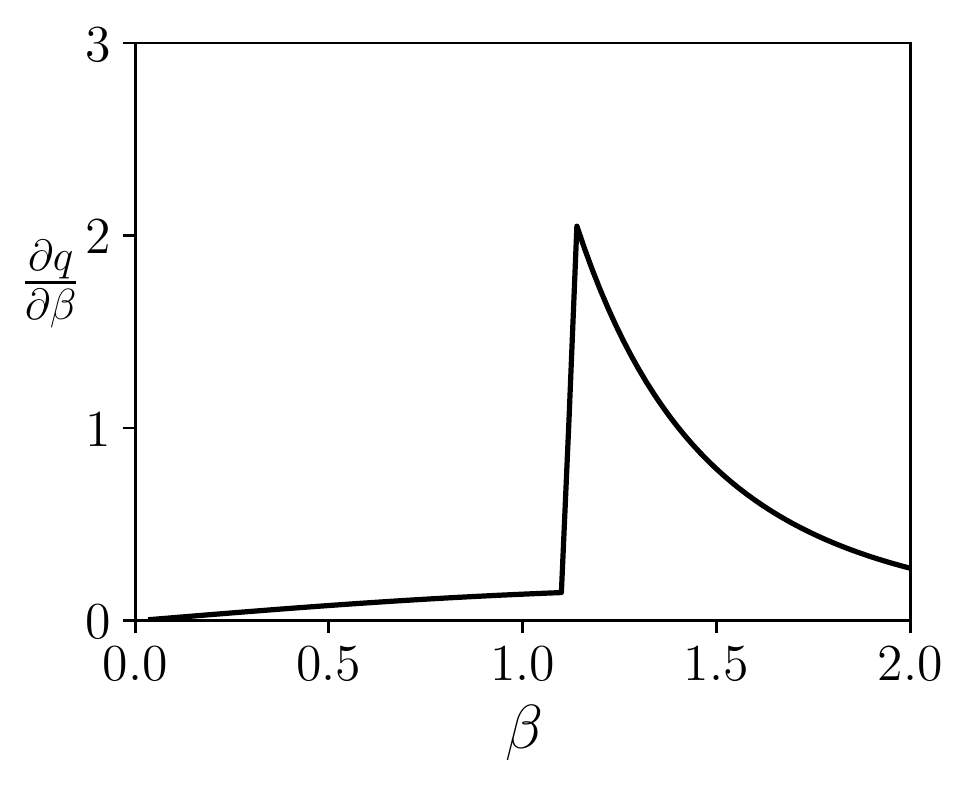}
\end{tabular}
\end{center}
\caption{\textbf{Analytical results of the behaviour of the infinite kinetic Ising model}. A critical point is found for approximately $\beta_c=1.1108$ with parameters $J_0=1$, $J_\sigma=0.1$ and $H_0=0.5$.
} 
\label{fig:infty-ising}
\end{figure}

Similarly, assuming a nonequilibrium steady state (NESS) in which $m_t=m_{t-1}=m$ we can obtain the critical point of the system by computing the non-zero solutions of the first order Taylor expansion around $m=0$ of the right hand part of Supplementary Eq.~\ref{app-eq:mean-activation-order-parameter},
\begin{align}
    m \approx& \frac{1}{2\beta H_0} \int \mathrm{D}z \log\pr{ \frac{\cosh \pr{\beta\pr{H_0 + J_\sigma z} }}{\cosh\pr{\beta\pr{-H_0 + J_\sigma z} }}}
    \\ +&\frac{1}{2\beta H_0} \int \mathrm{D}z \pr{\tanh \pr{\beta\pr{H_0  + J_\sigma z }} - \tanh \pr{\beta\pr{-H_0  + J_\sigma z }}} \beta J_0 m
    \\ =& \frac{1}{ H_0} \int \mathrm{D}z \tanh \pr{\beta\pr{H_0  + J_\sigma z }}  J_0 m,
\end{align}
that yields the same solution for $\beta_c$ solving the equation
\begin{align}
    \frac{1}{ H_0} \int \mathrm{D}z \tanh \pr{\beta_c\pr{H_0  + J_\sigma z }}  J_0 = 1.
\end{align}

We can characterize some critical exponents of the system using the reduced temperature $\tau=-\frac{\beta - \beta_c}{\beta_c}$
, and finding that near the critical point the first order Taylor expansion around $\beta_c$
\begin{align}
    \frac{1}{ H_0} \int \mathrm{D}z \tanh \pr{\beta\pr{H_0  + J_\sigma z }}  J_0  \approx& 1 +  \frac{1}{ H_0} \int \mathrm{D}z \pr{1 - \tanh^2 \pr{\beta\pr{H_0  + J_\sigma z }} } \pr{H_0  + J_\sigma z } J_0 \pr{\beta - \beta_c}
    \\ =& 1 - K' \tau
    \label{eq:taylor-expansion-reduced-temperature}
\end{align}
and computing the value of $m$ around $\beta=\beta_c$ with the third order Taylor expansion around $m=0$
\begin{align}
    m \approx &   \frac{1}{H_0} \int \mathrm{D}z \tanh \pr{\beta\pr{H_0  + J_\sigma z }} J_0 m 
    \\-&  \frac{1}{3 \beta  H_0} \int \mathrm{D}z \tanh \pr{\beta\pr{H_0 + J_\sigma z }} (1-\tanh^2 \pr{\beta\pr{H_0  + J_\sigma z }} ) \pr{\beta J_0 m}^3
    \\ =& (1 - K' \tau)m - K'' m^3,
    \\ m \propto& \pr{-\tau}^{-\frac{1}{2}}.
\end{align}
This yields a critical exponent $\beta'=\frac{1}{2}$ (note that this is different from the inverse temperature $\beta$), consistent with the mean-field universality class.

Similarly, we can compute the susceptibility to a uniform magnetic field $B$ added on top of $H_i$, having that the derivative of $m$ with respect to $B$ evaluated at $B=0$
\begin{align}
    \frac{\partial m}{\partial B} = \frac{1}{2\beta H_0} \int \mathrm{D}z \pr{\tanh \pr{\beta\pr{H_0  + J_0  m + J_\sigma z }} - \tanh \pr{\beta\pr{-H_0  + J_0  m + J_\sigma z }} } \pr{\beta + \beta J_0  \frac{\partial m}{\partial B} }.
\end{align}
The susceptibility evaluated at $m=0$ yields, for the limit $\tau\to 0$,
\begin{align}
    \frac{\partial m}{\partial B} =& \pr{1 - K' \tau} \pr{\frac{1}{J_0} + \frac{\partial m}{\partial B} },
    \\ \frac{\partial m}{\partial B} \propto & \frac{1 - K' \tau}{\tau} \approx \tau^{-1}.
\end{align}
The result retrieves the $\gamma =1$ exponent consistent with the mean-field universality class.

\subsubsection*{Finite-size fluctuations}

Finally, to evaluate the behaviour around $\beta_c$ in the finite network used in the article, we numerically simulated the behaviour of a network of size $N=512$, with parameters $H_0=0.5$, $J_0=1, J_\sigma = 0.1$ as in the numerical results from the main text. In Supplementary Fig.~\ref{fig:exp-ising} we show the values of $\*m_t$. $\*C_t$ and $\*D_t$, showing that correlations peak around the value of $\beta_c$ computed theoretically.

\bigskip

\begin{figure}[h]
\begin{center}
\begin{tabular}{lll}
 \textbf{A} & \textbf{B} &  \textbf{C}\\
  \includegraphics[width=5.0cm]{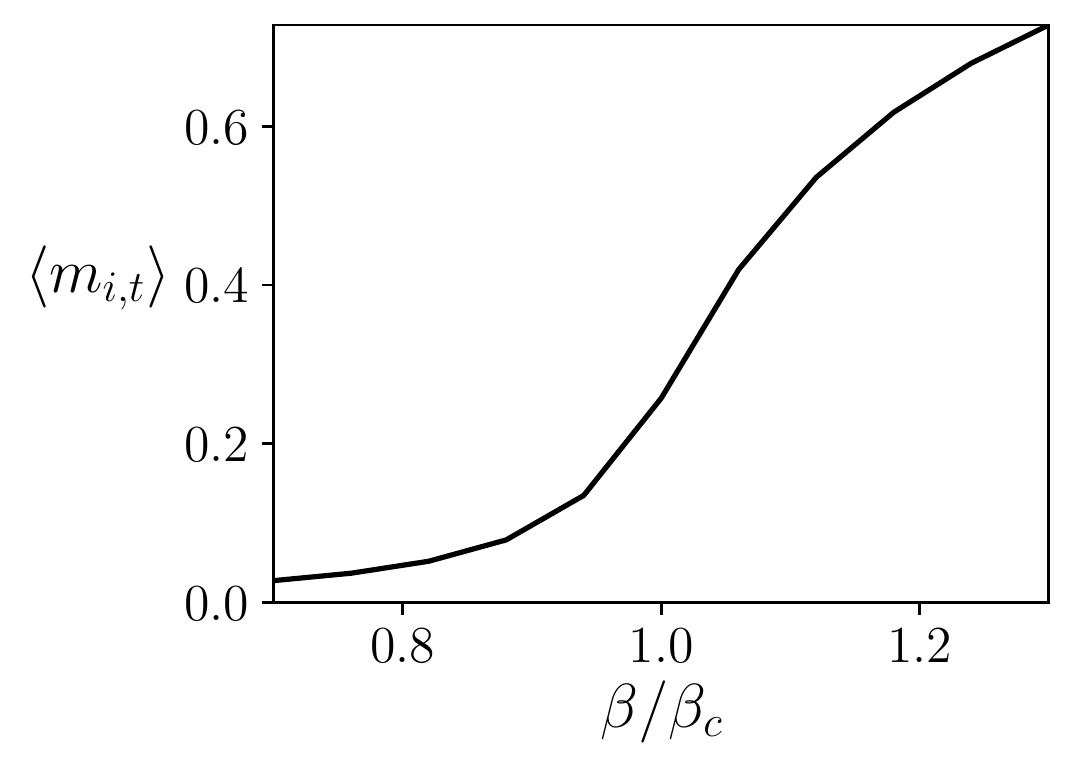}&
  \includegraphics[width=5.0cm]{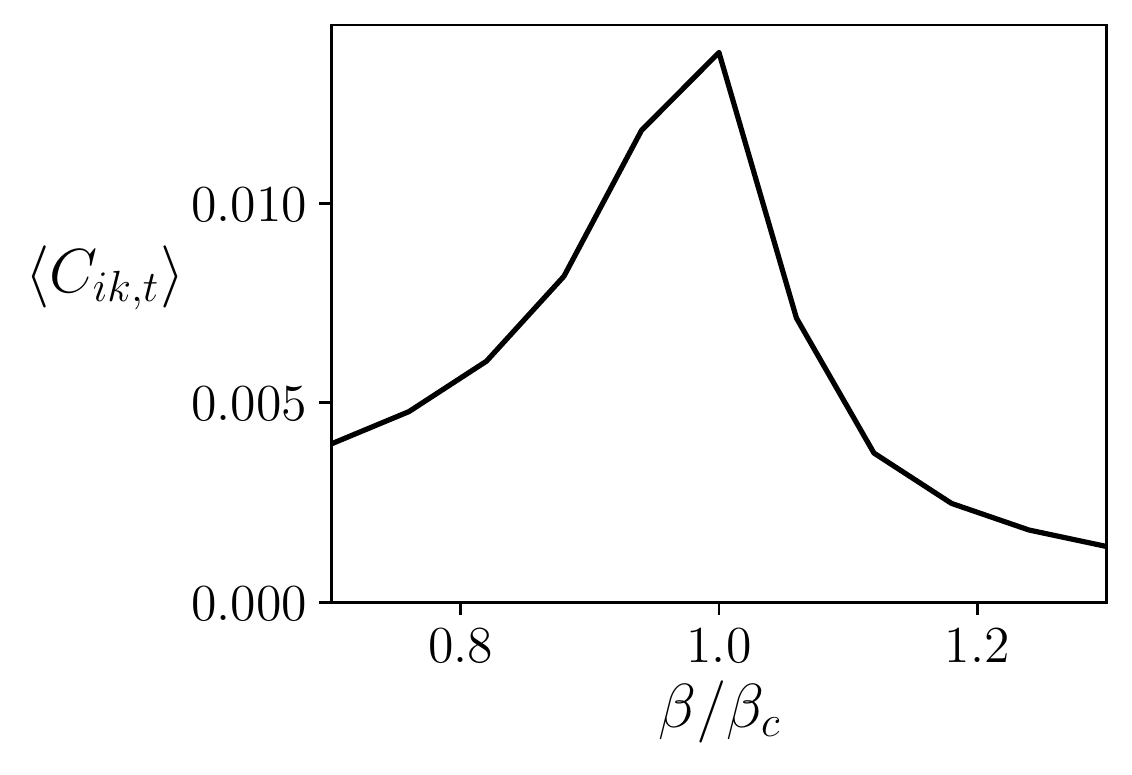} &
  \includegraphics[width=5.0cm]{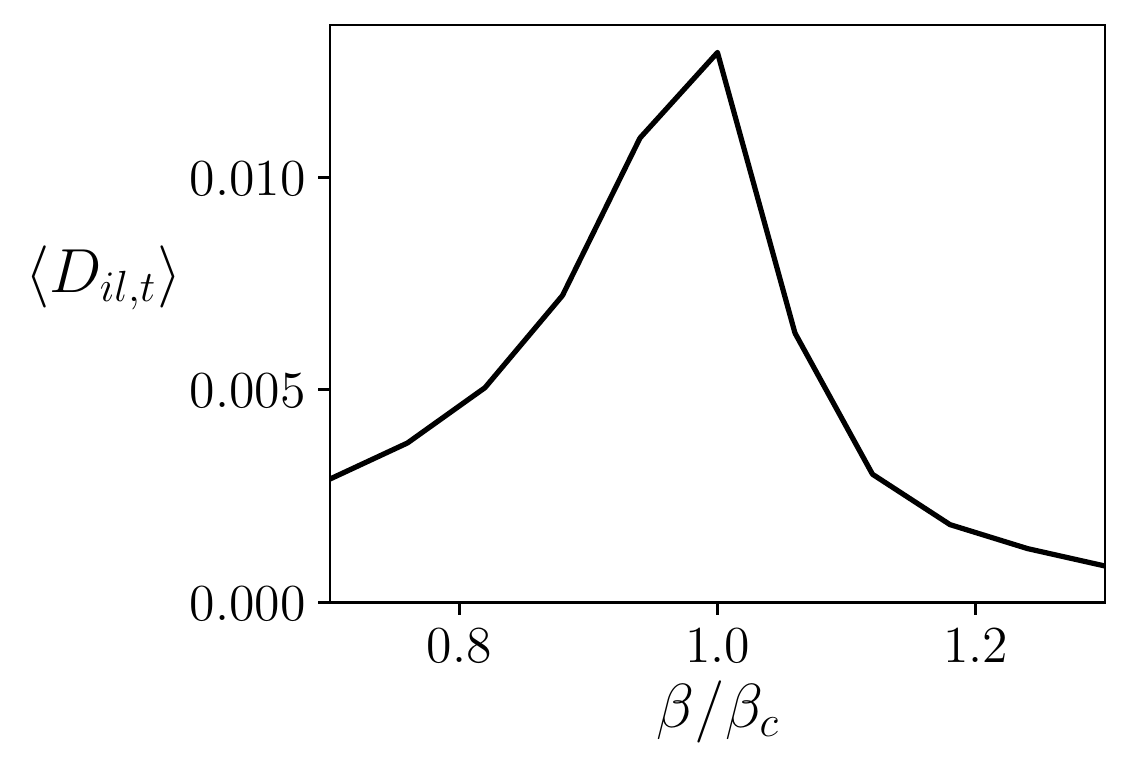} 
\end{tabular}
\end{center}
\caption{\textbf{Finite-size simulation of the asymmetric Ising model}. Experimental results of the behaviour around the critical point ($\beta_c=1.1108$) of an asymmetric Ising model with $N=512$ and parameters $H_0=0.5$, $J_0=1$ and $J_\sigma=0.1$.
} 
\label{fig:exp-ising}
\end{figure}

\newpage
\section{Complexity of the models and their computational costs}

Here we clarify the structure of each model, and quantitatively compare the relation between model complexity and their computational cost. We start by noting that the reference model in Plefka[$t-1,t$] is a submodel of the reference models in the other three approximation models used in Plefka[$t$], Plefka[$t-1$], and Plefka2[$t$]. The latter approximation models (Plefka[$t$], Plefka[$t-1$], Plefka2[$t$]) are not mutually inclusive, rather each model manifests a distinct assumption. 

To clarify the relation between the performance of each approximation and its model complexity, we provide the number of free and fixed parameters of each model. We then examine the relation between accuracy of the models in forward/inverse Ising problems and their computational time required to solve these problems. 

\subsection*{Number of free and fixed parameters in the models}

To compare the structure of each approximated model, let us first revisit the definition of each model, and introduce the number of their free or fixed parameters. Each approximation of $P$ (a marginal distribution of spins covering some time steps $t$ of the model) defines a reference manifold using $Q$ to perform a Plefka expansion. $Q$ is defined at each time point, and obtained by fixing some parameters (couplings) of $P$ at zero. The rest of the parameters are either free parameters that are fitted as an m-projection from $P$, or preserved as the original values of $P$. In the Supplementary Table \ref{tab:num_parameters}, we listed the number of free and fixed (either to zero or to their original value) parameters of reference models. For example, in Plefka[$t$], we preserve one field parameter ($H_i$) and $N$ couplings ($J_{ij}$) for each $i$-th spin at time $t-1$ with their original values, while fixing $N$ couplings of the $i$-th spin at time $t$ at zero. A field parameter of the $i$-th neuron at time $t$ ($\Theta_{i,t}$) is a free parameter obtained as an m-projection from $P$ (i.e., it is fitted so that $Q$ and $P$ have the same expectation $\*m$). We need to perform this projection for all neurons (i.e., $N$ times).

\bigskip

\begin{table}[ht]
    \centering
    \begin{tabular}{ m{2cm} || m{3.5cm}| m{3cm} | m{3cm}| m{3.5cm} |}

     & Number of times the approximation is computed
     & Free parameters to be fitted (at each approximation)
     & Parameters fixed at zero (at each approximation)
     & Parameters fixed at their original value (for each approximation)
    \\  \hline \hline
    Plefka[$t-1,t$]
     & $2N$ ($N$ neurons at time $t$ and $N$ neurons at time $t-1$)
     & $1$ field
     & $N$ couplings
     & $0$
     \\     \hline
    Plefka[$t$]
     & $N$ ($N$ neurons at time $t$)
     & $1$ field
     & $N$ couplings
     & $1$ field + $N$ couplings (for $N$ neurons at $t-1$)
     \\     \hline
    Plefka[$t-1$]
     & $N$ ($N$ neurons at time $t-1$)
     & $1$ field
     & $N$ couplings
     & $1$ field + $N$ couplings (for $N$ neurons at $t$)
     \\     \hline
    Plefka2[$t$]
     & $2 N^2$ (once for every pair of neurons at time $t$ once for every pair at $t,t-1$)
     & $2$ fields (time $t$ and $t-1$) and $1$ coupling
     & $N-1$ couplings at $t$, $N$ couplings at $t-1$
     & $1$ field and $N$ couplings (for $N-1$ neurons except for $l$ at $t-1$)
     \\     
     \hline 
    \end{tabular}
    \caption{\textbf{Comparison of methods}. The number of free and fixed parameters of each approximation model.}
    \label{tab:num_parameters}
\end{table}

\subsection*{Accuracy versus computational costs}

The structure of the reference model $Q$ influences both the accuracy of the approximation methods and their computational complexity. For example, Plefka[$t$] is computationally more demanding than Plefka[$t-1,t$] since the former requires to compute multiplications over the correlation matrices $\*C_{t-1}$, while the latter only makes use of the means $\*m_{t-1}$.  
Plefka[$t-1$] involves performing a Gaussian integral, which becomes computationally demanding, especially for the equal-time correlations, where one needs to perform a 2-dimensional integral. 
Plefka2[$t$] is computationally more expensive than Plefka[$t$] because, while the latter needs to be computed $N$ times (once per spin), the former needs to be computed $2N^2$ times (once per each pair of delayed and same-time spins).

In Supplementary Fig.~\ref{fig:results-execution-time-forward}, we show the trade-off between computational time and accuracy for the different methods. These results are obtained by simulating the model for $T=128$ steps at $\beta=\beta_c$ (in principle the most challenging point of our model). We computed the average squared error of its statistical moments and cumulants $\epsilon_{\*{m}}= \langle \langle (m^o_{i,t} - m^p_{i,t})^2 \rangle_{i} \rangle_t$, $\epsilon_{\*{C}}= \langle \langle (C^o_{ik,t} - C^p_{ik,t})^2 \rangle_{ik} \rangle_t $ and $\epsilon_{\*{D}}= \langle \langle (D^o_{ik,t} - D^p_{ik,t})^2 \rangle_{il} \rangle_t$.

As expected, Plefka[$t-1,t$] is the fastest method, although it yields the most inaccurate approximation. Plefka[$t$] significantly improves accuracy with a slight increase in computational time. Plefka2[$t$] increases further both computational cost and accuracy. Finally, Plefka[$t-1$] is relatively slow compared to the other methods, offering a poor performance in the forward Ising problem. In sum, we recommend to use Plefka2[$t$] if users pursue the accuracy, and Plefka[$t$] if users pursue the speed in the forward Ising problem. 

\bigskip 

\begin{figure}[ht]
\centering{
\includegraphics[width=18cm]{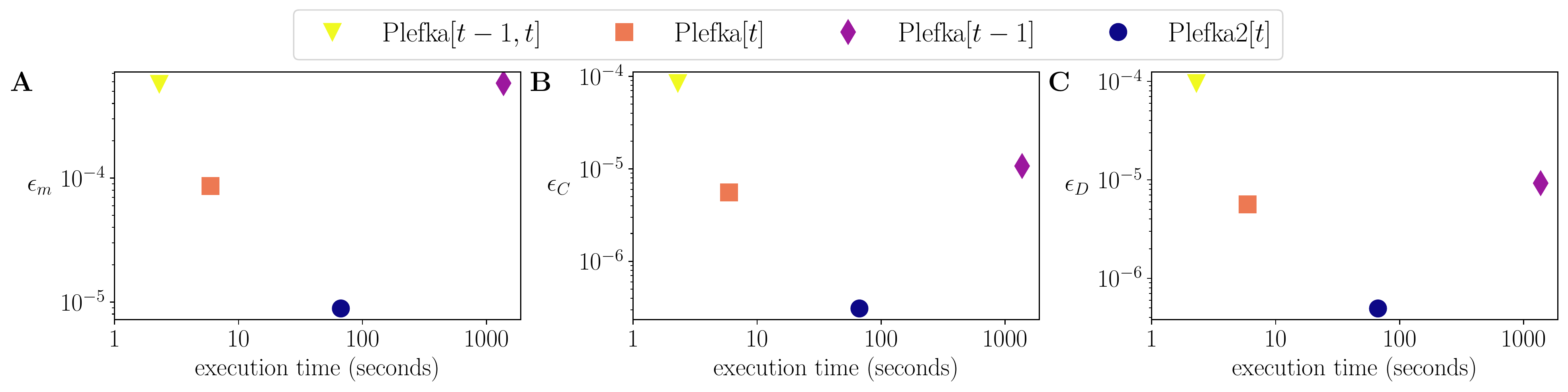}
}
\caption{\textbf{Accuracy vs computational time in the forward Ising problem}.
Average squared error in (A) the Ising model average magnetizations, (B) equal-time correlations and (C) delayed correlations at $t=1,\dots,128$ versus computation time in minutes for different mean-field models for $\beta=\beta_c$. Results were obtained using a 2-core Intel(R) Core(TM) i7-5500U CPU @ 2.40GHz processor.
} 
\label{fig:results-execution-time-forward}
\end{figure}

In Supplementary Fig.~\ref{fig:results-execution-time-inverse}, we show the results for the inverse Ising problem at $\beta=\beta_c$. The (in)accuracy is evaluated by  the average squared error between the inferred parameters and true parameters: $\epsilon_{\*{H}}= \langle \langle (H^o_{i,t} - H^p_{i,t})^2 \rangle_{i} \rangle_t$ and $\epsilon_{\*{J}}= \langle \langle (J^o_{ik,t} - J^p_{ik,t})^2 \rangle_{il} \rangle_t$. We found that the results are different from the forward Ising model. In the inference problem, Plefka[$t-1,t$] shows a poor performance in accuracy as in the forward problem, but it was no longer the fastest method. Plefka[$t$], Plefka[$t-1$] and Plefka2[$t$] offer similar performances.

There are several reasons why the results in the inverse Ising problem differ from the previous forward problem. First, the computational time in the inverse problem depends on how fast the learning dynamics reach convergence. These dynamics differ across the methods (e.g., due to the slowest speed of convergence, Plefka[$t-1,t$] is slightly slower than Plefka[$t$]). In addition, we need to compute only $\*m_t$ and $\*D_t$, and do not need to compute $\*C_t$ in the inverse problem. Since Plefka[$t-1$] is very slow in computing $\*C_t$, by freeing from its computational load, Plefka[$t-1$] becomes much faster in the inverse problem.

\bigskip

\begin{figure}[ht]
\centering{
\includegraphics[width=12cm]{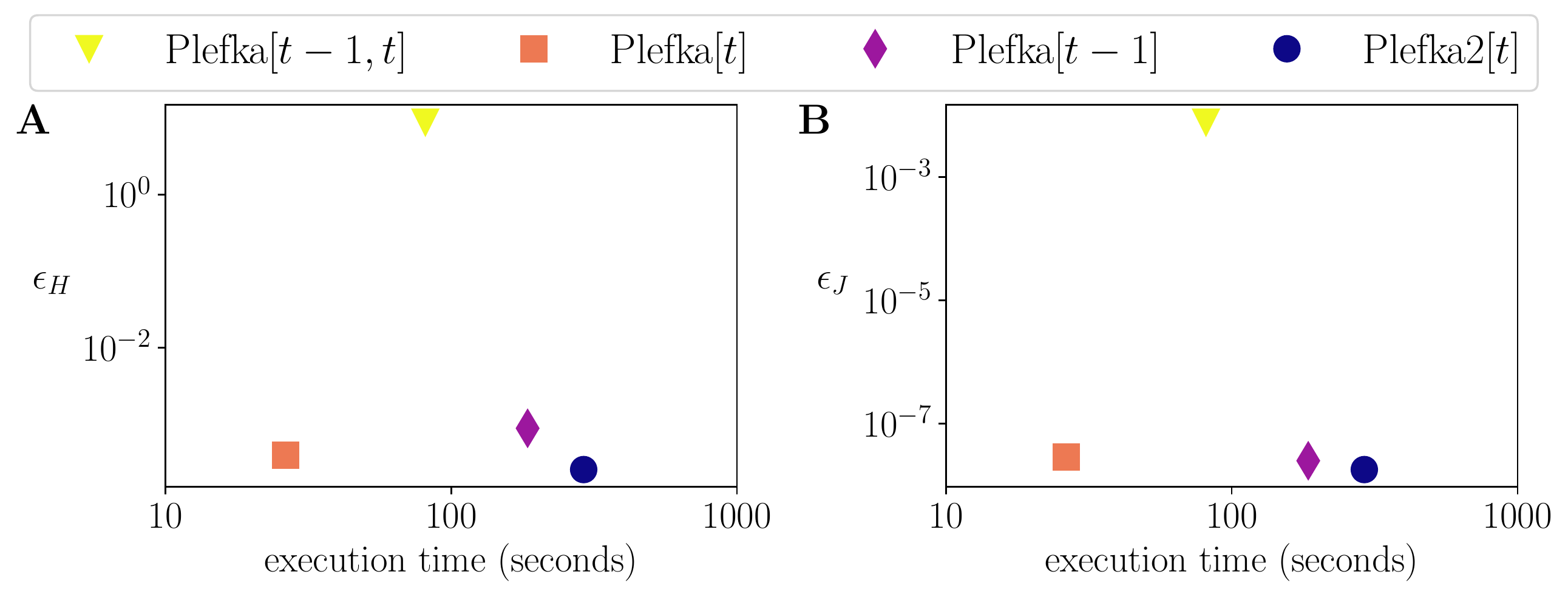}
}
\caption{\textbf{Accuracy vs computational time in the inverse Ising problem}. 
Average squared error in the inverse Ising inferred (A) external fields and (B) couplings versus computation time in minutes for different mean-field models for $\beta=\beta_c$. Results were obtained using a 2-core Intel(R) Core(TM) i7-5500U CPU @ 2.40GHz processor.
} 
\label{fig:results-execution-time-inverse}
\end{figure}


\end{document}